\documentclass{uofsthesis-cs}
\usepackage{graphicx}
\usepackage{booktabs}
\usepackage{multirow}
\usepackage{pgfplots, pgfplotstable}
\usepackage{scalefnt}
\usepackage{tikz}
\usetikzlibrary{shapes.gates.logic.US,trees,positioning,arrows}
\usetikzlibrary{shapes.multipart}
\usetikzlibrary{shapes,arrows}
\usetikzlibrary{patterns}
\usepgfplotslibrary{statistics}
\pgfplotsset{compat=1.8}
\usepackage{makecell}

\usepackage{amssymb,amsmath,amsthm}     
\usepackage{xcolor}
\usepackage{comment}
\usepackage{rotating}

\usepackage{listings}
\usepackage{color}

\definecolor{mygreen}{rgb}{0,0.6,0}
\definecolor{mygray}{rgb}{0.5,0.5,0.5}
\definecolor{mymauve}{rgb}{0.58,0,0.82}

\usepackage{xcolor}

\colorlet{punct}{red!60!black}
\definecolor{background}{HTML}{EEEEEE}
\definecolor{delim}{RGB}{20,105,176}
\colorlet{numb}{magenta!60!black}

\lstdefinelanguage{json}{
    basicstyle=\normalfont\ttfamily,
    numbers=left,
    numberstyle=\scriptsize,
    stepnumber=1,
    numbersep=8pt,
    showstringspaces=false,
    breaklines=true,
    frame=lines,
    backgroundcolor=\color{background},
    literate=
     *{0}{{{\color{numb}0}}}{1}
      {1}{{{\color{numb}1}}}{1}
      {2}{{{\color{numb}2}}}{1}
      {3}{{{\color{numb}3}}}{1}
      {4}{{{\color{numb}4}}}{1}
      {5}{{{\color{numb}5}}}{1}
      {6}{{{\color{numb}6}}}{1}
      {7}{{{\color{numb}7}}}{1}
      {8}{{{\color{numb}8}}}{1}
      {9}{{{\color{numb}9}}}{1}
      {:}{{{\color{punct}{:}}}}{1}
      {,}{{{\color{punct}{,}}}}{1}
      {\{}{{{\color{delim}{\{}}}}{1}
      {\}}{{{\color{delim}{\}}}}}{1}
      {[}{{{\color{delim}{[}}}}{1}
      {]}{{{\color{delim}{]}}}}{1},
}

\newtheorem*{remark}{Definition}

\title{An Intermediate Data-driven Methodology for Scientific Workflow Management System to Support Reusability}

\newcommand{\joy}[1]{\textcolor{black!100}{#1}}

\author{Debasish Chakroborti}

\degree{\MSc}         

\defencedate{August/2019}

\abstract{
Automatic processing of different logical sub-tasks by a set of rules is a workflow. A workflow management system (WfMS) is a system that helps us accomplish a complex scientific task through making a sequential arrangement of sub-tasks available as tools. Workflows are formed with modules from various domains in a WfMS, and many collaborators of the domains are involved in the workflow design process. 
Workflow Management Systems (WfMSs) have been gained popularity in recent years for managing various tools in a system and ensuring dependencies while building a sequence of executions for scientific analyses. As a result of heterogeneous tools involvement and collaboration requirement, Collaborative Scientific Workflow Management Systems (CSWfMS) have gained significant interest in the scientific analysis community.  In such systems, big data explosion issues exist with massive velocity and variety characteristics for the heterogeneous large amount of data from different domains. Therefore a large amount of heterogeneous data need to be managed in a Scientific Workflow Management System (SWfMS) with a proper decision mechanism. Although a number of studies addressed the cost management of data, none of the existing studies are related to real-time decision mechanism or reusability mechanism. 
Besides, frequent execution of workflows in a SWfMS generates a massive amount of data and characteristics of such data are always incremental. Input data or module outcomes of a workflow in a SWfMS are usually large in size. Processing of such data-intensive workflows is usually time-consuming where modules are computationally expensive for their respective inputs. Besides, lack of data reusability, limitation of error recovery, inefficient workflow processing, inefficient storing of derived data, lacking in metadata association and lacking in validation of the effectiveness of a technique of existing systems need to be addressed in a SWfMS for efficient workflow building by maintaining the big data explosion.
 
To address the issues, in this thesis first we propose an intermediate data management scheme for a SWfMS. 
In our second attempt, we explored the possibilities and introduced an automatic recommendation technique for a SWfMS from real-world workflow data (i.e Galaxy \cite{Afgan2016TheUpdate} workflows) where our investigations show that the proposed technique can facilitate 51\% of workflow building in a SWfMS by reusing intermediate data of previous workflows and can reduce 74\% execution time of workflow buildings in a SWfMS. Later we propose an adaptive version of our technique by considering the states of tools in a SWfMS, which shows around 40\% reusability for workflows. Consequently, in our fourth study, We have done several experiments for analyzing the performance and exploring the effectiveness of the technique in a SWfMS for various environments. 

The technique is introduced to emphasize on storing cost reduction, increase data reusability, and faster workflow execution, 
to the best of our knowledge, which is the first of its kind. Detail architecture and evaluation of the technique are presented in this thesis. We believe our findings and developed system will contribute significantly to the research domain of SWfMSs. 
}

\acknowledgements{
First and foremost, I have to thank my research supervisors, Dr. Chanchal K. Roy and Dr. Kevin A. Schneider for continuous support with their immense knowledge and patience during my MSc study. 

I express my heartfelt gratitude to Dr. Banani Roy, who made this work possible. Her guidance helped me in formulating and shaping up the thesis to its current form. I could not have imagined having a better advisor and mentor for my MSc study.

I would also wish to express my gratitude to Dr. Manishankar Mondal for extended discussions, valuable suggestions, passionate participation, and input, which have contributed greatly to the improvement of the thesis.

 Besides my advisors, I would like to thank the rest of my thesis committee: Dr. Mark Keil, Dr. Ralph Deters, and Dr. Aryan Saadat Mehr, for their willingness to take part in the advisement and evaluation of my thesis work.

I would like to express my special appreciation and thanks to the anonymous reviewers for their valuable comments and suggestions on improving the papers produced from this thesis. 

I thank my fellow labmates of the Software Research Lab for their stimulating discussions, important feedback on my work, and for all the fun we have had in the last two years. In particular, I would like to thank Shamima Yeasmin, CM Khaled Saifullah, Kawser Wazed Nafi, Amit Kumar Mondal, Masud Rahman, Muhammad Asaduzzaman, Avijit Bhattacharjee, Md Nadim, Saikat Mondal, Judith Islam, Tonny Kar, Hamid Khodabandeloo, Zonayed Ahmed, and Golam Mostaeen.

I am grateful to the Department of Computer Science of the University of Saskatchewan for their generous financial support through scholarships, awards, and bursaries that helped me to concentrate more deeply on my thesis work. 

I am also grateful to all the staffs of the Department for their constant support. In particular, I would like to thank Gwen Lancaster, Findlay Sophie, and Heather Webb.

Getting through my thesis required more than academic support, and I have many, many people to thank for listening to and, at times, having to tolerate me over the past two years. I cannot begin to express my gratitude and appreciation for their friendship. Paromita Sengupta, Mashrafi Haider Iqra and Sakib Mostafa, Mike SB have been unwavering in their personal and professional support during the time I spent at the University. For many memorable evenings out and in, I must thank everyone above as well as Sanchita Sen, Purnendu B. Sen and Asif Anjum Khan, Yeasir Jubaer. I would also like to thank Pritam Sen, who opened both his home and help to me when I first arrived in the city.

I would especially like to thank all my seniors who have been extremely helpful and encouraging throughout the period of my thesis work. In particular, I would like to thank Md. Sami Uddin and Md. Aminul Islam.

And finally, last but by no means least, I would like to thank my family: my father Rajendranath Chakroborti, my mother Protima Chakroborti, my wife Sristy Sumana Nath, my daughter Ushasi Srija Chakroborti, my elder brother Shamiron Chakroborti elder sister Mousumi Chakroborti Moumita Chakroborti for their unconditional support, encouragement, inspiration, and love at every stage of my life. Without their endless sacrifice, I would not have come this far.
}

\dedication{I dedicate this thesis to my mother, Protima Chakroborti, my wife, Sristy Sumana Nath, and my daughter, Ushasi Srija Chakroborti,  whose influence helps me to accomplish every step of my life.}

\loa{\abbrev{WfMS}{Workflow Management System}
\abbrev{SWfMS}{Scientific Workflow Management System}
\abbrev{CSWfMS}{Collaborative Scientific Workflow Management Systems}
\abbrev{RISP}{Recommending Intermediate States from Pipelines}
\abbrev{NASA-TLX}{The NASA Task Load Index}
\abbrev{DAG}{Directed Acyclic Graph}
\abbrev{RDD}{Resilient Distributed Datasets}
\abbrev{HDFS}{Hadoop’s Distributed Filesystem}
\abbrev{SHIPPI}{Spark High-throughput Image Processing Pipeline}
\abbrev{ID}{Intermediated Data}
\abbrev{P2IRC}{Plant Phenotyping and Imaging Research Centre}
\abbrev{LRWoI}{Leaves Recognition workflow without storing intermediate data}
\abbrev{LRWtI}{Leaves Recognition workflow with storing intermediate data}
\abbrev{LRSD}{Leaves Recognition workflow by skipping descriptor operation}
\abbrev{SWoI}{Segmentation workflow without storing intermediate data}
\abbrev{SWtI}{Segmentation workflow with storing intermediate data}
\abbrev{SSTA}{Segmentation workflow by skipping transformation and analysis}
\abbrev{CWoI}{Clustering workflow without storing intermediate data}
\abbrev{CWtI}{Clustering workflow with storing intermediate data}
\abbrev{CSTA}{Clustering workflow by skipping transformation and analysis}
\abbrev{PT}{Proposed Technique-RISP}
\abbrev{TSAR}{Technique that recommends Storing All Results}
\abbrev{TSPAR}{Technique that recommends Storing Previously Appeared Results}
\abbrev{TSFR}{Technique that recommends Storing the Final Result}
\abbrev{LR}{Likeliness of Reusing from previously stored results}
\abbrev{PSRR}{Percentage of Stored Results that were Reused}
\abbrev{FRSR}{Frequency of Reusing Stored Results}
\abbrev{PISRS}{Percentage of Intermediate State Results that were Stored}
\abbrev{GUI}{Graphical User Interface}
\abbrev{APM}{Application Performance Monitoring}
}

\begin{document}
\maketitle

\frontmatter


\chapter{Introduction}

In this chapter, a brief introduction of the thesis is presented. Section \ref{motivation} is presented with the main motivation of the thesis. 
In Section \ref{problemstatement} and \ref{ourcontribution}, we define the problems and describe our contributions towards mitigating the problems.
In Section \ref{publications}, all the publications related to the thesis are listed to show our contribution to the Scientific Workflow Management domain. Finally, in section \ref{thesisoutline}, outlines of the remaining chapters of the thesis are addressed. 

\section{Motivation}
\label{motivation}
A Workflow Management System (WfMS) gives a foundation to sequences of predefined tasks known as workflows for executing, monitoring, and performance-enhancing. Similarly, a Scientific Workflow Management System (SWfMS) is a specialized form of WfMS where a sequence of computational or data manipulation modules are composed and executed for scientific analyses (i.e., discussed details in chapter \ref{main_background}). Engaging and producing a large amount of data from various sources with proper dataflow for different technologies influence researcher of many domains to move towards analyses of data-driven methodologies of sequential processing. SWfMSs have gained popularity in recent years to manage such sequential tasks efficiently. Besides visualization, monitoring, and error recovery can be supported in a SWfMS as a compact solution with proper technologies. Frequently executed workflows with data or computationally intensive modules in a SWfMS require special care to reduce time consumption and complexity. Thus most of the SWfMS are evolved with the basic supports such as specification, modification, execution, failure recovery, and monitoring. Furthermore, for the incremental data from the frequently executed workflows need to be handled with proper data management scheme. For instance, in a study \cite{Woodman2015WorkflowProvenance} of cost management of SWfMSs entitled as ``Workflow Provenance: An Analysis of Long Term Storage Costs", authors showed that a significant cost is required for storing the data produced from a SWfMS. Besides, there are some other studies \cite{Yuan2011On-demandSystems} \cite{Koop2008VisComplete:Pipelines} that showed that special algorithms are required to manage data in a SWfMS. Plant Phenotyping and Genotyping where both image processing and bioinformatics analysis are mandatory, a SWfMS is required to establish a correlation among heterogeneous tools. In such a system, a proper data management scheme is essential to maintain the cost of the incremental heterogeneous data in a storage system. 

In the past few years, a significant number of research experiments and developments such as \textit{ Anduril,   Apache Taverna, Galaxy, Kepler, KNIME, Pegasus, VisTrails, and so on} have been done in the area of Scientific Workflow Management Systems for bioinformatics, image analysis, astronomy, biodiversity, and genomic domain. However, none of them considered the management of intermediate data produced from executed workflows. Furthermore, there is no study that considered data reusability in a SWfMS. Besides, consideration of the predefined cost of data for a data management technique of a SWfMS might not be useful for future reuse. 

A great number of studies \cite{article34636} \cite{4404805}  \cite{4534284} \cite{Simmhan:2005:SDP:1084805.1084812} \cite{10.1007/11890850_2} \cite{Gray:2005:SDM:1107499.1107503} \cite{Davidson:2008:PSW:1376616.1376772} have been done in recent years for supporting a data management technique in a SWfMS. Several studies \cite{4534284} \cite{Woodman2015WorkflowProvenance} \cite{Yuan2011On-demandSystems} explored the possibilities of optimal cost of data in a management technique for a SWfMS. In the era of big data, researchers are focusing more on the optimal cost of data management techniques and other areas of big data applications, which is also considered in our proposed management technique with increased reusability of data in a SWfMS.

\section{Problem Statement}
\label{problemstatement}
In a SWfMS, repetitive nature of most of the tasks makes researchers speculate on reusable units. Most of the tasks are divided into modules to make them reusable in a SWfMS. Typically, SWfMSs provide both local scripts and web services as modules, and they are reusable in different workflows. However, outcomes of the modules in a workflow are not considered as reusable units in the existing studies, which cause performance degradation. In the following, we discuss the shortcoming in details.

\textbf{Problem Statement 1} (Lack of data reusability and limitation of error recovery in workflow composition and execution): Data reusability in a SWfMS is not reflected significantly by researchers except some data provenance studies.  Many provenance mechanisms \cite{4404805} \cite{4534284} \cite{Simmhan:2005:SDP:1084805.1084812} \cite{Davidson:2008:PSW:1376616.1376772} \cite{Software_Architecture} use storing procedure for streamed data in a SWfMS to keep traces for checkpoints. However, error recovery and rollback are possible up to a certain level (i.e., for a specific session) by using those provenance model, but reusability of the data (i.e., intermediate data) is not possible after a certain period. Hence processed data storing mechanism in a SWfMS is essential and one of the important research problems in the domain. For addressing the problem, two research questions need to be answered before going to design a data management scheme in a SWfMS such as ``Does the system support data reusability that ultimately reduces time and efforts?" and ``How the system is supporting fast processing for frequently used workflows?". Furthermore, a number of studies \cite{crawl2008provenance} \cite{sindrilaru2010fault} \cite{mouallem2010fault} \cite{zhao2011opportunities} \cite{talia2013workflow} showed that errors are common in a SWfMS. Scientific workflows are incorporated with different active components such as associated module ports, scripts, datalink, intermediate data, provenance information, and so on \cite{10.1007/11890850_2}. Errors can frequently occur in workflows for their complex structure, association, compatibility among various components. Various mechanisms have been proposed \cite{crawl2008provenance} \cite{sindrilaru2010fault} \cite{mouallem2010fault} for error recovery and rollback with checkpoints for workflows in a SWfMS. However, many things need to be addressed, and a few important research questions need to be answered while designing a management system. Such as ``Does it have any error recovery mechanism that could construct workflows quicker if errors occur?", and ``How long the technique persists for other users in a system?". 

\textbf{Problem Statement 2} (Inefficient processing of frequent workflows and storing of modules outcomes): A great number of studies \cite{Altintas:2006:PCS:2165554.2165572} \cite{Davidson:2008:PSW:1376616.1376772} \cite{10.1007/11890850_2} \cite{scheidegger2008querying} on efficient processing show the needs for a desired techniques of execution in a SWfMS. However, the existing techniques are not persistent after a certain period. Consequently, there is no study on optimal data storing of a decision mechanism in a SWfMS. Although a number of studies of the provenance model follow a fully reproducible workflow mechanism, we need to store all of the generated data for using the model in a SWfMS. There is no study on storing a minimum amount of data for maximum reuse in future. This problem is severe when input data sets are large, and modules are computationally expensive. Thus a few questions need to be answered in this context such as ``What is gain in execution time by using a technique?", ``How often can we reuse components for efficiency?", ``How can we measure the efficiency of a technique?", ``Is the system costly in terms of storage?". 

\textbf{Problem Statement 3} (Lacking in metadata apposition with tool outcomes in a SWfMS): Although taxonomy of metadata and usage of metadata for raw data management are considered in number of studies \cite{4534284} \cite{belhajjame2008metadata} \cite{4782949} \cite{Big_Metadata} \cite{Distributed_metadata} \cite{Towards_Multi-site} \cite{Managing_hot_metadata} \cite{Unlocking_the}, intermediate data management is never been proposed with metadata management for a SWfMS. How tool states can be formed in a SWfMS is discussed in existing studies, but the effect of them on intermediate data is not discussed in the study. Configuring and tuning parameters in a workflow can have a significant impact on certain types of model that can change the states of processes (i. e. modules). Different outputs might be produced for a sequence of modules with identical input datasets in different workflows for a different set of parameter configuration. ``How can we handle the recommendation considering tool states in a SWfMS?", and ``How much data are reusable if we consider tool state?" are need to be answered before designing a recommendation technique with tool states.     

\textbf{Problem Statement 4} (Lacking in validation of the effectiveness of a methodology): Scientists from different areas are working on SWfMS, i.e., from Bioinformatics to Image Processing. Various tools and technologies are used in a SWfMS, and various methodologies  \cite{crawl2008provenance} \cite{Altintas:2006:PCS:2165554.2165572} \cite{Davidson:2008:PSW:1376616.1376772} \cite{10.1007/11890850_2} \cite{scheidegger2008querying} \cite{sindrilaru2010fault} \cite{mouallem2010fault} are proposed by researchers to increase efficiency. However, the existing studies are not discussed with quantitative analyses rather than some qualitative analyses to validate the effectiveness of the techniques. To test the effectiveness of a new methodology in a SWfMS, some factors need to be considered with user evaluation such as ``What is the overall performance of the methodology in a SWfMS?", ``How is the methodology being used by users in the system?", ``Whether designed tool using the methodology is helping to compose workflows efficiently?". In addition, research experiments on Collaborative Scientific Workflow Management Systems (CSWfMSs) \cite{zhang2012confucius} \cite{lu2009collaborative} are gained significant interest among researchers for real-time collaboration from different domain experts. However, for time-constrained, sometimes it may not be possible to do the workflow composition in real-time, how the situation can be handled need to be addressed and what percentages of components in a SWfMS are reusable or shareable for future use need to be explored.

\section{Our Contribution}
\label{ourcontribution}
To introduce an automated data management technique for SWfMSs by concentrating on the research problems, our studies carried on that contribute towards the following four sections. All the research problems and their solutions are summarized in Table \ref{tab:problemSolution} with the research questions. Major four contributions from our studies are briefly discussed in this section. 


\begin{table}[]
\caption{Research problems and associated solutions}
\label{tab:problemSolution}
\resizebox{\textwidth}{!}{
\begin{tabular}{|p{3cm}|p{8cm}|p{3cm}|}
\hline
Problem & Research Question & Solution \\ \hline
Lack of data reusability and limitation of error recovery in workﬂow composition and execution & Does the system support data reusability that ultimately reduces time and efforts?, How the system is supporting fast processing for frequently used workflows?, Does it have any error recovery mechanism that could construct workflows quicker if errors occur?, How long the technique persists for other users in a system?                     & Data Management Scheme for Micro-Level Modular Computation-intensive Programs\\ \hline
Inefficient processing of frequent workflows and storing of modules outcomes & What is gain in execution time by using a technique?, How often can we reuse components for efficiency?, How can we measure the efficiency of a technique?, Is the system costly in terms of storage? & A technique for optimized Storing of Workflow  Outputs\\ \hline
Lacking in metadata apposition with tool outcomes in a SWfMS & How can we handle the recommendation considering tool states in a SWfMS?, How much data are reusable if we consider tool state? & Optimal Storing Modes of Workflow Considering Tool State \\ \hline
Lacking in validation of the effectiveness of a methodology & What is the overall performance of the technique/scheme in a SWfMS?, How is the technique being used by users in the system?, Whether designed tool using the methodology is helping to compose workflows efficiently? & Usability Study of Recommending Intermediate States \\ \hline
\end{tabular}
}
\end{table}

\subsubsection{Data Management Scheme for Micro-Level Modular Computation-intensive Programs} A great number of studies have been done on data management in SWfMs for efficient handling of data while building workflows. In the studies of data management in SWfMSs, cost management is a vital issue and how users can store various data from workflows efficiently for associating them in later is studied in many studies with efficient algorithms \cite{4534284} \cite{Woodman2015WorkflowProvenance} \cite{Yuan2011On-demandSystems}. These management techniques help users to manage incremental data automatically in a SWfMS at low cost.  A number of studies have been done on metadata management, execution log management, and raw data management \cite{4534284} \cite{belhajjame2008metadata} \cite{4782949}. However, intermediate data from different modules of workflows in a scientific workflow management system are huge in size and incremental. Only for data provenance and introducing reproducibility, some of the systems manage them by storing all. There is no automatic process to store them in a SWfMS in an optimal way. In order to introduce such a management technique and explore the possibility of intermediate data for computationally expensive distributed programs, we propose a data management scheme that can manage intermediate data for efficient execution, error recovery and increased reusability. Using the proposed technique, user can gain a significant amount of time in execution while building workflows. We conducted several experiments in both distributed environment and local environment by executing workflows from our SWfMS to compare and explore the time gain possibilities. Our experiments show that the proposed technique can gain 87\% in execution time in a SWfMS while building workflows. Also, few significant research questions and questions in Problem Statement 1 of Section \ref{problemstatement} are answered in the study for future direction and an automatic approach of intermediate data management. We present this study in details in Chapter \ref{ManagementScheme}.

\subsubsection{A technique for optimized Storing of Workflow  Outputs} Although a data management scheme is essential in a SWfMS for efficient associations and using the data in various workflows. However, an automatic technique is profoundly needed for decision making to store and retrieve data in the runtime of workflows in a SWfMS. Furthermore, the technique should suggest an optimal solution for workflow outcomes to store as there can be multiple outcomes from a particular module of different workflows. For optimal storing of workflow outputs, few systems consider storing workflow output with minimum cost \cite{4534284} \cite{Woodman2015WorkflowProvenance} \cite{Yuan2011On-demandSystems}. In this study, we have emphasized on reusability to store outcomes using a technique that ultimately works towards an optimal solution. We used mining association rules for the technique to associate data and modules in a SWfMS.  We present our study by exploring real-life Galaxy \cite{Afgan2016TheUpdate} workflows to find an optimal solution for increasing data reusability in a SWfMS and answered the research questions of Problem Statement 2 in Section \ref{problemstatement}. Details of this study are presented in Chapter \ref{optstoring}.

\subsubsection{Optimal Storing Modes of Workflow Considering Tool State} By analyzing different workflows from various commonly used SWfMSs of MyExperiment \cite{goble2010myexperiment}, we realize that tool state needs to be considered while recommending intermediate states for modules. The reason behind that different parameter configurations of a tool while composing workflows in a SWfMS make different states of the tool that could generate varying outcomes. Considering the tool state, we propose an adaptive version of the RISP that can work with parameter configuration and tuning in a SWfMS. In previous studies \cite{10.1007/11890850_2} \cite{Deelman2008DataWorkflows} \cite{4782949} \cite{SEFFINO1999105} tool state is considered to introduce proper reproducibility. Storing of parameter configuration and management of them are also addressed in such studies.  In our proposed approach, we mainly considered the effects of parameter configuration for outcomes (i.e., intermediate data) of modules as we are proposing a technique of data management for the outcomes of modules. We present different experimental studies from 534 workflows of a SWfMS (i. e., Galaxy) for evaluating the technique. We present the studies with different metrics to ensure reusability in a SWfMS and to answer the research question presented in Problem Statement 3 of Section \ref{problemstatement}. We present the details of this study in Chapter \ref{OptimalMode}.

\subsubsection{Usability Study of Recommending Intermediate States} We conducted several user studies to see the user behavior with the technique in a SWfMS and ensure the validity of the technique in a SWfMS. In our user studies, participants were asked to compose and execute workflows where Google Analytics and Tobii Eye Tracing device are used to trace all the activities of the participants. From the data, several interesting facts are revealed that help us to answer the research questions in Problem Statement 4 of Section \ref{problemstatement}. Details of this study are presented in chapter \ref{DesignRISP}.

\section{Publications}
\label{publications}
Below is the list of publications and other works that are prepared for submission (with collaborator) from this thesis. 
\begin{itemize}
	\item Debasish Chakroborti, Banani Roy, Amit Mondal, Golam Mostaeen, Chanchal Roy, Kevin Schneider, and Ralph Deters, ``A Data Management Scheme for Micro-Level Modular Computation- intensive Programs in Big Data Platforms", International Symposium on Big Data Management and Analytics (BIDMA), University of Calgary, 2018.
	\item Debasish Chakroborti, Manishankar Mondal, Banani Roy, Chanchal K. Roy, Kevin A. Schneider, ``Optimized Storing of Workflow Outputs through Mining Association Rules", 2018 IEEE International Conference on Big Data (IEEE Big Data), pp. 508-515, Seattle, WA, USA, 2018.
	\item Debasish Chakroborti, Sristy Sumana Nath, Banani Roy, Chanchal K. Roy, Kevin A. Schneider, ``Optimal Storing Modes of Workflow in A Scientific Workflow Management System Considering Tool State", IEEE Journal - Transactions on Services Computing (to be submitted).
	\item Debasish Chakroborti, Banani Roy, Sristy Sumana Nath, Manishankar Mondal, Chanchal K. Roy, Kevin A. Schneider, ``Designing for Recommending Intermediate States in A Scientific Workflow Management System", IEEE Journal - Transactions on Services Computing (to be submitted).
\end{itemize}

\section{Thesis Outline}
\label{thesisoutline}
The thesis outline is described by giving some background knowledge of Workflows, Scientific Workflows, Scientific Workflow Management System and the required technologies (i.e., distributed environment, mining association rules and so on.) in Chapter \ref{main_background}.  We present our Data Management Scheme for SWfMS in Chapter \ref{ManagementScheme}. Chapter \ref{optstoring} is described with the technique (i. e. RISP) for optimized Storing of Workflow Outputs in a SWfMS. In Chapter \ref{OptimalMode}, an adaptive version of the RISP is presented by considering tool states in workflows with reusability analyses. We also developed a prototype of the technique and integrated the technique in a SWfMS. To ensure the effectiveness and efficiency of the technique, we conducted user studies in a SWfMS. Details of the user studies are presented in Chapter \ref{DesignRISP}. Lastly, Chapter \ref{main_conclusion} is presented with an overall summary of our work and concluded with some future directions.

\chapter{Background}
\label{main_background}

A brief discussion of the background and the required technologies of this thesis are discussed in this chapter. Section \ref{workflowwfms} defines the Workflow and Workflow Management System (WfMS) in generalized ways. In Section \ref{SwF} and \ref{SWfMS}, we present the general overview of Scientific Workflows and Scientific Workflow Management System (SWfMS). Section \ref{toolstech} is discussed with the required tool and technologies for this study.

\section{Workflow and Workflow Management System}
\label{workflowwfms}

Workflow is composed for the automation of a sequence of jobs where information or responsibilities are passed among shareholders by a predefined set of rules to reach or provide a particular goal \cite{hollingsworth1995workflow}. 
Workflow theory is developed from the concept of process in manufacturing and office work.  Work activities are separated into well-defined tasks, roles, rules, and procedures which control most of the work in manufacturing or office by a workflow \cite{georgakopoulos1995overview}. However, processes were previously carried out by human beings, but nowadays those are automated by the information system. So, the concept of workflow is similar to reengineering and automating business processes and information processes in an environment. Such a workflow comprises of cases, resources, and triggers that are related to a particular task. A workflow can be represented with the above-mentioned content as a Petri net with a specific workflow state \cite{van2004workflow}.

Workflow Management System is an arrangement that can fully define, manage, and execute “workflows” by means of software execution and the order of execution is operated by a computer representation of the workflow implementation \cite{hollingsworth1995workflow}. Workflow Management System (WfMS) is a foundation that supports reengineering and automation of the processes for a particular environment. I other words, a workflow management system (WfMS) is a software solution for the implementation of a workflow system that can support workflows to be executed in a particular environment. A WfMS is not a specific business solution, by configuring the system and its workflows, it can be turned into one that can be used for other purposes. In other words, a workflow management system is a generic application to support various workflows \cite{van2004workflow}. Additionally, WfMS coordinates the executions of its processes and supports fast design, interoperability among heterogeneous resources, autonomous components, distributed systems, ensures reliability, concurrency, and failures.

\section{Scientific Workflow}
\label{SwF}
Processing of different logical data processing activities by a set of rules and automating them make a workflow \cite{Liu:2015:SDS:2884709.2884750}. In scientific workflows, scientific data processing activities are assembled and automated to execute the activities to reduce the execution time \cite{Liu:2015:SDS:2884709.2884750}. Scientific Workflow is usually defined in a system as a DAG (Directed Acyclic Graph), where the nodes depict individual computational activities (i.e., known as modules) and the edges denote data flow and conditions among the modules \cite{Deelman:2015:PWM:2775768.2776457}. Scientific Workflow is gaining interest in scientific research communities that can be seen from various events such as Workshop, Forum, Research Projects Dealing, and organization funding \cite{ludascher2006scientific}. In addition,  for the opportunity of data-intensive operation, SWfMSs (e.g., Galaxy \cite{Afgan2016TheUpdate}, myGrid/Taverna \cite{oinn2004taverna}, Kepler \cite{ludascher2006scientific}, VisTrails \cite{callahan2006vistrails}, and so on) are becoming increasingly common \cite{Davidson:2008:PSW:1376616.1376772} for scientific data analysis. As well as the capabilities of handling heterogeneity of data made scientific workflows traditional to analyze scientific problem \cite{4404805}. 

\section{Scientific Workflow Management System}
\label{SWfMS}

A Scientific Workflow Management System (SWfMS) is a software solution to execute scientific workflows and manage scientific data sets in various computing environments \cite{Liu:2015:SDS:2884709.2884750} \cite{10.1007/978-3-540-68111-3_78}. Such a SWfMS can manage a scientific workflow all along its life cycle \cite{Liu:2015:SDS:2884709.2884750}. In other words, a Workflow Management System (WfMS) is an arrangement that can define, create, and manage the execution of workflows and a SWfMS is a WfMS that can handle and manage scientific workflow execution \cite{Liu:2015:SDS:2884709.2884750}. 
In addition, in a SWfMS automated technique and provenance tracking commit advantages for both composition and execution \cite{Davidson:2008:PSW:1376616.1376772} \cite{4404805}. By using such automate computational activities in SWfMS, many disciplines are benefiting such as astronomy, biology, chemistry, environmental science, engineering, geosciences, medicine, physics, and social sciences \cite{4404805}. Furthermore, a SWfMS needs to be reliable and efficient to coordinate and automate data flow and task execution in various environment such as in distributed computational, clusters, cyberinfrastructures, commercial and academic clouds environments \cite{Deelman:2015:PWM:2775768.2776457}. In a system of workflow management translation of tasks to jobs and execute them is not the only thing to be counted; data management, monitoring executions, and failures handling need to be considered for the reliability and efficiency \cite{Deelman:2015:PWM:2775768.2776457} \cite{4404805}

\section{Tool and Technologies}
\label{toolstech}
In this Section, we provide a short description of the tools and technologies we have used to implement the various prototypes of the thesis. Rest of the Subsection of this Section, are discussed with the tools and technologies required to implement our Scientific Workflow Management System.

\subsubsection{Hadoop} 
Apache Hadoop \cite{white2012hadoop} is a combination of tools that facilitate using a cluster of computers to solve computationally expensive problems requiring huge amounts of data. The software package provided by the Hadoop framework supports both distributed storage and processing of data. To take advantage of the parallel processing that Hadoop provides, we need to express our query as a MapReduce job. Hadoop supports MapReduce, MapReduce is a distributed data processing model and execution environment that runs on large clusters of computer systems.

\subsubsection{Spark}
Spark is a distributed framework that assists iterative processes to reuse data while retaining the scalability and fault tolerance of MapReduce \cite{zaharia2010spark}. To introduce the reusability, Spark offers a concept named resilient distributed datasets (RDDs). The RDDs in a distributed system are read-only groups of objects distributed over different machines and can be reconstructed if necessary.

\subsubsection{Yarn}
For the tight coupling and centralized job handling nature of Apache Hadoop MapReduce,  YARN computing platform is introduced in  Hadoop ecosystem. YARN decouples the programming model from the resource management infrastructure and substitutes many scheduling duties in a cluster \cite{vavilapalli2013apache}.

\subsubsection{HDFS}
Hadoop’s Distributed Filesystem (HDFS) is a filesystem that runs on large clusters of physical or virtual machines and manages data storing and retrieving in the cluster. HDFS is designed to work efficiently in conjunction with MapReduce model \cite{white2012hadoop}. Furthermore, HDFS is intended for storing a vast amount of data sets reliably, and for making sure of the flow of those data sets at high bandwidth to their respective applications \cite{shvachko2010hadoop}

\subsubsection{Association Rules}
Association rules \cite{agrawal1993mining} is introduced to identify uniformity between goods in large-scale transaction data recorded by a system in a store. Association rule mining is a rule-based technique for finding interesting relationships between items in large databases. Strong rules can be discovered in large databases by satisfying some constraints. 

\subsubsection{Bioinformatics Tools}
In our various experiment, we have considered some well-known tools of bioinformatics domain. Mainly PEAR \cite{zhang2013pear}, FastQC \cite{bioinformatics2011fastqc}, FLASH \cite{magovc2011flash} are used in our experiments while composing workflows.

\subsubsection{Deep Learning Based Image Processing Tools}
Effective functionality of a wide variety of learning settings and several basic statistical, optimization, and linear algebra fundamentals can be found in the MLlib library \cite{meng2016mllib} of Spark. As a Spark’s distributed machine learning library, MLlib supports various languages and grants a high-level API that can be used with Spark-based programs to simplify the implementation of machine learning or deep learning pipelines. Deep Learning Pipelines of the MiLIb presents high-level APIs also for the deep learning methodologies in Python Environment. Currently, the pipelines support \textit{inceptionV3, Xception, ResNet50, VGG16, VGG19} deep learning models of Keras \cite{chollet2018keras}.

\subsubsection{Python}
We chose Python language for our development because it has a wide range of libraries for scientific analysis. As an interpreted language with expressive syntax structure, it is also easy to implement. As stated in \cite{oliphant2007python}, Python is an extraordinary language to extend scientific codes written in different languages.  Furthermore,  Python can transform fundamental implementations into a high-level language that can be accommodated for scientific analysis for fast, immediate use and flexible for additional extensions.

\subsubsection{Servers}
Our web server is configured with Apache and Nginx together. In such configuration, direct communication with the outside world will be in two steps whereas Apache will be running as a backend server and will only exchange data with Nginx. In the second step, a uWSGI module is required to communicate with our application. A uWSGI project is implemented by default in a Python application. Where a uWSGI module of Nginx allows Nginx to communicate with python applications through the uwsgi protocol \cite{nedelcu2010nginx}. 

CouchDB is a comparatively modern database management system, designed from scratch to satisfy the requirements of modern software that are intended to be web-based, document-oriented, and distributed in type \cite{lennon2009introduction}. In our application CouchDB server is used to metadata management. 

\subsubsection{Web UI Technologies}
Interactive flowcharts, organizational hierarchies, tree structures, and other complex diagrams of nodes can be implemented with the JavaScript library GoJS \cite{SHAHZAD2016100}. For supporting interactive diagram and visualization in a web platform and facilitating custom construction of components make this technology a good choice for our SWfMS implementation. GoJS library allows some interactive features such as dragging-and-dropping, copying-and-pasting, text editing, tool-tipping, templating, data binding and modeling, event handling of its component on a web interface. Such features help us to implement our interactive feature of the SWfMS. 
 
\subsubsection{Google Analytics}
As a web analytics service, Google Analytics provides APIs that allows collecting user usage data \cite{clifton2012advanced}. 

\subsubsection{Apache Jmeter}
As a load testing tool, Apache JMeter is used for analyzing and measuring the performance of web applications \cite{halili2008apache}.

\section{Web Services vs. Local Scripts vs. Distributed Scripts} 
\begin{figure}[htbp]
    \centerline{\includegraphics[width=\textwidth]{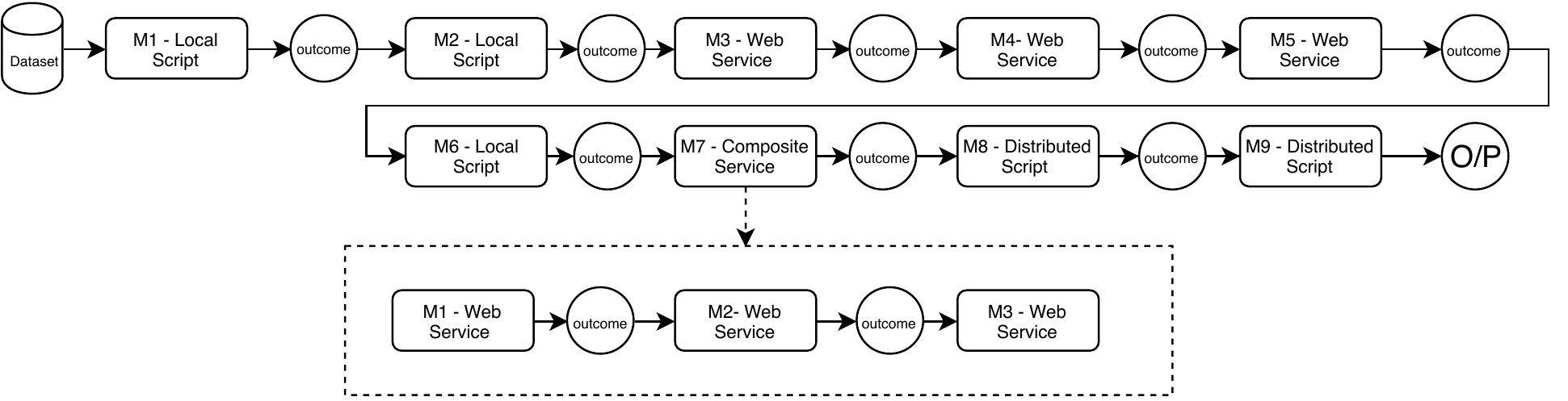}}
    \caption{Local scripts, web services, and distributed computing scripts in a workflow.}
    \label{fig_nestedworkflow}
    \end{figure}

All computational scripts and resources should have the plug and play characteristics in a SWfMS for their efficient reproducibility \cite{ludascher2006scientific}. Computational jobs as modules in workflows in a SWfMS could be local-scripts, distributed-scripts, and web-services (Figure \ref{fig_nestedworkflow}) and could reside in different servers.  In our various experiments of data management, we use local scripts, distributed computing scripts, and their outcomes are managed with our proposed data management techniques in a SWfMS. Local-scripts are plain scripts for some modularized tasks, and in our SWfMS, Python language is used to write the scripts. If a user executes a local-script module in a workflow, usually, a request is sent to the webserver of the SWfMS to run a python script for the particular module as a subprocess in the same server. The outcomes of the local-scripts as intermediate data are suggested to store in the same system of the webserver from our SWfMS with the proposed data management technique. In the case of the distributed computing job, a client request is processed on a Spark/YARN distributed cluster where the script of the requested job resides in the master node of the cluster. The outcomes of the distributed computing scripts (i.e., intermediate data of modules) are stored in a Hadoop Distributed File System (HDFS) of the same distributed cluster. Another important job type in workflows in a SWfMS is web-service. Web services are deployed by Service Oriented Architecture (SOA) to process responses for the demands of clients. For the availability of APIs and SOA architecture, web-services also can be incorporated easily in workflows as modules. In a SWfMS, incorporated web-services can be both composed and micro-services. Composite web service also serves as a module in a workflow with a nested structure of the workflow (i.e., composed web services as a workflow inside a workflow) in a SWfMS \cite{oinn2004taverna}. 

    \begin{figure}[htbp]
    \centerline{\includegraphics[width=\textwidth]{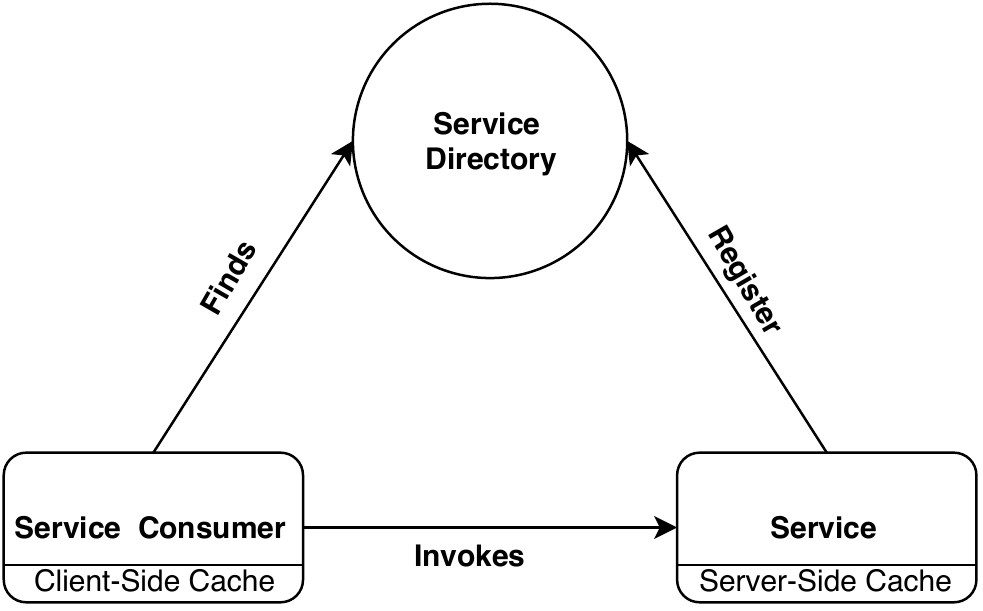}}
    \caption{Service oriented architecture.}
    \label{fig_soa}
    \end{figure}

With the help of service-oriented architecture, service provider also follow caching in servers of the services for efficient user experience \cite{truty2006systems} \cite{Friedman:2002:CWS:584490.584508} \cite{terry2003caching}. In addition, Liu and Ralph \cite{liu2007efficient} considered caching in both client-side and server-side to overcome the service issues for the temporary loss of connectivities to introduce seamless user experience. The suggested techniques in the cashing studies also designed to store service outcomes by various algorithms in active memory for rapid access. The major difference from those storing to our storing scheme is the scope of their usage and purpose. For instance, web service caching techniques store data or responses in a way that serves the objective of efficient and fast responses. On the other hand, our technique stores outcomes of the modules from workflows for a long time in persistent memory. Intermediate data are stored in the persistent memory respect to a particular dataset to be reusable in the future in other workflows for the same dataset in a SWfMS.

Other differences from the existing storing technique of the service composition to our technique is discussed in the following paragraph.
Web services caching can be introduced in both the service and service-consumer side of the SOA architecture (i.e., Figure \ref{fig_soa}).   In the studies of caching techniques of either side, data are stored for a particular web service or composite service, and the lifetime of the data depends on the request and response cycles of the service. In our case, the data management technique in a SWfMS is presented to store the module outcome for future workflows. Thus, the data management technique stores a set of outcomes optimally in a SWfMS with the highest probability of reuse. The set of outcomes are considered from all of the outcomes of modules of all the workflows for a particular timeline. In the data provenance study of SWfMS, Woodman et al. \cite{Woodman2015WorkflowProvenance} stated, It would not be a feasible solution to store all of the generated data from workflows in a SWfMS for the provenance and other purposes. Therefore an optimization is required in a SWfMS for the sets of module outcomes, and the proposed data management technique works for the optimized sets. Although, web services are considered from the log of workflow execution to explore the probability of data reusability in our analyses, explicitly web services are not tested with the data management technique in our SWfMS. As web services can be composite services and micro-services and selection of the intermediate data of the services is another research problem with the user accessibility, we planned to research this area in the future for managing the web services data for a SWfMS. In the rest of the thesis, the term services are used for referring both the local and distributed services.

\chapter{A Data Management Scheme for Micro-Level Modular Computation-intensive Programs in Big Data Platforms}
\label{ManagementScheme}

Big-data analytics or systems developed with parallel distributed processing frameworks (e.g., Hadoop and Spark) are becoming popular for finding important insights from a huge amount of heterogeneous data (e.g., image, text and sensor data). These systems offer a wide range of tools and connect them to form workflows for processing Big Data. 
Independent schemes from different studies for managing programs and data of workflows have been already proposed by many researchers and most of the systems have been presented with data or metadata management. However, to the best of our knowledge, no study particularly discusses the performance implications of utilizing intermediate states of data and programs generated at various execution steps of a workflow in distributed platforms. In order to address the shortcomings, we propose a scheme of Big-data management for micro-level modular computation-intensive programs in a Spark and Hadoop based platform. In this chapter, we investigate whether management of the intermediate states can speed up the execution of an image processing pipeline consisting of various image processing tools/APIs in Hadoop Distributed File System (HDFS) while ensuring appropriate reusability and error monitoring. From our experiments, we obtained prominent results, e.g.,  we have reported that with the intermediate data management, we can gain up to 87\% computation time for an image processing job.

\section{Motivation}
\label{ManagementSchemeMotivation}

\label{sec_intro}
With the rapid advancement of Big Data platforms, software systems  \cite{Bottleneck}, \cite{Plant_phenotyping}, \cite{Development_of_a_knowledge}, \cite{iPlant_atmosphere}, \cite{Development_of} are being developed to provide an interactive environment for large-scale data analysis to the end users in the area of scientific research, business, government, and journalism. Big Data platforms such as Hadoop, Spark, Google Data-flow, and so on provide us a high-level abstract interface for implementing distributed-cluster processing of data. Recently, many researchers \cite{Analysis_of_Six}, \cite{Development_of_a_knowledge}, \cite{Big_Metadata}, \cite{Hadoop_high} have focused on developing architectures and frameworks for large-scale data analysis tools utilizing these platforms. Most of these architectural frameworks are adopting workflows or pipelines \cite{Towards_a_Reference_Architecture}, \cite{iPlant_atmosphere}, \cite{Towards_Multi-site} for data analysis to provide flexible job creation environment. Workflows, in general, connect a sequence of interoperating tools to accomplish a task. For example, in order to find features from images of a large crop field, different image processing tools such as \textit{image registration,  stitching, segmentation, clustering and feature finding tools} are needed to be connected in order to form a workflow. In such workflow management systems, modularization is important to support scalability and heterogeneity. Thus, a wide range of tools is incorporated where each tool is treated as an independent process or module or service and can be implemented using different programming languages. Traditionally, researchers and data analysts require to run same workflows frequently with different algorithms and models that causes overwhelming efforts even with the moderate size of data by running all of the modules in the workflows. Utilization dimensions of those modules can be increased by storing their outcomes as intermediate states in such systems. In other words the possibility to use the transformed data in later stages to reduce computation time, to enhance reusability and to ensure error recovery can be increased by a proper data management.

Modular programming is a software design approach that draws attention to uncoupling the functionalities of a script towards independent, interchangeable, reusable units \cite{Micro-level}. On the other hand, data modularity is always anticipated for quick and flexible access to a dataset. In this chapter , the intermediate states of data are being recognized as modular data which are outcomes of modular programs. The dynamic management of datasets is another ground to have intermediate states for a huge amount of data \cite{Big_Data_Usage}. Some other common benefits of intermediate data are sustainability, scalability, quick construction and cost savings \cite{Big_Data_Usage}, \cite{Data_Multiverse}, \cite{Software_Architecture}, \cite{Signal_Processing}. In many circumstances of a data-intensive distributed application, image processing is an inevitable part where contemporary technologies of image analysis are pertinent for processing a large set of image data. However, special care is needed for both image data and program to process a huge amount of data with such emerging technologies. To make users' tasks more realistic in real time for Big Data analytics in image processing, intermediate modular data states can be an appreciable settlement to design and observe pipelines or workflows. In image-based tasks, the common execution operations (such as pre-processing, transformation, model fitting, and analysis) can be wrapped up as modules, and their outputs can be used for both observation and post-processing. Another important aspect of Big Data analytics is to store lots of data in a way that can be accessed in a low-cost \cite{Cloud_Resource_Management}, thus data management with the reusable intermediate data and program states might be a great deal to restore program state by reflecting the lower cost.

A distributed image processing system requires huge storage and data processing time for high-resolution large files. For such a system, stored intermediate states can be fruitful rather than transmitting raw files for a paralleled task. Spark itself generates some intermediate states of data in its processing engine to reduce IO cost \cite{Visualization_and_Adaptive}. Hence, in our modular programming paradigm, if we introduce a mechanism of storing intermediate states and manage them, we can reuse our data from various modules without computation which eventually minimizes total cost of processing. Many researchers and organizations are now pointing up on long time data processing with distributed parallel environments for low cost and convenient system to handle a large amount of data \cite{Parallel_Processing}. Following the trends, a machine/deep learning-based algorithm with heterogeneous data is another emerging practice to analyze a large amount of data for agricultural and other data-intensive tasks \cite{Machine_Learning}. Considering the current technologies and trends, our scheme would be a contemporary package to analyze a large amount of image data.

Although a few studies  \cite{Big_Data_Usage}, \cite{Data_Multiverse}, \cite{Software_Architecture}, \cite{Signal_Processing}, \cite{statistical_models} focused on propagated results in the intermediate states in Big Data management, to the best of our knowledge none reflected the performance implications of reusability of intermediate data and model in distributed platforms. Moreover, these studies do not show how the intermediate states across various image processing workflows can be reused. To fill the gap, in this chapter  we investigated whether the intermediates states generated from modular programs can speed up the overall execution time of workflows in Hadoop/Spark based distributed processing units by assuring enhanced reusability and error recovery. Here our idea was that loading intermediate states from HDFS (Hadoop Distributed File System) might take longer than generating them in memory during the execution of a workflow. In order to figure out the actual situations, we developed a web-based interactive environment where users can easily access intermediate states and associated them across various workflows, e.g., some outcomes of the image segmentation pipeline are reusable to the image clustering and registration pipelines. We found that even though it takes some time to load intermediate states from an HDFS based storage, a workflow can be executed faster in a system of handled intermediate states in persistent memory than a system of no intermediate states in persistent memory. Our finding is described in Figure \ref{fig_performance} for various workflows. In the figure, we illustrate if we can skip some modules with the help of stored intermediate states, we can get better performance than the case of without storing intermediate states. Details of the performance and re-usability have been discussed in Section \ref{result_discussion}.
The rest of the chapter  is organized as follows. Section \ref{sec_relatedwork} discusses the related work, Section \ref{proposedmodel} presents our proposed scheme of intermediate data management, Section \ref{exp_setup} describes our experiment setup, Section \ref{result_discussion} compares our proposed scheme with the usual technique of intermediate data handling by using various image progressing pipelines, Section \ref{discussion_lesson} describes some valuable findings towards data management for modular programming, and finally, Section \ref{conclusion} concludes the chapter .

\section{Related Work}
\label{sec_relatedwork}

A number of studies \cite{Experiences_on}, \cite{Development_of_a_knowledge}, \cite{Hadoop_high}, \cite{Testing_of_several} have investigated recent techniques of filesystem to process Big Data and some studies \cite{Big_Metadata}, \cite{Distributed_metadata}, \cite{Towards_Multi-site}, \cite{Managing_hot_metadata}, \cite{Unlocking_the} were presented with application of data or image processing in such file systems with metadata management. Similarly, various databases, algorithms, tools, technologies, and storage systems are acknowledged in a number of studies while solving the problem of big data processing for image analysis and plant-based research in distributed environments.  Studies related to phenotyping, image processing and intermediate states in distributed environments are considered to compare with our work, and some of them are presented below in three subsections.

\subsection{Study on file systems of Big Data processing}
Blomer \cite{Experiences_on} investigates various file systems of computer systems and compared them with parallel distributed file systems. Benefits of both distributed and core file systems of computer systems are discussed. As well as, a distributed file system as a layer of the main file system has been addressed with common facilities and problems. In their work, data and metadata management’s key points are also discussed for map-reduce based problems and fault tolerance concept. Luyen et al. \cite{Development_of_a_knowledge} experiment on different data storage systems with a large amount of data for phenotyping investigation. Their work on a number of heterogeneous database systems and technologies such as MongoDB, RDF, SPARQL, and so on. explores possibilities of interoperability and performance. Similar to the above studies, Donvito et al. \cite{Testing_of_several} discuss and compare various popular distributed storage technologies such as HDFS, Ceph, and GlusterFS. Wang et al. \cite{Hadoop_high} propose a technique of metadata replication in Hadoop based parallel distributed framework for failure recovery. By emphasizing metadata replications in slave nodes for reliable data access in big data ecosystem, the proposed metadata replication technique works in three stages such as initialization, replication and failover recovery. Similarly, Li et al. \cite{Distributed_metadata} propose a log-based metadata replication and recovery model for metadata management in a distributed environment for high performance, balanced load, reliability, and scalability. In their work, caching and prefetching technologies are used for the metadata to enhance performance. While above studies focus on metadata management in various file systems, our study is fundamentally different with an investigation of a scheme of intermediate data management for distributed file systems considering reusability, error recovery and execution at a lower cost.

\subsection{Big Data in terms of Image Processing}
Minervini et al. \cite{Bottleneck} discuss different challenges in phenotyping by addressing the current limitations in image processing technologies, where collaboration is emphasized from diverse research group to overcome the challenges.  Although collaboration is an important part of a scientific workflow management system (SWfMS) and we are proposing a scheme for a SWfMS, but the main focus in this study is consideration of intermediate states with micro-level modular computational-intensive programs while managing workflows in a distributed environment for increasing efficiency. Some other studies on phenotyping concentrate on the costs of processing and appropriate usage of current technologies to reduce the costs such as Minervini et al. \cite{Application-aware} propose a context-aware based capturing technique for JPEG images that will be processed by distributed receivers in a Phenotyping environment.  Singh et al. \cite{Machine_Learning}  provide a comprehensive study on ML tools in phenotyping area for appropriate and correct application of them in plant research. Coppens et al. \cite{Unlocking_the} emphasize data management for automated phenotyping and explore the image metadata potentiality for synergism (Usage of metadata with main features). Sensor's metadata of imaging technique could be used by appropriate filtering for the synergism is also addressed in their work. Smith et al. \cite{Big_Metadata} address the importance of metadata in Big Data ecosystems to migrate and share data or code from one environment to another. Data consolidation, analysis, discovery, error, and integration also have been scrutinized from metadata usage perspective in the Big Data environment. Pineda-Morales et al. \cite{Managing_hot_metadata} emphasize on both file metadata and task metadata to propose a model for filtering hot metadata. After categorizing and analyzing both hot and cold metadata, they recommend for distributing hot metadata (frequently accessed) and centralizing cold metadata in a system of data management.  Although the above works show the behavior and direction of metadata, none of them analyze the processed data or intermediate data as metadata for efficient management. In our work, modular outcomes of computational intensive programs are analyzed, and a scheme is proposed to store those intermediate outcomes for efficiency in a SWfMS. 

\subsection{Study on Intermediate-data}
Becker et al. \cite{Big_Data_Usage} survey on current Big Data usage and emphasize on intermediate data behavior for iterative analysis at every stage of processing to minimize the overall response time with an algorithmic perspective. Besides, intermediate data can be saved in persistent storage at checkpoints was considered for error recovery. Same as Heit et al. \cite{statistical_models} explained the nature of intermediate data while giving a statistical model for Big Data as well as intermediate data should be independent was reported in their work. Intermediate data management is considered by Tudoran et al. \cite{Data_Multiverse}, where they pointed out storing of intermediate-data for rollback purposes from the application level. Intermediate data for checkpoints analysis are also considered by Mistrik and Bahsoon in their work \cite{Software_Architecture}. Likewise, how intermediate data are dispatched in parallel distributed systems such as Spark, Hadoop is discussed by Han and Hong \cite{Signal_Processing} in their book by bringing together and analyzing various areas of Big Data processing.
 
Although some of the above works were discussed with the fault-tolerant concept using intermediate states, none of them discussed the possibility and necessity of storing intermediate data for Big data analytics in a distributed environment. Reviewing the above works, we realized that in a distributed environment efficient data access mechanism is a big concern and time implications for both data storing and loading are needed to be observed in a distributed environment. Focusing these, we investigate the possibility of a data management mechanism by storing intermediate states with the help of modularized programs for enhancing processing time, reusability and error recovery.

\section{Proposed Method}
\label{proposedmodel}
Modular programming for computation-intensive tasks is already a proven technology for reusability of code \cite{Exploration_of_modularity}, \cite{ROSLab}, \cite{USING_MODULAR_PROGRAMMING}, \cite{Euro-Par}, but for large raw data transaction and loading from distributed systems, we cannot fully use the proficiency of modularity. On the other hand, processed data require less memory, transaction and loading time. Considering both the cases, we store processed intermediate data with other program data and want to explore three research questions in a SWfMS, such as
\begin{itemize}
    \item \textbf{RQ1:} Does the system support data reusability that ultimately reduces time and efforts? 
    \item \textbf{RQ2:} Does it have any error recovery mechanism that could construct workflows quicker if errors occur?
    \item \textbf{RQ3:} and How the system is supporting fast processing for frequently used workflows? 
\end{itemize}
Answers to these three research questions have been presented in the section \ref{result_discussion}. To use both of these intermediate data and modular programs, and explore the usability of modular outcomes in our full system a modularized data access mechanism is introduced. We have an interactive web interface, which is used for accessing datasets of various modules’ input and output. A dedicated distributed data server where images and intermediate data are stored was used in our system for parallel processing. From the web interface, image processing workflows can be built, and modules of the workflows can be submitted to a distributed processing master node as jobs, by this way distributed image data are processed with specified modules from the workflows.

\begin{figure}[htbp]
\centerline{\includegraphics[width=\textwidth]{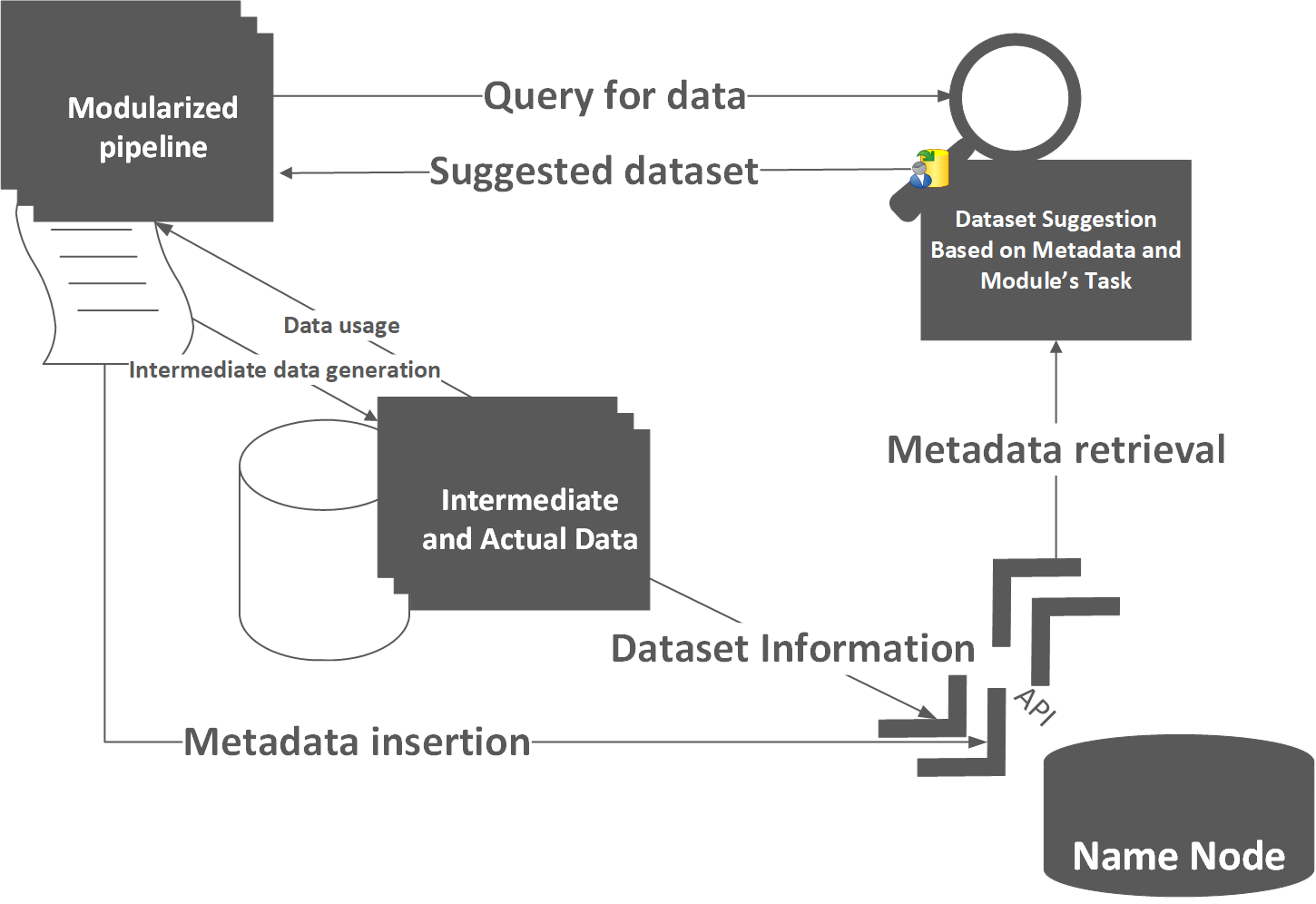}}
\caption{Proposed data management scheme.}
\label{fig_proposedmodel}
\end{figure}

Two major types of metadata of image processing environment are file-structure-metadata and result-metadata \cite{Managing_hot_metadata}.  In our system, we managed both kinds of metadata using APIs that retrieve and store information on a dedicated Hadoop server. This dedicated server is designed to facilitate various parallel processing from different modules of pipelines or workflows. When a SHIPPI’s (Spark High-throughput Image Processing Pipeline) module needed to be executed for a task, available data or intermediate data states are suggested from a web interface with the help of the metadata and HDFS APIs  (e.g., Figure \ref{fig_proposedmodel}). SHIPPI is our designed library that provides cloud-based APIs for the modular functionality of image processing \cite{Micro-level}. Details of the SHIPPI has been described in the experimental setup section, i.e., Section \ref{exp_setup}. Our system at first tries to find suitable datasets for a modular task with task information and information from the metadata server, suggested datasets are presented on a web interface to be selected for a particular task from our distributed storage system, i.e., HDFS. Suggested datasets can be an intermediate or raw data, but the intermediate states will always have a higher priority to be selected if available. Another important task of our system is to keep track of metadata and intermediate data. After every execution, there are a few outputs from a single module such as analysis results, intermediate data, and so on. Intermediate data are stored in a distributed file system with actual data, and analysis results are stored in a server as metadata for future use.

\begin{figure}[htbp]
\centerline{\includegraphics[width=\textwidth]{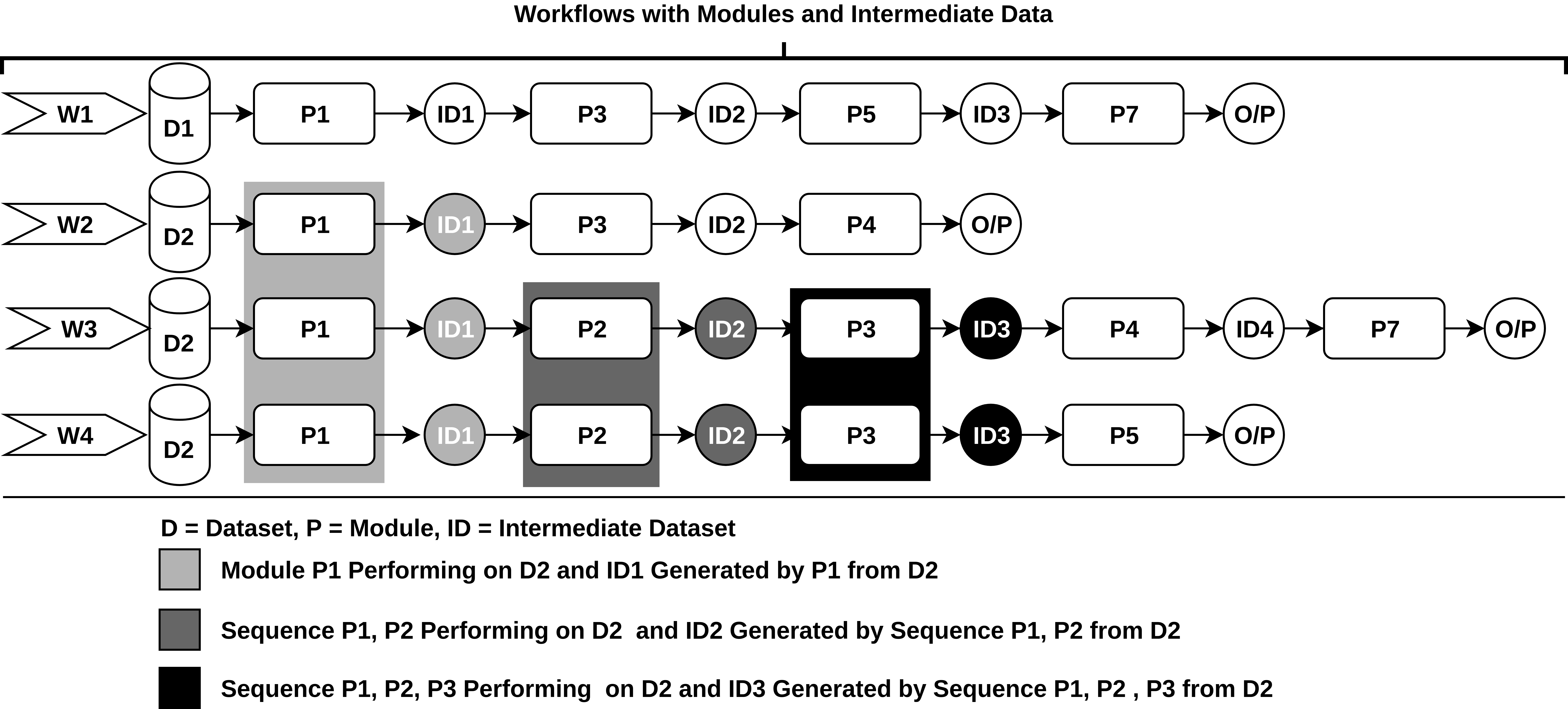}}
\caption{Reusable intermediate data in workflows.}
\label{fig_intermediatestages}
\end{figure}

How the proposed scheme of data management by modular programs and intermediate data can introduce reusability, error recovery and faster processing in a Scientific Workflow Management System (SWfMS) is described with simple workflows (Figure \ref{fig_intermediatestages}). In the figure, four workflows are illustrated with modules and intermediate data. All black and grey marked processes (Ps) and intermediated data (IDs) represent more than the one-time existence of them in the four workflows and possibility to be reused of them. This reusability of them means they are likely to be reused later in either for the same workflow with different tools or the other workflows in a SWfMS. Different intensities of black and grey colours represent the same modules or intermediate states from different workflows in different groups. A particular group is formed for the same operations on the same dataset or for the intermediate states that are being generated for the same operation of module sequences on them with same input dataset. In workflow 1, we have four modules (P1, P3, P5, and P7). These modules perform operations sequentially on dataset D1 and produce three intermediate datasets (ID1, ID2, and ID3), where the last outcome is desired output (O/P). All workflows have some processing modules to perform operations on specific datasets or intermediate datasets for generating desired intermediate outcomes or final results. The intermediate states are usually used by intermediate modules sequentially (N. B. For simplicity we are considering only sequential module processing in workflows) in workflows. Workflow 2 has three modules (P1, P3, and P4) that generate two intermediate datasets (ID1 and ID2). Workflow 3 and 4 have five and four processing modules with four and three intermediate stages respectively. Last three workflows in the figure are working on the same dataset D2, and a few modules such as P1, P2, and P3 are frequently used for building those workflows. This scenario opens the possibility of reusability of intermediate states, as it is common in a SWfMS that same modules can be used in different workflows. As well as, the same sequence of operations on the same or different dataset can be occurred in various workflows by supporting the possibility of program reusability. In the workflow 2, an operation by module P1 is performed on dataset D2 and generates intermediate dataset ID1. This operation and the outcome (ID1) are the same in the other three workflows for the same module sequence and same dataset. Same sequence $P1 \rightarrow P2 \rightarrow P3$ in workflow 3 also occurs in workflow 4 with the same input data set D2. Thus, ID3 (intermediate dataset) from workflow 3 can be directly used in workflow 4 for skipping the first three module operations to analyze with only the last module or increase the performance. To introduce this type of efficiency and performance enhancement in a system of workflow management, modules outcome is needed to be stored with appropriate access mechanism. Furthermore, if failures occur in any workflow,  stored intermediate states (IDs) can be used to recover the workflow by addressing the issues only in faulty modules. In a distributed environment, data are partitioned and replicated on different nodes, and the transition time of data is not as simple as it is a single file system. Data transfer time from slave nodes to master nodes is needed to be considered for both storing and retrieving. So, for storing and retrieving intermediate data from such distributed systems, experiments are necessary regarding transition and computation time. Our experimental section (Section \ref{result_discussion}) is designed in that way to investigate the actual scenario of  I/O operation and computation time for the micro level modular computation-intensive programs in a distributed environment and explore the possibility of reusability. 

\section{Experimental Setup}
\label{exp_setup}

\begin{figure}[htbp]
\centerline{\includegraphics[width=\textwidth]{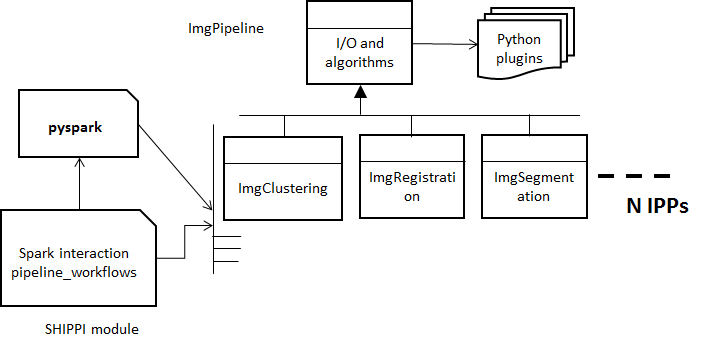}}
\caption{Block diagram of components of SHIPPI.}
\label{fig_shippi}
\end{figure}

In our experiments, for evaluating a full system performance with the proposed scheme, a web platform is used, a web interface is designed with current web technologies and Flask (A Python micro-framework) web framework. From the web interface, SHIPPI’s APIs are called to execute a job in an OpenStack based Spark-Hadoop cluster (e.g., five-node, 40 cores, and a total of 210 GB RAM). It provides cloud-based APIs for the modular functionality of image-processing pipeline and automatically parallelizes different modules of a program. Each API call is modularized into four sections such as conversion, estimation, model-fitting, analysis-transform. Figure \ref{fig_shippi}, shows the block diagram of SHIPPI, which represents the relationship among different libraries and modules of the framework. In our testing environment, HDFS is used to store our image data for parallel processing. To store intermediate states, we used the Pickle library of Python, and we designed three image processing pipelines - Segmentation, Clustering and Leaves Recognition following our proposed model discussed in Section \ref{proposedmodel}. Every pipeline was tested at least five times and an average of their execution time was recorded. For testing purposes, we mainly used three datasets - \textit{Flavia} \cite{A_Leaf_Recognition}, \textit{2KCanola}, \textit{4KCanola}. Each of those datasets has more than 2000 images, where \textit{Flavia} is a known dataset used for leaves recognition system, and \textit{2KCanola} and \textit{4KCanola} datasets are from the P2IRC (Plant Phenotyping and Imaging Research Centre) project of University of Saskatchewan used for various purposes of image analysis.

\begin{figure}[htbp]
\centerline{\includegraphics[width=\textwidth]{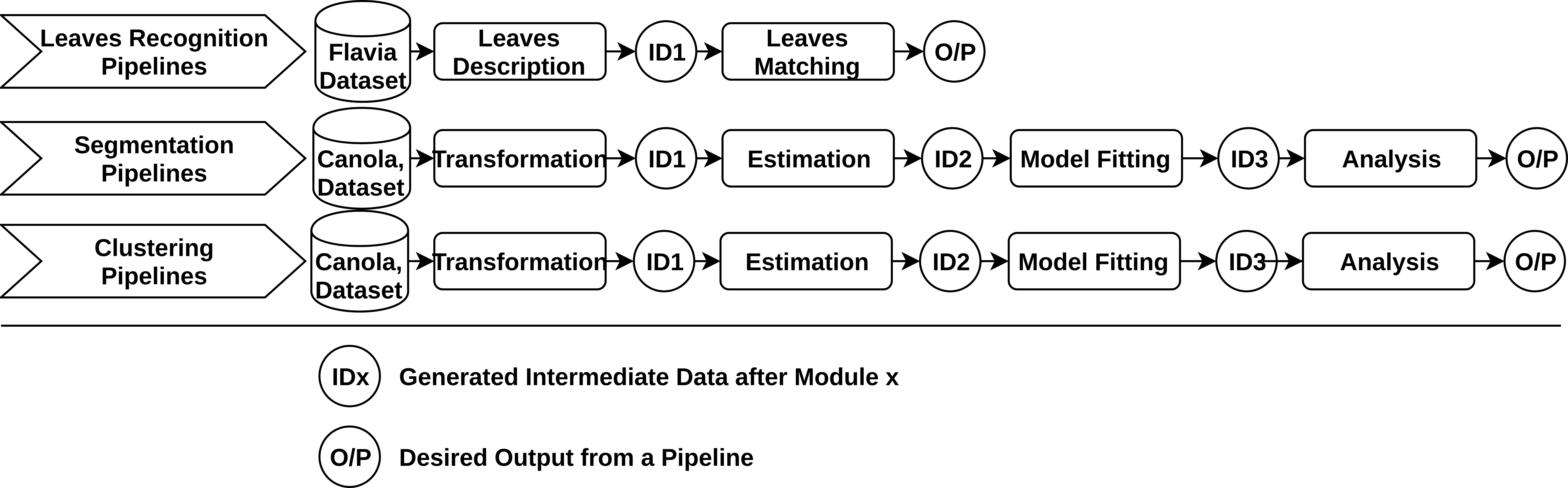}}
\caption{Pipeline building with image processing tools.}
\label{fig_intermediatestagesREAL}
\end{figure}

The three pipelines are illustrated in Figure \ref{fig_intermediatestagesREAL}, where a single job such as leaves recognition, segmentation or clustering is built by their respective modularized parts. The first module (Leaves Description) of the leaves recognition pipelines extracts features and prepares data for the next module leaves matching. The next module (Leaves Matching), which is the final module in the leaves recognition pipelines does the job of comparing those features and reports a classification result of leaves. Similarly, image segmentation and clustering pipelines are modularized by four common modules such as \textit{transformation, estimation, model fitting, and analysis}. Transformation is used for mainly colour conversion, estimation is used for feature extraction, model fitting is used for training an algorithm and setting parameters, and the final module analysis is used for testing and result preparation. Pipelines can be built by selecting modules and parameters for submitting a desired task in the distributed environment on the web interface, where each module works as a job in the distributed environment and their outcomes are gathered from a master node and distributed file system for further analysis.    

\section{Results and Evaluation}
\label{result_discussion}

We developed a Spark High-throughput Image Processing Pipeline framework known as SHIPPI, which is an image processing pipeline library to abstract map reduce, data loading, and low-level complexity of image operations. SHIPPI was written in Python 2.7 and Spark 2.0.1, four modules of our modular program paradigm of a pipeline using SHIPPI can be modified based on image processing operations concerning the desired purpose. Furthermore, these modules are parallel with the help of PySpark library, and each output of a module is used in the next module. In a spark-based solution all data are in main memory, so to reuse those processed data, we store them as intermediate states in Hadoop distributed file system from those modules that give reusable data. This system not only designed for reusability but also for error recovery (for the higher failure rate of a long-running program or pipeline with a huge amount of data). Thus, if we store intermediate data, there is a chance for us to recover a system at low cost.

\begin{table}
\caption{Time gain for the reusable intermediate data}
\centering
\begin{tabular}{|l|l|l|l|}
\hline
\textbf{Pipeline Tool}   & \textbf{Step 1}         & \textbf{Step 2} & \textbf{Gain} \\ \hline 
Leaves Description & 1163.7 sec + 35.7 sec & -      &   -   \\ \hline
Leaves Matching    & Can be skipped              & 175.9 sec     &  1199.5 sec    \\ \hline
\end{tabular}

\label{table_time}
\end{table}

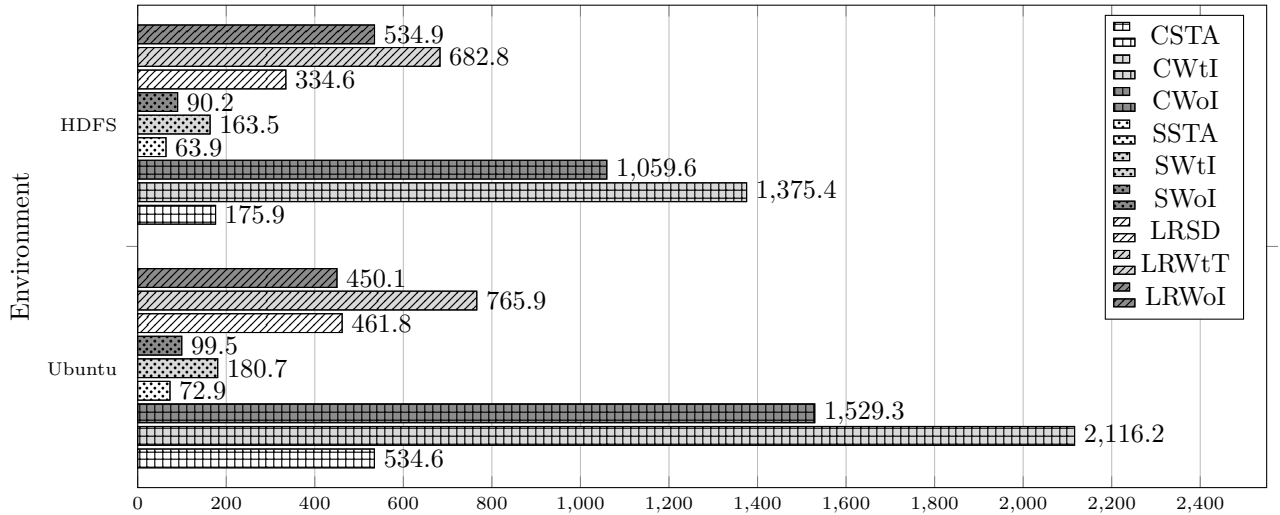
\begin{figure}[htbp]
\begin{tikzpicture}
  \begin{axis}[
    xbar=0.05cm,
    bar width=0.25cm,
    width=\textwidth,
    height=8cm,
    xmin=0,
	xmax=2550,
    xticklabel style={font=\scriptsize},
    yticklabel style={yshift=0ex, rotate=0, font=\scriptsize},
    major y tick style = {opacity=0},
    minor y tick num = 1,
	minor tick length=1ex,
	xmajorgrids = true,
    enlarge y limits  = 0.49,
    ylabel={Environment},
    symbolic y coords = { Ubuntu, HDFS},
    nodes near coords,
    ytick distance=1
  ]
 \addplot [draw=black, fill=darkgray!00, postaction={pattern = grid}] coordinates {(175.9,HDFS) (534.6,Ubuntu)};
 \addplot [draw=black, fill=darkgray!20, postaction={pattern = grid}] coordinates {(1375.4,HDFS) (2116.2,Ubuntu)};
 \addplot [draw=black, fill=darkgray!60, postaction={pattern = grid}] coordinates {(1059.6,HDFS) (1529.3,Ubuntu)};
 \addplot [draw=black, fill=darkgray!00, postaction={pattern = crosshatch dots}] coordinates {(63.9,HDFS) (72.9,Ubuntu)};
 \addplot [draw=black, fill=darkgray!20, postaction={pattern = crosshatch dots}] coordinates {(163.5,HDFS) (180.7,Ubuntu)};
 \addplot [draw=black, fill=darkgray!60, postaction={pattern = crosshatch dots}] coordinates {(90.2,HDFS) (99.5,Ubuntu)};
 \addplot [draw=black, fill=darkgray!00, postaction={pattern = north east lines}] coordinates {(334.6,HDFS) (461.8,Ubuntu)};
 \addplot [draw=black, fill=darkgray!20, postaction={pattern = north east lines}] coordinates {(682.8,HDFS) (765.9,Ubuntu)};
 \addplot [draw=black, fill=darkgray!60, postaction={pattern = north east lines}] coordinates {(534.9,HDFS) (450.1,Ubuntu)};
  \legend{
  CSTA,
  CWtI,
  CWoI,
  SSTA,
  SWtI,
  SWoI,
  LRSD,
  LRWtT,
  LRWoI
  }
  
\end{axis}
\end{tikzpicture}

\caption{Performance comparison of workflows with-intermediate and without-intermediate data}
\label{fig_performance}
\end{figure}

Total eight distinct modules of the three image processing pipelines have been implemented with current technologies, and all of them are designed for both without intermediate and with intermediate data concept. Both transformation and estimation modules in the pipelines use the same technology, and they are similar in the last two pipelines. Figure \ref{fig_performance} is the execution-time comparison graph of both of the concepts for the three pipelines in distributed and ubuntu file systems. Although saving intermediate states is a memory overhead task for all of the pipelines in the figure, but there are possibilities to use stored data and skip some processing steps for executing a pipeline at low cost. Skipping procedure eventually increases the flexibility and reusability to analyze fractions of pipelines in low cost. Pipelines with skipped modules by using intermediate data are presented with the improved performance in the figure. Another major concern of our investigation was to explore the trade-off scenario between computation time and data loading/storing time of modules in distributed systems. For investigating this, previously we assumed that a data and metadata transition time among slave and master nodes would be more than a module computation time in a distributed environment. But the experimental results for the pipelines of skipped modules show improved performance and contradict with our assumption regarding time implication of data transfer. Hence, a system of workflow management with distributed processing unit is capable of handling intermediate states, and a scheme is possible to introduce in such systems for reusability, error recovery and performance enhancement. 
We organize our experiment into three sections to answer the research questions of our system's usability. All of the three pipelines of image processing are considered in each section for their respective illustrations. Below are the three main subsections of our experiments.

\subsection{Experiment for reusability}

\begin{figure}[htbp]
\centerline{\includegraphics[width=\textwidth]{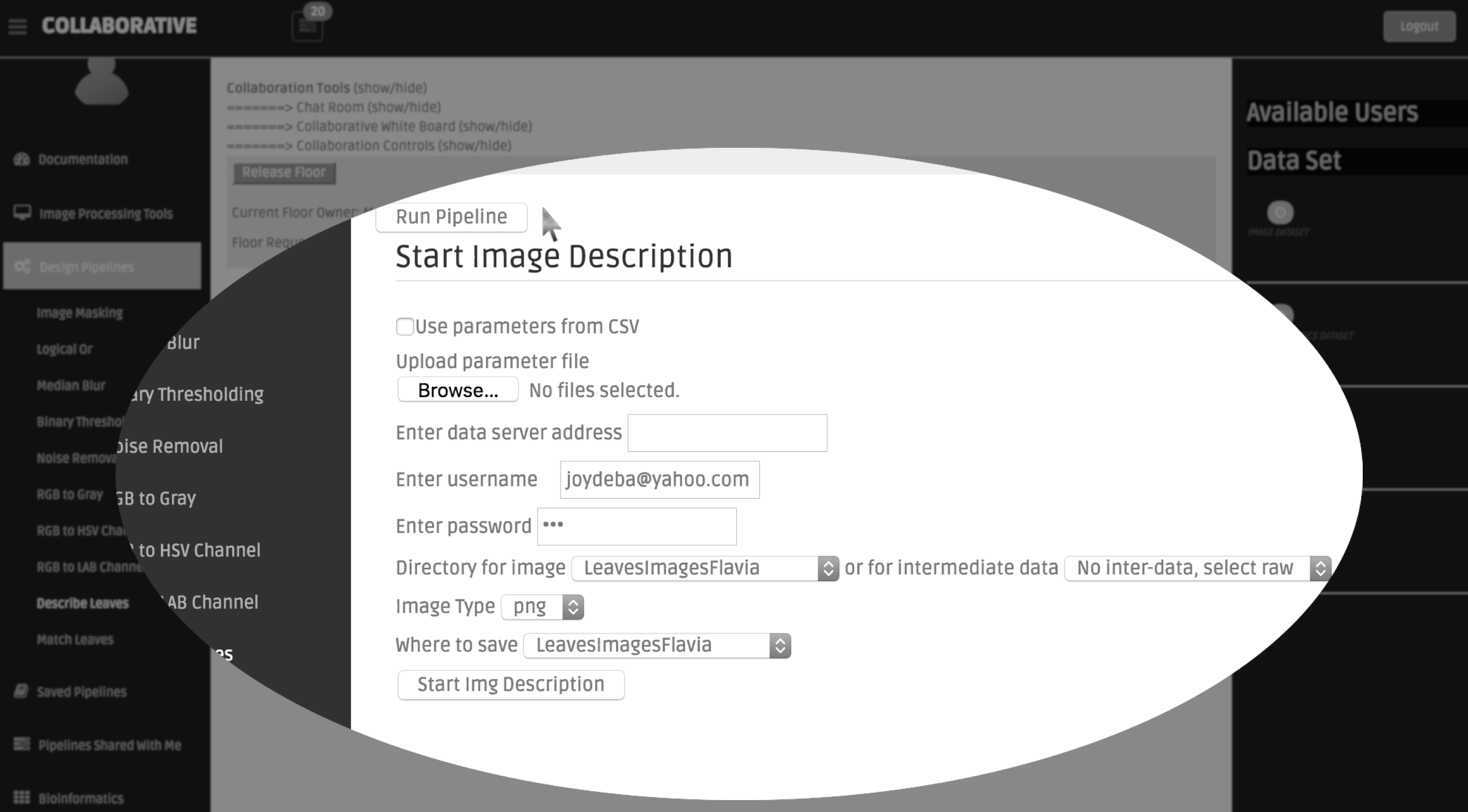}}
\caption{Image description module in our interactive workflow composition environment.}
\label{fig_descroption}
\end{figure}

\begin{figure}[htbp]
\centerline{\includegraphics[width=\textwidth]{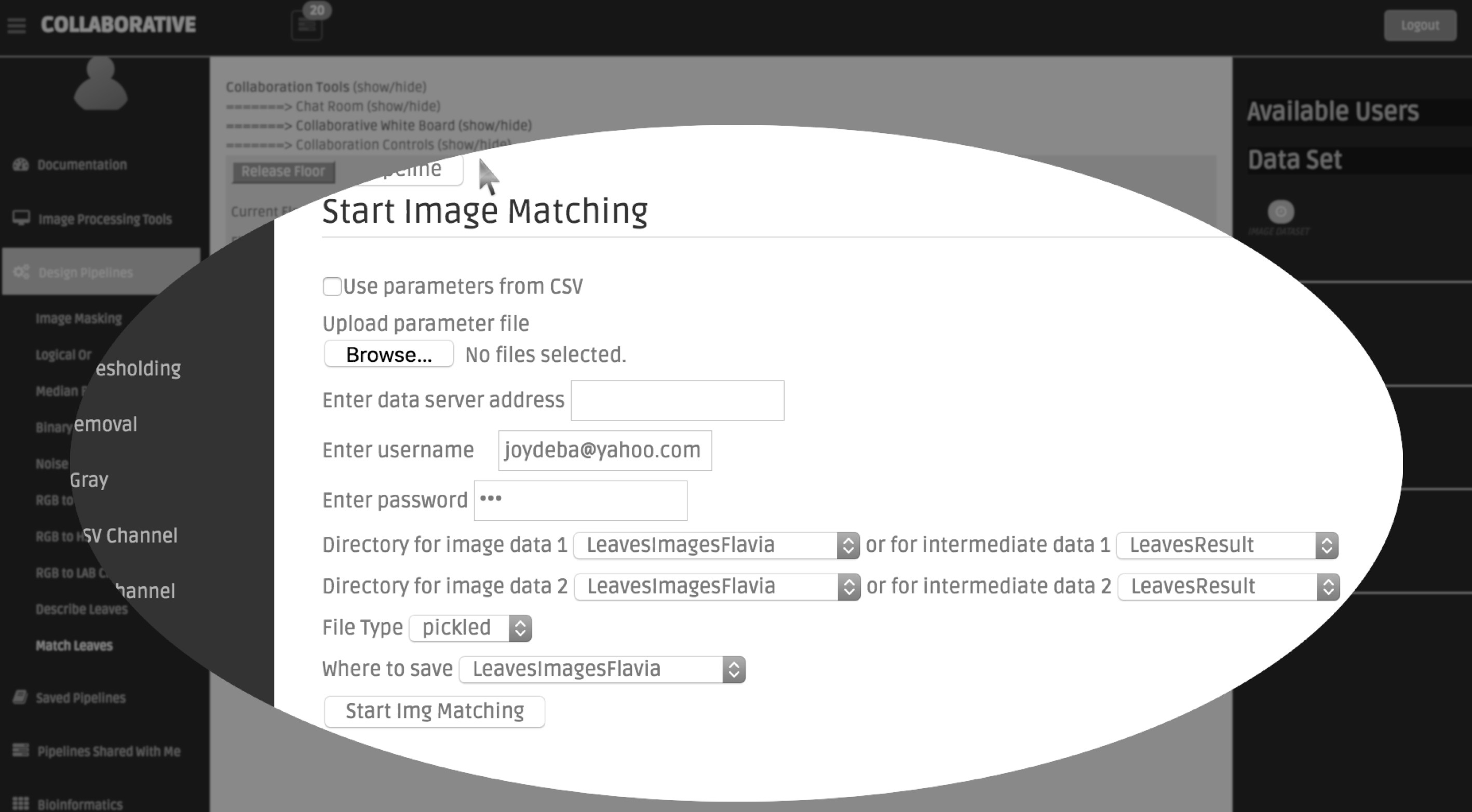}}
\caption{Image matching module in our interactive workflow composition environment.}
\label{fig_matching}
\end{figure}

A collaborative SWfMS is used for design and job submission of the image processing pipelines to a distributed parallel processing system. Figure \ref{fig_descroption} and  \ref{fig_matching} show the configuration interface for the image recognition pipelines, which is implemented with SHIPPI’s four-phase modular architecture where only two modules are considered for simplicity. Intermediate state from the first module of this pipeline can be reusable to image matching modules. So, an outcome of the image description stage is important for an image matching technique and needed to be stored for reuse. In Figure \ref{fig_matching} of the image matching technique, there is an option to use intermediate states for the reuse. Possibility to use intermediate states is the reusability model we are using in our system, an outcome of a module can be used in other modules if available and appropriate for those modules. Consider another situation, when image files are huge in size for high resolution; then pipelines take a large portion of total execution time for only data transferring and loading in a distributed parallel processing system. Thus, stored features or descriptors as intermediate states can help to reduce the transfer time up to a certain level. From table \ref{table_time}, we can see that module at step 1 (Image Descriptor) required more than 1199.4 seconds to accomplish the feature extraction part. In a pipeline design, we can skip this step if we can use intermediate states and can gain up to 1199.4 seconds.




\subsection{Experiment for error recovery}

\begin{figure}[htbp]
\centerline{\includegraphics[width=\textwidth]{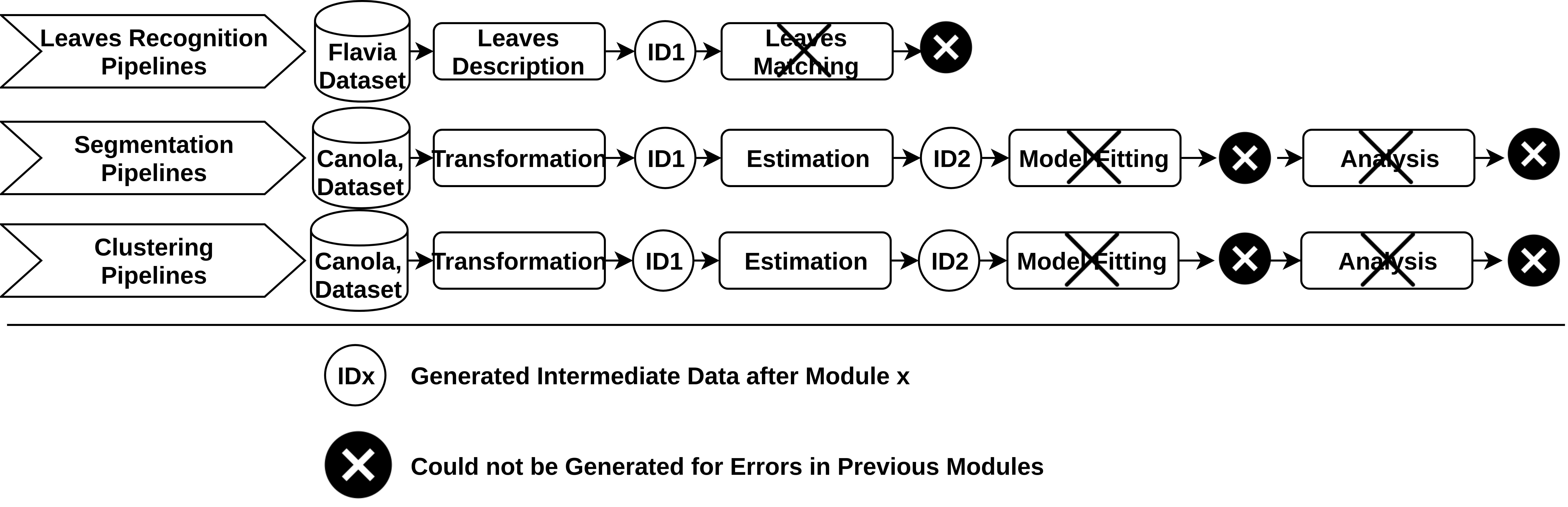}}
\caption{Common error occurrences in workflow execution.}
\label{fig_error}
\end{figure}

While a pipeline is running in a cluster, it is common that all the steps/modules may not be succeeded and the pipeline may fail to produce the desired output. But in a system of workflow management stored intermediate states can be helpful to recover a workflow execution by applying appropriate module configuration. If we store intermediate data and necessary parameters for modules and if there is an error in a module operation; still, data and configuration up to the previous module can be used for stored intermediate data and parameters. So, configuring the failed module with the stored parameters and intermediate data, next run of the pipeline will only take time for the failed modules. For example, in our case (Figure \ref{fig_error}), image recognition pipeline is divided into two modules using SHIPPI library, the second stage sometimes fails for too many comparisons and I/O operations, but using intermediate states we can quickly re-run the second stage by only reconsidering the error issues in that module. Similarly, both segmentation and clustering pipelines are error-prone at the model fitting stage for lots of computation, in that case, up to two stages of computation can be saved if we can reuse stored data with appropriate parameters, and this is possible from our web interface. Besides, in a distributed environment when computation and data are enormous, the failure rate is high for various tasks of a job due to memory overhead and access limitations of different libraries. So, our proposed approach can be a feasible solution in such type of computation-intensive tasks to avoid errors or failures.

\subsection{Experiment for fast processing}

Both distributed and ubuntu file systems are considered for the execution time experiments of our scheme, where pipeline design is considered with both with-intermediate and without-intermediate data. Table \ref{table_time}, Figure \ref{fig_performance} and Figure \ref{fig_performance_line} are presented with those execution time experiments. Examining the table, we can say that for the intermediate states, possibilities of skipping execution steps can make pipeline execution time shorter than the normal execution.
The Figure \ref{fig_performance} illustrates the execution time comparisons and performances for the three image processing pipelines in our SWfMS. These image processing pipelines are categorized by three techniques of workflow building for two file systems. The techniques are - execution of workflows by storing intermediate data, by not storing intermediate data and by skipping some module operations.  So, there are nine experiments in each file system that makes total of eighteen execution scenarios in our system for both file systems. Short forms of the above techniques are given below.

\begin{itemize}
    \item Leaves Recognition workflow without storing intermediate data (LRWoI)
    \item Leaves Recognition workflow with storing intermediate data (LRWtI)
    \item Leaves Recognition workflow by skipping descriptor operation (LRSD)
    
    \item Segmentation workflow without storing intermediate data (SWoI)
    \item Segmentation workflow with storing intermediate data (SWtI)
    \item Segmentation workflow by skipping transformation and analysis (SSTA)
    
    \item Clustering workflow without storing intermediate data (CWoI)
    \item Clustering workflow with storing intermediate data (CWtI)
    \item Clustering workflow by skipping transformation and analysis (CSTA)
\end{itemize}

The grided dark-grey, grey, and white bars in the Figure \ref{fig_performance} representing leaves recognition system performance in two different file systems. If we skip the descriptor calculation part using intermediate state, it performs better, the grided white bars in the figure presents skip of the descriptor part and performances. Same goes for the clustering and segmentation pipelines, dotted and lined bars in the figure represent segmentation and clustering algorithms’ performances respectively, and white bars of these patterns represent the skipping of modules and improved performance. Furthermore, from the Figure \ref{fig_performance}, we can have an idea of quantitative values of performance for three different pipelines where grey colours represent all modules computation time of pipelines with or without intermediate states, and white colours represent the computation time of the pipelines that can skip some modules’ operation for the availability of intermediate states. So, it can be interpreted from the chart that, if we can skip some module operations for the intermediate states, system or pipeline can be executed at a low cost in both of the file systems.
However, there are some delays to store data in a storage system, but for the reusability, a system can really work fast for already processed data. In a SWfMS, amount of data is increased frequently for processing of various workflows. So, without proper data management, volume and velocity of big data cannot be addressed efficiently. As well as, different types of data exist in a SWfMS without proper annotation and association variety of big data processing cannot be granted. All of the three V’s of big data can be appropriately handled in a system of scientific workflow management by using a data-centric component management. Data-centric component development for large-scale data analysis is a current trend to reduce error rates and increase usability, and our system can provide such kind of facilities using stored intermediate data.

\section{Discussion and Lesson Learned}
\label{discussion_lesson}
\begin{figure}[htbp]
\begin{tikzpicture}
\begin{axis}[
    width=.99\linewidth,
    legend style={at={(0.5,-0.1)},anchor=north},
    legend cell align={left},
    ylabel = {Seconds},
    title = ID -- Intermediate data,
    symbolic x coords={Without  ID, With  ID, Skipping Modules},
    xtick=data]
    \addplot[mark=triangle*,thick,black] coordinates {(Without  ID,1529.3)         (With  ID,2116.2)         (Skipping Modules,534.6) };
    \addlegendentry{Leaves Recognition in Ubuntu File System}
    \addplot[mark=triangle*,mark options={solid},black,thick,dashed] coordinates {(Without  ID,1059.6)         (With  ID,1375.4)         (Skipping Modules,175.9)};
    \addlegendentry{Leaves Recognition in HDFS}
    \addplot[mark=square*,thick,black] coordinates {(Without  ID,450.1)        (With  ID,765.9)        (Skipping Modules,461.8)    };
    \addlegendentry{Clustering in Ubuntu File System}
    \addplot[mark=square*,mark options={solid},black,thick,dashed] coordinates {(Without  ID,534.9)        (With  ID,682.8)        (Skipping Modules,334.6)        };
    \addlegendentry{Clustering in HDFS}
    \addplot[mark=o,thick,black] coordinates {(Without  ID,99.5)        (With  ID,180.7)        (Skipping Modules,72.9)        };
    \addlegendentry{Segmentation in Ubuntu File System}
    \addplot[mark=o,mark options={solid},black,thick,dashed] coordinates {(Without  ID,90.2)        (With  ID,163.5)        (Skipping Modules,63.9)        };
    \addlegendentry{Segmentation in HDFS}
\end{axis}
\end{tikzpicture}
\caption{Performance comparison in both environments}
\label{fig_performance_line}
\end{figure}
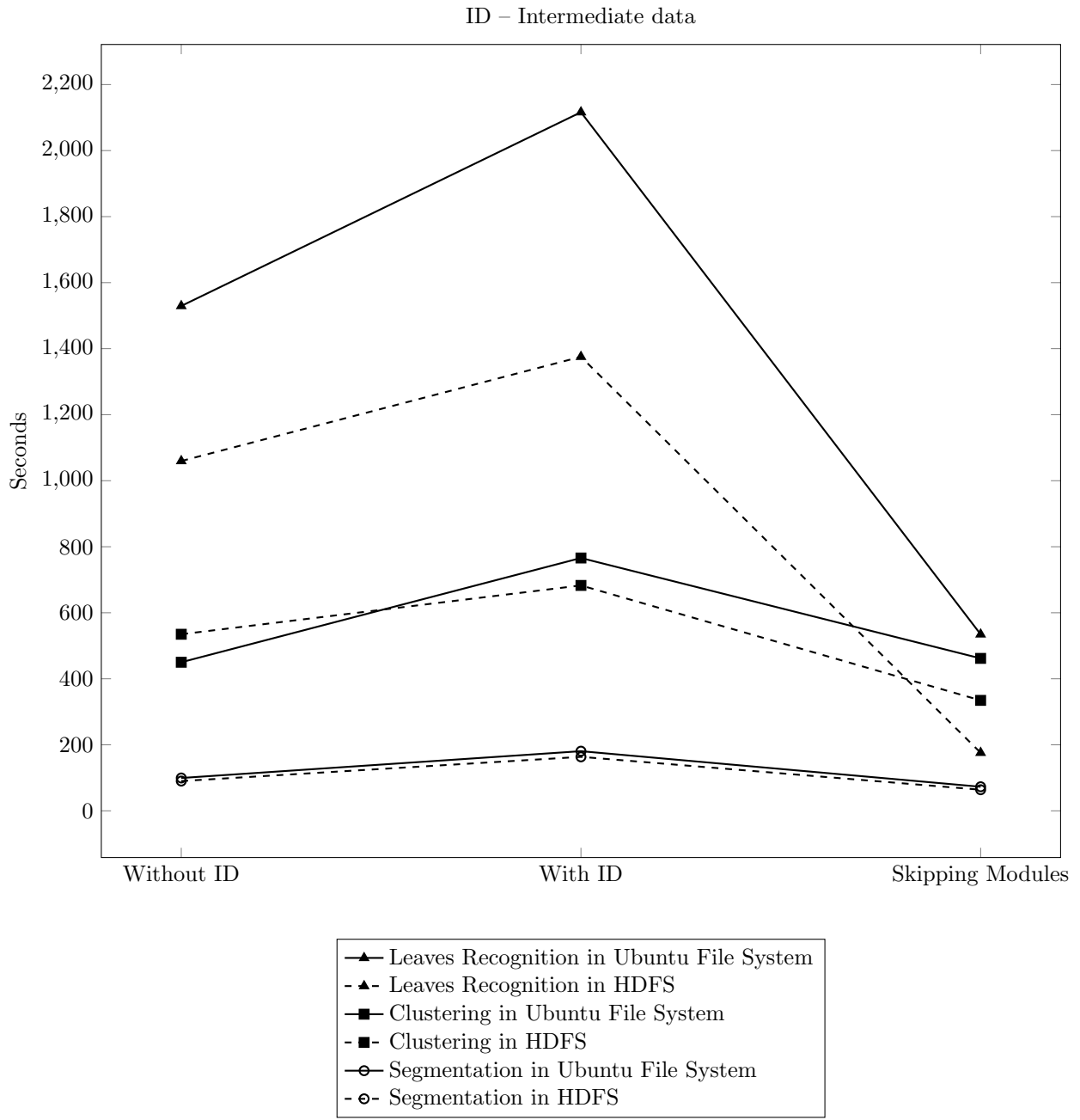

From our experimental studies, we figured out that although loading intermediate states from HDFS takes some extra time, the solution and possibility of using intermediate states shows better performance than loading raw large image files for the execution of image processing workflows. In Figure \ref{fig_performance_line}, normal execution to modules skipping performances have been illustrated for all of the three image processing pipelines.  For each pipeline, gain by skipping modules is directly proportional to their normal execution time, as much as the computation time increases, the gain is increased for the time too. This increment directly portraits that the proposed scheme is more effective for computationally intensive tasks than the average workload of pipelines.

We also experienced that discovering data reusability across various image processing tasks are challenging, and significant analysis is required in this regard. In this situation, breaking up an image processing task into micro-level tasks and offering APIs for each task with the micro-level modular programming concept is a viable solution. We learned in a system of modular programming a user has more control over a task configuration such as if a user wishes, certain steps can be used, or some can be skipped with respect to the required functionalities and data availability. Some other common benefits of modular program design that we experienced are - easy bug detection, less computation error, easy program handling, and modular testing. Although we noticed this kind of modularity hampers performance, with the appropriate reusability of intermediates states, it is possible to recover the loss. In order to figure out matching intermediate states across various image processing tasks, applying machine learning algorithms might be another solution. In this case, we need to tag intermediate states with metadata to associate them  with the tasks they might be used. A system should automatically suggest users for using intermediate states during compositions of workflows.

While we are working in our cluster, we gathered some valuable knowledge of a distributed processing framework such as for parallel processing the cores of different nodes are busy with different tasks of a specific job. So, if anyone wants to store any processed data, it needs to happen on a master node. To store any states, some data transition time from different nodes will be added to the storing procedure, which eventually increases a program's execution time. Another common problem of a distributed system is memory overhead exception. If a master node does not have enough main memory, programs cannot execute operations on a large dataset. For example, in our case, 30GB RAM of the master node could not process 4K images, later which was increased to 90GB and then 4K images were processed with a clustering algorithm. Another problem of a Spark based distributed system is, for rapid file system access program it throws IO exceptions or SSH library file loading exceptions. Although there are various challenges to process a large dataset in a distributed environment but to facilitate the real-time experience for Big Data, distributed processing environments are inevitable.

\section{Conclusion}
\label{conclusion}
We devised a data management scheme for our image processing pipeline framework, where we considered to store intermediate data and program state to reuse processed data and recover a pipeline failure with the data. When a dataset has a large number of images and associate programs need long computational time, there is a chance of high error rate. In spite of that, existing models have not concentrated on an organization of intermediate states except raw data and metadata. Considering the above statements, we proposed a model of intermediate state and showed that using intermediate state, it is easy to recover failures quicker and increase reusability in a SWfMS. In addition, intermediate states can be used in other pipelines to reduce the time cost, which eventually increases data reusability. To provide end users an interactive environment for processing Big Data in a parallel distributed framework and at the same time give them reusability with more control is a challenging job. Only a modularization of a program or task might not be a feasible solution in many cases. So, our model with data modularization or intermediate data state at every stage of a modular program will give a user more control of usability. Above all, our study contributes to Volume, Velocity and Variety creation in Big image processing. In the current practice of workflow or pipeline design, high computation capability resources may not facilitate users if users do not have more control over data usability. Hence, using the proposed model of intermediate states of data, resource utilization can be increased up to a certain level. Our experimental results were presented with such utilization, which makes possible the practical use of our system. In the next chapter, we have discussed the technique of automatic data storing and recommendation of our proposed data management scheme. 

\chapter{Optimized  Storing  of  Workflow  Outputs  through  Mining  AssociationRules}
\label{optstoring}

\joy{While an intermediate data management scheme of modularized computational-intensive programs can increase the reusability and performance in a SWfMS, an automatic decision mechanism is essential for storing and recommending data of workflows in a SWfMS. Workflows are frequently built and used to systematically process large datasets using workflow management systems (WMS). A workflow (i.e., a pipeline) is a finite set of processing modules organized as a series of steps that is applied to an input dataset to produce a desired output.} In a workflow management system, \joy{users} generally create workflows manually for their own investigations. However, workflows can sometimes be lengthy and the constituent processing modules might often be computationally expensive. \joy{In this situation, it would be beneficial if users could reuse intermediate stage results generated by previously executed workflows for executing their current workflow.}

In this chapter, we propose a novel technique based on association rule mining for suggesting which intermediate stage results from a workflow that a user is going to execute should be stored for reusing in \joy{the future.} \joy{We call our proposed technique, RISP (Recommending Intermediate States from Pipelines)}. According to our investigation on hundreds of workflows from two scientific workflow management systems, our proposed technique can efficiently suggest \joy{intermediate state results to store for future reuse}. \joy{The results that are suggested to be stored have a high reuse frequency.} Moreover, for creating around 51\% of the entire pipelines, we can reuse results suggested by our technique. Finally, we can achieve a considerable gain (74\% gain) in execution time \joy{by reusing intermediate results stored by the suggestions provided by our proposed technique.}
We believe that our technique (RISP) has the potential to have a significant positive impact on Big-Data systems, because it can considerably reduce execution time of the workflows through appropriate reuse of intermediate state results, and hence, can improve the performance of the systems.

\section{Motivation}

A scientific workflow management system (SWfMS) is a special type of WfMS (workflow management system) that lets its users perform computationally expensive and time-consuming tasks by decomposing them into sequentially organized and inter-dependent modules. Our research in this chapter  deals with workflows (i.e., pipelines) in scientific workflow management systems. \joy{Currently, scientific workflow management systems are commonly used in various scientific, engineering, and business organizations for conducting investigations, improving system efficiency, increasing productivity by reducing costs, and improving information exchange.} In Big-data analytics, when a large volume of heterogeneous data needs to be processed with various mechanisms, scientific workflow management systems (SWfMS) should be considered for efficiency.

In a SWfMS, users can manually build their workflows by selecting and sequentially adding processing modules from a finite set of modules available in the SWfMS for performing their desired investigations. Each workflow or pipeline works on a particular dataset provided by a user. The processing modules in a workflow are ordered in such a way that the output produced by a particular intermediate module can be used as input by the next module in the workflow. The output that we obtain from the last module is considered the final output from the workflow. Users often create workflows for processing large datasets. The intermediate state results produced by the modules in such workflows can also be very large. Moreover, each of the processing modules might require a considerable amount of time for data processing. \joy{In such a situation, it would be beneficial if a user could reuse results produced by previously executed workflows when the user plans to execute a workflow on the same dataset.} 

In order to provide automatic support for reusing results from \joy{previously} executed workflows, we need to have a mechanism for determining which of the intermediate state results obtained from a particular workflow have a high possibility of being reused in \joy{the future}. In our research presented in this chapter , we propose such a mechanism (i.e., technique) which we call RISP (Recommending Intermediate States from Pipelines). \joy{RISP} provides suggestions for storing intermediate state results by analyzing association rules between data and processing modules from the pipelines in the history. To the best of our knowledge, our study is the first one to investigate providing suggestions for storing intermediate state results from pipelines.

\joy{Previous studies considered storing all intermediate state results from pipelines. While storing all intermediate state results is good for provenance, it is not suitable from the perspective of reusability and resource utilization. To store} all intermediate state results from all pipelines, we need a significant amount of storage space. As new pipelines will be created, the size of the stored results will continuously increase. Also, it might be seen that many of the stored intermediate state results are not being reused at all. On the other hand, if we do not store any of the intermediate state results, we might need to build and execute the same workflows again and again. This might have a significant negative impact on the efficiency when the processing modules in the workflows are time-consuming. In this situation, our proposed technique, RISP, can be useful. It suggests intermediate state results for storing by analyzing their reuse possibility through mining association rules. The following example will explain our idea.

Let us assume that a SWfMS without any mechanism for storing intermediate state results has been used three times for executing three workflows as shown in Fig. \ref{workflows}. A user is now going to execute the fourth workflow. In such a situation, if we integrate RISP with this SWfMS, then it will suggest \joy{storing} the result that will be obtained from module, M2, of the fourth workflow. The reason behind making this suggestion is that the dataset D1 that is going to be processed in the fourth workflow was also processed in the first and third workflows and the modules M1 and M2 were executed serially in both of these workflows. Thus, there is a high possibility that when a user will attempt to create a workflow in the future using dataset D1, \joy{the user} will first apply the two modules M1 and M2 serially on D1. Moreover, \joy{when} executing the fourth workflow, if the result from M2 is stored in the system, then in \joy{the future when a user attempts to create a workflow using dataset D1, RISP will notify the user about the presence of the result that was stored from the fourth workflow. Section \ref{proposedtechnique} describes our suggestion technique.}

We implement our proposed technique, RISP, as a prototype tool and apply it on hundreds of pipelines created and used by the researchers and users in two scientific workflow management systems. We have the following findings:

\begin{figure}[t]
      \centering
      \includegraphics[width=1\textwidth]{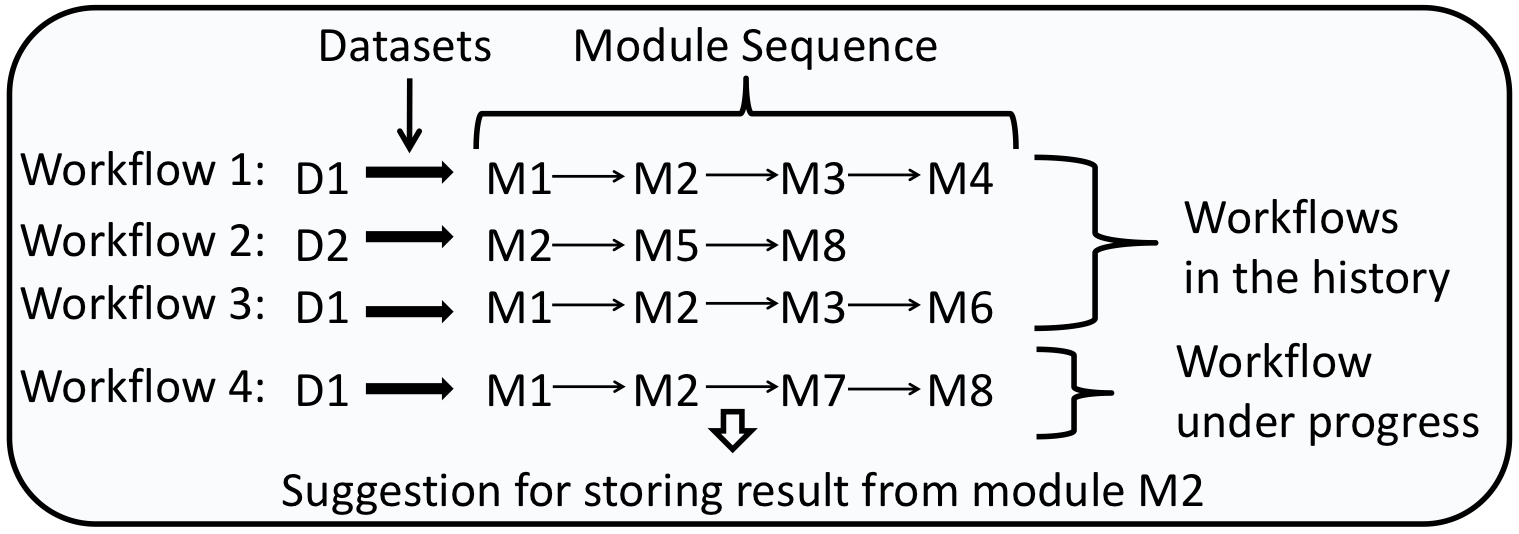}
      \caption{Automatically suggesting intermediate state results for storing from a workflow under progress.}
      \label{workflows}
\end{figure}

\textbf{(1)} By analyzing association rules from the previously executed pipelines, our proposed technique can automatically suggest which intermediate state results from a pipeline under progress should be stored for \joy{future reuse}.

\textbf{(2)} The intermediate state results that can be stored according to the suggestions from our proposed technique have a high frequency of being reused. The 508 pipelines that we investigated had 7165 possible intermediate state results. However, our technique suggests storing only 49 of these. Each of these 49 results can be reused 5 times on an average. The stored results can be reused for creating and executing around 51\% of the entire set of pipelines.

\textbf{(3)} By storing intermediate state results according to the suggestions from our proposed recommendation technique we can have a considerable gain (around 74\% gain) in execution time while executing pipelines.

Our findings indicate that our proposed technique (RISP) can significantly improve the performance of Big-Data systems through appropriate reuse of the intermediate state results from the workflows.

The rest of the chapter  is organized as follows. Section \ref{cha2_background} defines and describes association rules, Section \ref{proposedtechnique} presents our proposed technique for suggesting intermediate state results to store for reuse, Section \ref{experimentsetup} describes our experiment setup, Section \ref{comparison} compares our proposed technique with three other techniques in suggesting intermediate state results to store, Section \ref{threatsToValidity} describes possible threats to validity, Section \ref{relatedWork} discusses the related work, and finally, Section \ref{conclusions} concludes the chapter by mentioning future work.

\section{Background}
\label{cha2_background}
Association rules have been used in software engineering research and practice for performing impact analysis tasks. 
We define an association rule in the following way.

\textbf{Association Rule.}
An association rule \cite{Agrawal:1993:MAR:170036.170072} is an expression of the form $X => Y$ where $X$ is the antecedent, and $Y$ is the consequent. Each of $X$ and $Y$ is a set of one or more program entities. 
The meaning of such a rule is that if $X$ gets changed in a particular commit operation, $Y$ also has the tendency of being changed in that commit.

\textbf{Support and Confidence.}
As defined by Zimmermann et al. \cite{Zimmermann:2004:MVH:998675.999460}, \emph{support is the number of commits in which an entity or a group of entities changed together}.
The support of an association rule is determined in the following way.
\begin{equation}
	\mathit{support}(X=>Y)=\mathit{support}(X, Y)
\end{equation}
Here, $(X, Y)$ is the union of $X$ and $Y$, and so $\mathit{support}(X=> Y) = \mathit{support}(Y => X)$.
\emph{Confidence of an association rule, $X => Y$, determines the probability that $Y$ will change in a commit operation provided that $X$ changed in that commit operation}. We determine the confidence of $X => Y$ in the following way.

\begin{equation}
	\mathit{confidence}(X => Y) = \mathit{support}(X, Y) / \mathit{support}(X)
\end{equation}

In our research, we derive association rules between datasets and modules from pipelines and investigate those for providing suggestions regarding which intermediate state result from a pipeline under progress should be stored.

\begin{figure}[t]
      \centering
      \includegraphics[width=1\textwidth]{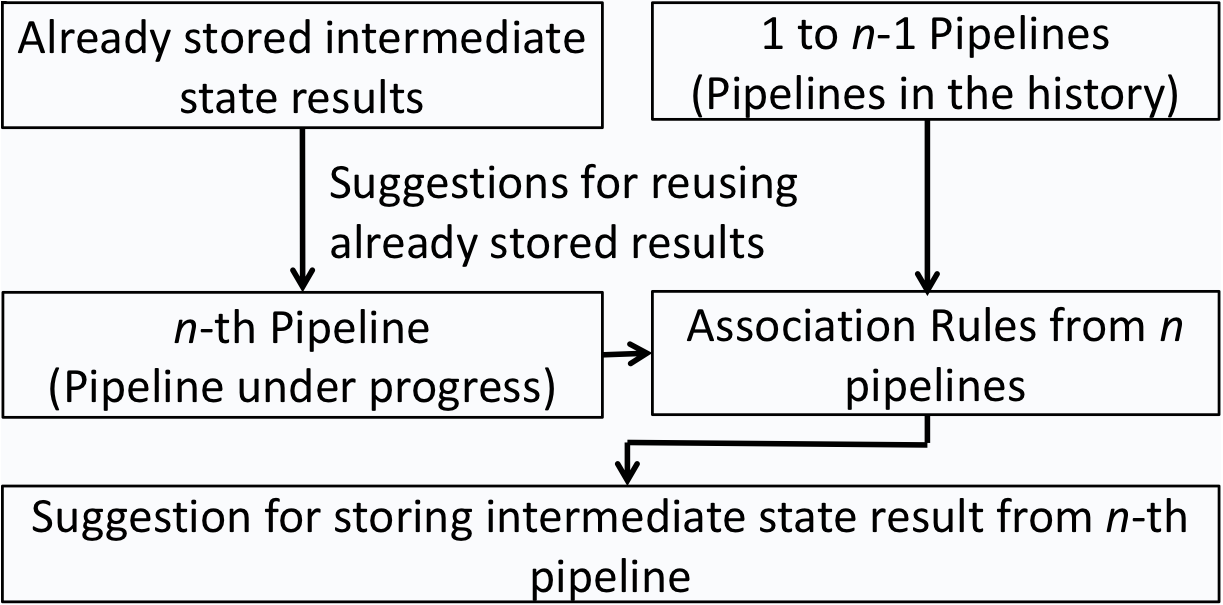}
      \caption{Our proposed technique for suggesting intermediate state results to store for \joy{future reuse}}
      \label{ourtechnique}
\end{figure}

\section{Recommending intermediate stage results to store for reusing}
\label{proposedtechnique}
Let us assume that a user is going to create the $n$-th pipeline in a scientific workflow management system (SWfMS) as shown in Fig. \ref{ourtechnique}. Let us further assume that during the creation of each of the previous $n-1$ pipelines, our recommendation system (RISP) recommended particular intermediate state results to store. Thus, we already have some stored results. The top left rectangle in Fig. \ref{ourtechnique} denotes these already stored results. When the user attempts to create the n-th pipeline, our recommendation system observes which dataset she is going to use and determines whether there are any stored intermediate state results using this dataset. These results are shown to the user so that she can decide whether she wants to reuse any of these already stored results. Let us assume that the user completes her pipeline with or without using the already stored intermediate state results. Just after she completes her pipeline (i.e., the $n$-th pipeline), our recommendation system determines all the distinct association rules from all the $n$ pipelines (i.e., $n-1$ previous pipelines and the $n$-th pipeline that the user is going to execute), calculates their supports and confidences, and analyzes these support and confidence values for recommending which of the intermediate state results from the newly created pipeline (i.e., $n$-th pipeline) should be stored. On the basis of this recommendation, one intermediate state result from the $n$-th pipeline might be stored if that was not stored previously.  In this way, when a user attempts to create a new pipeline in SWfMS, our recommendation technique performs the following two tasks:
\begin{itemize}
\item Recommending a number of already stored intermediate stage results to the user so that she can reuse it.
\item Recommending an intermediate stage result for storing from the pipeline she created by analyzing association rules from the pipelines in the history.
\end{itemize}
\subsection{Determining association rules from pipelines}
We derive association rules among datasets and modules by analyzing existing pipelines in the following way. Let us consider the first pipeline in Fig. \ref{workflows}. It consists of four modules (M1, M2, M3, and M4) that sequentially work on the dataset D1. The results generated from each of these modules except M4 are intermediate state results. The result that we obtain from M4 is the final result from the pipeline. It is possible to store results from each of these four modules. If we store result from a particular module, then it might discard the necessity of executing any previous modules including that particular one whenever a user needs to execute a similar pipeline in \joy{the future} with the same dataset. For example, if we store the result obtained from module M3, then for executing the possible pipeline (D1$\,\to\,$M1$\,\to\,$M2$\,\to\,$M3$\,\to\,$M7) in \joy{the future}, a user does not need to execute the first three modules. He can just reuse the previously stored result from module M3 and can only execute module M7 on the stored result. 

We determine association rules from a pipeline on the basis of how many results can be stored from it. For example, from the first pipeline in Fig. \ref{workflows}, we determine the following four association rules: \textbf{(1)} D1$=>$M1, \textbf{(2)} D1$=>$[M1, M2], \textbf{(3)} D1$=>$[M1, M2, M3], and \textbf{(4)} D1$=>$[M1, M2, M3, M4]. For the first rule, D1 is the \textbf{antecedent} and M1 is the \textbf{consequent}. For the second rule, the consequent is the module sequence (M1, M2). An association rule, for example the second one, means that if a user attempts to make a pipeline using the dataset D1 in \joy{the future}, then he has a possibility of applying the modules M1 and M2 sequentially on D1 as the first two modules. From all four pipelines in Fig. \ref{workflows}, we get ten distinct association rules: (1) D1$=>$M1, (2) D1$=>$[M1, M2], (3) D1$=>$[M1, M2, M3], (4) D1$=>$[M1, M2, M3, M4], (5) D2$=>$M2, (6) D2$=>$[M2, M5], (7) D2$=>$[M2, M5, M8], (8) D1$=>$[M1, M2, M3, M6], (9) D1$=>$[M1, M2, M7], and (10) D1$=>$[M1, M2, M7, M8].
\subsection{Determining the supports and confidences of the  association rules obtained from the pipelines}
After determining all the distinct association rules from all the pipelines, we determine the support and confidence of each of the rules. For example, we will now determine the support and confidence of the association rule D1$=>$M1. 

\textbf{Support of an association rule.} Support of an association is the number of times it can be generated from the pipelines. The support of the association rule, D1$=>$M1, is 3 because, from all the four pipelines in Fig. \ref{workflows}, we can generate this association rule three times (i.e., from the first, third, and fourth pipelines). In the same way, the supports of the association rules, D1$=>$[M1, M2] and D1$=>$[M1, M2, M3], are 3 and 2 respectively. We express the support of an association rule formally in the following way.

\begin{equation}
support (D1=>M1) = 3
\end{equation}

\textbf{Confidence of an association rule.} We determine the confidence of the association rule D1$=>$M1 in the following way from its support value:

\begin{equation}
confidence (D1=>M1) = \frac{support (D1=>M1)}{support(D1)}
\end{equation}
    Here, support(D1) is the number of times D1 was used in the pipelines. Thus, support (D1) = 3, because D1 was used in three pipelines according to Fig. \ref{workflows}. Finally, confidence (D1$=>$M1) = 3/3 = 1. The highest confidence of an association rule is 1. Such a confidence for the association rule D1$=>$M1 implies that if some one attempts to make a pipeline in \joy{the future} using the dataset D1, then there is a high probability that he will choose M1 as the first module in the pipeline. The confidence of the association rule D1$=>$[M1, M2, M3] is 0.66 (confidence(D1$=>$[M1, M2, M3]) = support(D1$=>$[M1, M2, M3])/support(D1) = 2/3). Thus, if a user attempts to build a pipeline in \joy{the future} using the dataset D1, then there is a little probability (0.66) that he will use M1, M2, and M3 respectively as the first three modules in his pipeline.
\subsection{Recommending an intermediate state result for storing}
In Fig. \ref{workflows}, we see that the fourth pipeline is the one that is going to be executed (under progress). For recommending which intermediate state result should be stored from this pipeline, we sort the association rules made from this pipeline on the basis of their confidence values and determine the highest confidence rules. For example, from the fourth pipeline in Fig. \ref{workflows} we get four association rules: D1$=>$M1, D1$=>$[M1, M2], D1$=>$[M1, M2, M7], and D1$=>$[M1, M2, M7, M8] with confidences 1, 1, 0.33, and 0.33 respectively. The highest confidence rules are D1$=>$M1, D1$=>$[M1, M2]. We select the longest of these highest confidence rules for making suggestion because it helps us skip the highest number of modules. Thus, from the fourth pipeline, we recommend to store the result obtained from module M2.

\section{Experiment Setup}
\label{experimentsetup}
For conducting our experiment, we downloaded 508 workflows from Galaxy public server \cite{Afgan2016TheUpdate} at \emph{usegalaxy.org}. These workflows were created and executed on the SWfMS of Galaxy during the time between February, 2010 and August, 2018. The downloaded workflows were text-based files with a specific JSON format. We automatically retrieved the module execution sequence and dataset details of each workflow from its corresponding file using our implementation. 
The workflows were executed mostly for investigations related to bioinformatics with biological datasets as the inputs. We applied our recommendation technique (RISP) on these workflows to investigate how efficiently it can recommend intermediate state results for storing so that the results can be reused in \joy{the future}. Section \ref{comparisonresults} describes our investigation.

We also integrated our proposed technique with the SWfMS in the Plant Phenotyping and Imaging Research Center called P2IRC in USASK. This SWfMS runs on an OpenStack based Spark-Hadoop cluster having 6 nodes, 40 cores, and 40 GB RAM. It is frequently used for making pipelines to analyze and investigate large volumes of image data. The image data is stored in the Hadoop Distributed File System (HDFS) of the cluster. 
For investigating how much gain in execution time can be achieved by storing intermediate state results according to the suggestions provided by our proposed technique, we first integrated our technique (RISP) with the SWfMS, and then executed 32 image processing pipelines. The input dataset of each of these 32 pipelines consists of 4000 to 10000 images. 
The pipelines were created for different purposes such as image segmentation, image registration, counting the number of flowers, and image clustering. The Canola datasets (4KCanola, 10KCanola) of the P2IRC project of USASK were used as inputs for the pipelines. The findings of our investigation regarding execution time gain will be described in Section \ref{executiontimegainlabel}.
\section{Evaluating our proposed technique} 
\label{comparison}
\subsection{Candidate techniques for comparison}
We compared our proposed technique (RISP) with three other candidate recommendation techniques by doing investigation on 508 pipelines that we downloaded from Galaxy public server \cite{Afgan2016TheUpdate}. All four recommendation techniques (including our proposed one) are listed below.
\begin{itemize}
	\item \textbf{PT} \joy{(Proposed Technique-RISP): RISP recommends} storing intermediate state results indicated by the association rules with highest confidence.
	\item \textbf{TSAR} (Technique that recommends Storing All Results): This second technique recommends storing each of the intermediate state results of each of the pipelines.
	\item \textbf{TSPAR} (Technique that recommends Storing Previously Appeared Results): The third technique recommends storing those intermediate state results that were produced previously at least once.
	\item \textbf{TSFR} (Technique that recommends Storing the Final Result): This technique recommends storing the final result (output) from a pipeline.
\end{itemize}

The following paragraphs describe how we evaluate these techniques for making a comparison among them.

\textbf{Procedure of investigation using PT (Proposed technique).}
For evaluating this technique, we analyze each of the pipelines in the history serially beginning from the first one.
While examining the $n$-th pipeline, we analyze our technique's recommendation capability in the following way.

First, we apply our technique to determine which intermediate state results might have already been stored if our technique was integrated with the SWfMS from the very beginning of the history. For this purpose, our system analyzes association rules from the previous $n$-1 pipelines. If we see that any of the intermediate state results of the $n$-th pipeline have already been stored previously, then we can reuse that intermediate state result in this $n$-th pipeline, and also, we realize that our system previously took a correct decision regarding storing those intermediate state results. 

After completing the first step, we make association rules from this $n$-th pipeline and determine their supports and confidences by dealing with the association rules from the previous $n-1$ pipelines. We select the highest confidence rule(s) from the $n$-th pipeline and assume that we have stored the intermediate state result denoted by the longest of these highest confidence rules. In this way, we examine all the pipelines in the history and determine the number of cases where we could reuse previously stored intermediate state results if we could use our proposed technique. 

\textbf{Procedure of investigation using TSAR (Technique that recommends Storing All Results).}
By applying this second technique we analyze the pipelines from the very beginning one serially. When analyzing the $n$-th pipeline, we first check the database to see if any of the stored intermediate state results can be useful for skipping some processing modules in the pipeline. If several results are available, then the result that helps us skip the highest number of processing modules is used. After executing the pipeline, we store each of its intermediate state results in the database. If a particular result is already in the database, we do not need to store it again. 

\textbf{Procedure of investigation using TSPAR (Technique that recommends Storing Previously Appeared Results).}
The third candidate technique is a variant of our proposed technique. While our proposed technique depends on the confidence values of the association rules for deciding which of the intermediate state results should be stored from the pipeline under progress, the third technique solely depends on the support values of the association rules. When analyzing the $n$-th pipeline, this technique first checks which of the already stored intermediate state results can be the most appropriate one for reusing for this pipeline. Then, it determines the association rules from the pipeline. It identifies which of the rules have a support of at least one in the previous usage history (i.e., first $n-1$ pipelines). The intermediate state result indicated by the longest one of these association rules is considered for storing.

\textbf{Procedure of investigation using TSFR (Technique that recommends Storing the Final Result).} 
The fourth technique stores the final outcome of each of the pipelines. Thus, if the same pipeline is attempted to be re-executed in the future, then we can just reuse the result from the first execution. We do not need to again execute any module in the pipeline. We consider such a technique for comparison because we wanted to understand how often the same pipelines get executed.
\subsection{Investigated measures}
\label{fourmeasures}
We calculate four measures for our investigation for each of the candidate systems. The measures and their calculation mechanisms have been discussed below.
\begin{itemize}
	\item \textbf{LR} (Likeliness of Reusing from previously stored results). This measure determines how often we can reuse intermediate state results that can be stored according to the suggestions from a candidate system.
	\item \textbf{PSRR} (Percentage of Stored Results that were Reused). The second measure determines what percentage of the intermediate state results that can be stored according to the suggestions from a candidate recommendation technique can be reused in future.
	\item \textbf{FRSR} (Frequency of Reusing Stored Results). This measure determines how many times a stored intermediate state result was reused on an average.
	\item \textbf{PISRS} (Percentage of Intermediate State Results that were Stored). This measure calculates what percentage of the intermediate state results were stored for reusing according to the suggestions from a candidate recommendation technique.
\end{itemize}
\textbf{Calculation mechanism for LR.} By sequentially analyzing all the pipelines from the history, we determine two quantities: (1) the total number of pipelines that we have analyzed, (2) the number of pipelines for which we could reuse previously stored intermediate state results. From these quantities we calculate LR using the following equation.

\begin{equation}
\label{measure1}
LR = \frac{\mathit{\begin{split}Number~of~pipelines~for~which~we\\could~reuse~previously~stored~results\end{split}}}{Total~number~of~pipelines} \times 100
\end{equation}
\textbf{Calculation mechanism for PSRR.} 
If we see that only a very small percentage of the previously stored results get reused (most of the stored results remain unused) during creating pipelines, then the corresponding recommendation mechanism should not be considered as an efficient one. Thus, when comparing the efficiencies of the candidate recommendation mechanisms, we should also focus on the second measure (PSRR). It determines what percentage of the stored results get reused during creating pipelines. By examining all the pipelines in the history we determine two quantities: (1) the total number of intermediate state results that were stored by a candidate mechanism and (2) the number of intermediate state results that could be reused. We then calculate PSRR by the following equation.

\begin{equation}
PSRR = \frac{No.~of~reused~results}{No.~of~stored~results} \times 100
\end{equation}

\textbf{Calculation mechanism for FRSR.} The measure FRSR (Frequency of Reusing Stored Results) focuses on how many times a previously stored result was used during creating pipelines. If it is observed that the intermediate state results stored according to the recommendation from a candidate technique were used rarely during creating pipelines, then the technique should not be regarded as an efficient one. For calculating FRSR, we determine two quantities by examining the entire set of pipelines: (1) the total number of intermediate state results that were stored from the recommendations of a candidate technique and (2) the total number of times the stored results were used for making pipelines. We then calculate this measure according to the following equation.

\begin{equation}
	\mathit{FRSR} = \frac{No.~of~times~ stored~ results~ were~ reused}{No.~ of~ stored~ results}
\end{equation}

\textbf{Calculation mechanism for PISRS.} The measure PISRS (Percentage of Intermediate State Results that were Stored) calculates what percentage of the intermediate state results that were produced during the execution of the entire set of pipelines were stored according to the recommendation of a candidate technique. A candidate technique that stores comparatively less number of results than other techniques but ensures higher re-usability should be considered an efficient one. For measuring PISRS, we examine the whole set of pipelines and determine the following two quantities for each recommendation technique: (1) the total number of possible intermediate states (including the final states), (2) the total number of intermediate state results that were stored. We then calculate PISRS using the following equation.
\begin{equation}
	PISRS = \frac{No.~ of~ stored~ results}{No.~ of~ intermediate~ states} \times 100
\end{equation}
\subsection{Comparing the candidate recommendation techniques on the basis of the measures}
\label{comparisonresults}
\begin{table}[]
	\caption{Workflow information for the comparison}
	\label{workflowinfo}
		\begin{tabular}{|p{90mm}|p{6mm}|p{12mm}|p{14mm}|l|} \hline
			\textbf{Description of calculated measures}&\textbf{PT}&\textbf{TSAR}&\textbf{TSPAR}&\textbf{TSFR}\\ \hline
			Number of investigated pipelines considering each candidate technique & 508& 508& 508 &508\\ \hline
			Number of pipelines for which we could reuse previously stored data &264	&314	&261	&70\\ \hline
			Number of intermediate state results that were saved according to the suggestion from a candidate technique &49	&7165	&159	&457\\
			\hline
			\multicolumn{5}{|l|}{\scriptsize  \textbf{PT} = Proposed Technique \textbf{TSAR} = Technique that Recommends Storing  All intermediate Results}\\
			\multicolumn{5}{|l|}{\scriptsize  \textbf{TSPAR} = Technique that Recommends Storing Previously Appeared Results \textbf{TSFR} = Technique that Recommends Storing the Final Result}\\
			\hline
			
		\end{tabular}
\end{table}

\begin{figure}[t]
	\pgfplotstableread{
		1	51.968502
		2	61.811024
		3	51.377953
		4	13.779528
	}\dataset
	\begin{tikzpicture}
	\begin{axis}
	[ybar=0.12cm,
	bar width=0.45cm,
	width=1\textwidth,
	height=4.5cm,
	ymin=0,
	ymax=62,        
	yticklabel style={font=\scriptsize},
	xtick=data,
	xticklabels = { 
	PT, 
	TSAR,
	TSPAR, 
	TSFR},
	xticklabel style={yshift=0ex, rotate=0, font=\scriptsize},
	major x tick style = {opacity=0},
	minor x tick num = 1,
	minor tick length=1ex,
	ymajorgrids = true,
	legend entries={LR (\% of pipelines that could be made by reusing previously stored intermediate state results)
	},			
	legend style={
		at={(0.5,1.02)},
		anchor=south,
		legend columns=1,
		font=\scriptsize,
		text width=2.909in,
		minimum height=0.28in,
	},
	]
	\addplot[draw=black, fill=black!70] table[x index=0,y index=1] \dataset;
	\end{axis}
	\end{tikzpicture}
	
	\caption{Comparing the candidate techniques according to the percentage of pipelines that could be made by reusing previously stored results.}
	\label{lr}

\end{figure}
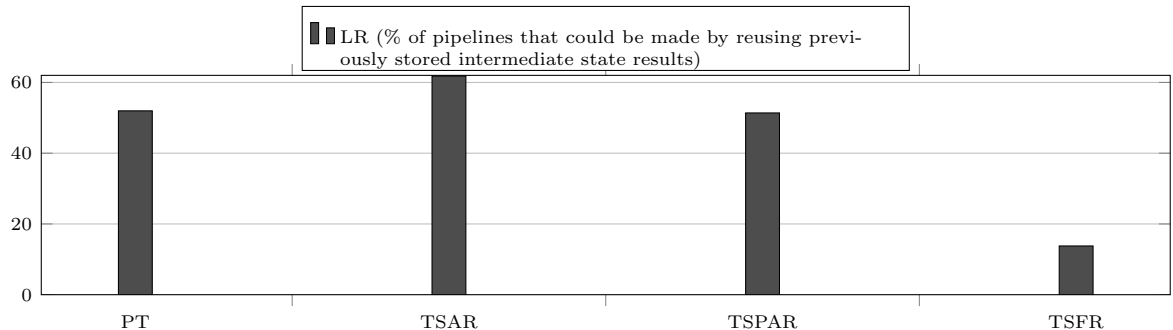
The following paragraphs compare the four candidate techniques (PT, TSAR, TSPAR, and TSFR) on the basis of the four measures (LR, PSRR, FRSR, and PISRS).

\textbf{Comparison regarding LR (Likeliness of Reusing from previously stored results)}
For each candidate technique, Fig. \ref{lr} shows the percentages of pipelines such that while building those pipelines we could reuse previously stored intermediate state results.
We can see that for the second candidate technique (TSAR), the percentage of pipelines (61.81\%) that could be built using previously stored results is the highest among the four techniques.
However, to achieve this percentage, we need to store all 7165 intermediate state results from 508 workflows (as indicated in Table \ref{workflowinfo}), and a SWfMS can face a huge storage overhead for storing such a big amount of results. For our proposed technique (PT), the likeliness of reusing from stored results is around 51.97\%. Although this percentage is smaller compared to TSAR, we can achieve this by storing only 49 intermediate state results. We also see that the percentage regarding our proposed technique is higher than the percentages regarding the third and fourth candidate techniques. From Table \ref{workflowinfo} we see that the third and fourth techniques suggest storing 159 and 457 intermediate state results respectively. Thus, even by storing a considerably smaller number of intermediate state results, our technique ensures a higher likeliness of reusing.

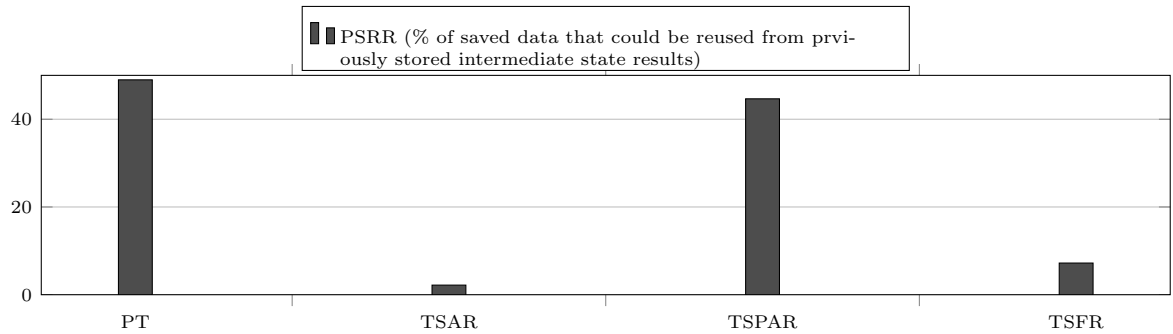
\begin{figure}[t]
	\pgfplotstableread{
1	48.97959
2	2.1912072
3	44.65409
4	7.2210064
	}\dataset
	\begin{tikzpicture}
	\begin{axis}
	[ybar=0.12cm,
	bar width=0.45cm,
	width=1\textwidth,
	height=4.5cm,
	ymin=0,
	ymax=50,        
	yticklabel style={font=\scriptsize},
	xtick=data,
	xticklabels = { PT, TSAR, TSPAR, TSFR},
	xticklabel style={yshift=0ex, rotate=0, font=\scriptsize},
	major x tick style = {opacity=0},
	minor x tick num = 1,
	minor tick length=1ex,
	ymajorgrids = true,
	legend entries={PSRR (\% of saved data that could be reused from prviously stored intermediate state results)
	},			
	legend style={
		at={(0.5,1.02)},
		anchor=south,
		legend columns=1,
		font=\scriptsize,
		text width=2.909in,
		minimum height=0.28in,
	},
	]
	\addplot[draw=black, fill=black!70] table[x index=0,y index=1] \dataset;
	\end{axis}
	\end{tikzpicture}
	\caption{Comparing the candidate techniques on the basis of the percentage of saved data that could be reused from previously stored results}
	\label{psrr}
\end{figure}

\textbf{Comparison regarding PSRR (Percentage of Stored Results that were Reused)}
Fig. \ref{psrr}  makes a comparison among the candidate techniques by considering the PSRR measure.
We see that the PSRR value regarding our proposed technique (PT) is the highest among all the techniques. The second candidate technique TSAR (that saves all the results) exhibits the lowest value (2.19\%). Thus, only 2.19\% of the intermediate state results stored according to the suggestion from TSAR can be reused. The remaining 97.81\% of the stored results (i.e., around 7008 of the 7165 stored results) stay unused. Finally, our proposed technique ensures the highest reuse of stored results.

\begin{figure}[t]
	\pgfplotstableread{
1	5.387755
2	0.043824144
3	1.6415094
4	0.15317287
	}\dataset
	\begin{tikzpicture}
	\begin{axis}
	[ybar=0.12cm,
	bar width=0.45cm,
	width=1\textwidth,
	height=4.5cm,
	ymin=0,
	ymax=6,        
	yticklabel style={font=\scriptsize},
	xtick=data,
	xticklabels = { PT, TSAR,TSPAR, TSFR},
	xticklabel style={yshift=0ex, rotate=0, font=\scriptsize},
	major x tick style = {opacity=0},
	minor x tick num = 1,
	minor tick length=1ex,
	ymajorgrids = true,
	legend entries={FRSR (Frequency of reusing previously stored intermediate state results)
	},			
	legend style={
		at={(0.5,1.02)},
		anchor=south,
		legend columns=1,
		font=\scriptsize,
		text width=2.909in,
		minimum height=0.24in,
	},
	]
	\addplot[draw=black, fill=black!70] table[x index=0,y index=1] \dataset;
	\end{axis}
	\end{tikzpicture}
	\caption{Comparing the candidate techniques on the basis of the frequency of reusing previously stored intermediate state results}
	\label{frsr}
\end{figure}
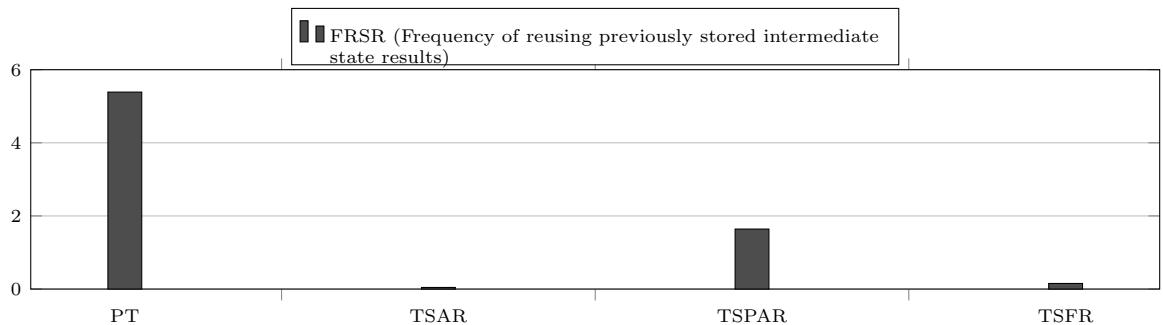

\textbf{Comparison regarding FRSR (Frequency of Reusing Stored Results).}
Fig. \ref{frsr} illustrates how frequently the intermediate state results stored according to the suggestion from a candidate technique get reused during making pipelines. From the figure we again realize that our proposed technique (PT) exhibits the highest reuse frequency (5.39) among all four techniques.  
Finally, our comparison in Fig. \ref{frsr} establishes our proposed technique to be the best one.

\textbf{Comparison regarding PISRS (Percentage of Intermediate State Results that were Stored)}\
Fig. \ref{pisrs} compares the candidate techniques on the basis of the PISRS measure.
According to our definition of PISRS, the technique with the lowest PISRS value should be regarded as the most efficient one. Fig. \ref{pisrs} shows that our proposed technique (PT) suggests storing the smallest percentage (0.68\%) of all 7165 intermediate state results. According to Table \ref{workflowinfo}, this percentage (0.68\%) indicates storing of 49 results only. The percentage regarding candidate technique TSAR is 100\% because it recommends storing all 7165 results. The PISRS value for each of the other two techniques is higher compared to our proposed technique. Thus, even according to this last measure (PISRS) our proposed technique performs the best.

\begin{figure}[t]
	\pgfplotstableread{
1	0.68
2	100
3	2.21
4	6.378227495
	}\dataset
	\begin{tikzpicture}
	\begin{axis}
	[ybar=0.12cm,
	bar width=0.45cm,
	width=1\textwidth,
	height=4.5cm,
	ymin=0,
	ymax=100,        
	yticklabel style={font=\scriptsize},
	xtick=data,
	xticklabels = { PT, TSAR, TSPAR, TSFR},
	xticklabel style={yshift=0ex, rotate=0, font=\scriptsize},
	major x tick style = {opacity=0},
	minor x tick num = 1,
	minor tick length=1ex,
	ymajorgrids = true,
	legend entries={PISRS (\% of intermediate state results that were saved from total generated intermediate states)
	},			
	legend style={
		at={(0.5,1.02)},
		anchor=south,
		legend columns=1,
		font=\scriptsize,
		text width=2.909in,
		minimum height=0.28in,
	},
	]
	\addplot[draw=black, fill=black!70] table[x index=0,y index=1] \dataset;
	\end{axis}
	\end{tikzpicture}
	\caption{Comparing the candidate techniques on the basis of the percentage of intermediate state results that were saved from all intermediate states}
	\label{pisrs}
\end{figure}


\begin{figure} [tp]
	\includegraphics[width=1\textwidth]{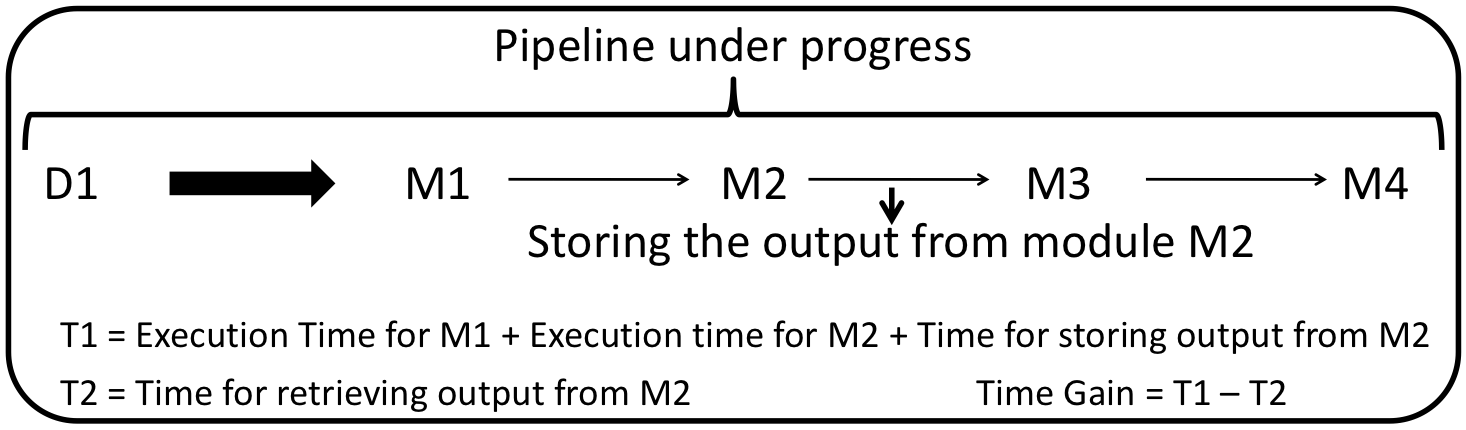} 
	\caption{Calculation of execution time gain}
	\label{executiontimegain}
\end{figure}


    

\subsection{Considering module execution time for evaluating our proposed recommendation technique}
\label{executiontimegainlabel}
Execution times of the processing modules in a pipeline should be considered important when making decisions about which intermediate state results we should store from a pipeline. We consider module execution time to evaluate our proposed technique in the following way.

Let us consider the pipeline in Fig. \ref{executiontimegain}. A recommendation technique recommends to store the result obtained from module M2. Let us assume that T1 is the time which is required to execute M1 and M2 and store results from M2 to HDFS (Hadoop Distributed File System). T2 is the time to retrieve the result of M2 from HDFS. Now, only if T1 is greater than T2, then storing the result from M2 is beneficial. 
Eq. \ref{eq9} calculates the execution time gain in this case.

\begin{equation}
\label{eq9}
Execution~Time~Gain = T1 - T2
\end{equation}
We consider this timing factor for determining whether storing an intermediate state result from a pipeline according to the suggestion from our technique is beneficial.

\textbf{Our investigation regarding module execution time.} For conducting this investigation, we executed 32 pipelines in the scientific workflow management system (SWfMS) of a plant phenotyping based image research center (P2IRC). This SWfMS runs on a parallel and distributed environment empowered by a SPARK-Hadoop cluster consisting of 6 nodes, 40 cores and 40GB RAM. 
While executing the 32 pipelines, we recorded the following three things from each pipeline:
    \textbf{(1)} The time that was needed to execute each of the modules in the pipeline,
    \textbf{(2)} The time that was needed to store the output from a particular module in the HDFS.
    \textbf{(3)} The time that was needed to retrieve the previously stored output of a particular module from the HDFS.
    
We apply our recommendation mechanism on the 32 pipelines that we executed on the SWfMS and determine which modules could be skipped from which pipelines if we stored the intermediate state results according to the suggestions from our recommendation technique. We finally determine the possible gain in execution time if we could reuse previously stored results.

\begin{figure*}
	\includegraphics[width=1.0\textwidth, height=40mm]{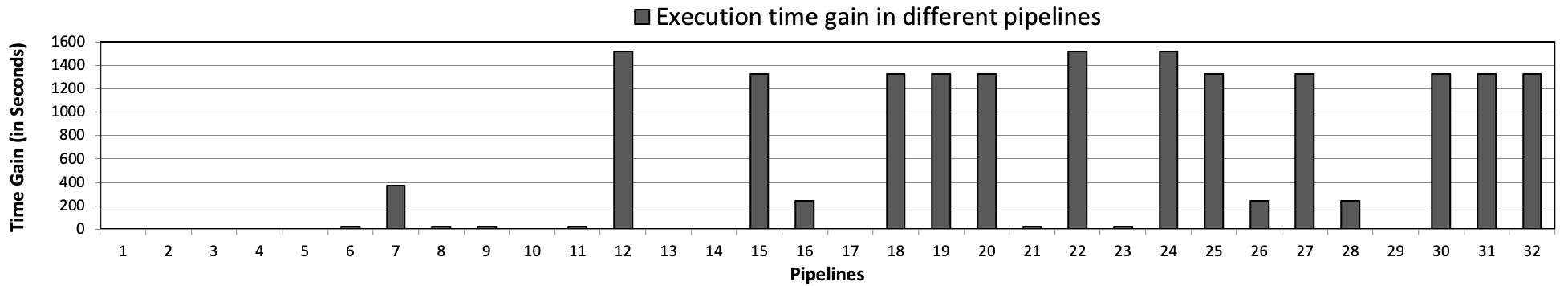}    
	\caption{Execution time gain for different pipelines by reusing previously stored intermediate state results according to our proposed technique}
	\label{timegain}
\end{figure*}

Fig. \ref{timegain} shows the execution time gain (in seconds) calculated using Eq. \ref{eq9} for each of the 32 pipelines. For the first few pipelines, we did not have a gain in execution time because while creating these pipelines there were no stored results for reusing. However, for the remaining ones, we often got a considerable gain. 
After executing all 32 pipelines by reusing results stored according to our proposed technique's suggestions, we can have a total gain of 17720 seconds (4 hours 55 minutes and 20 seconds). 

We also calculated time for executing these 32 pipelines without reusing stored results and found that 23865 seconds were required for executing all pipelines. However, if we reuse the results stored according to the suggestions from our proposed technique, we can execute all the pipelines in only 6145 seconds (i.e., we can save 17720 seconds). In other words, we can save around 74\% of total execution time by reusing results. Thus, our proposed technique can help us achieve a considerable gain in execution time. 
\section{Threats to Validity}
\label{threatsToValidity}

We analyzed 540 (508 pipelines from Galaxy public server \cite{Afgan2016TheUpdate} and 32 pipelines from the SWfMS of P2IRC) pipelines in our experiment for analyzing our proposed technique's efficiency in making suggestions. While a higher number of pipelines could make our findings more generalized, we see that our technique exhibits efficient performance on pipelines from two different workflow management systems. \joy{Thus, we believe that our findings cannot be attributed to chance.} Our proposed technique can be considered an important contribution towards managing pipelines.
\section{Related Work}
\label{relatedWork}
A great many studies \cite{Missier:2016:PDD:2915589.2915594} \cite{zhao2004semantically} 
\cite{Yuan:2010:DPS:1838759.1838814} \cite{Deelman2008DataWorkflows} \cite{Deelman2008TheExample} \cite{Hampton2013BigEcology} \cite{Goodman2014TenData} \cite{White2013NineData}
have been conducted on analyzing, reproducing, linking, visualizing, and managing workflows in a scientific workflow management system. We discuss these studies in the following paragraphs.

Woodman et al. \cite{Woodman2015WorkflowProvenance} proposed an algorithm for determining which subset of the intermediate state results from a pipeline can be stored with the lowest cost.
While their study solely focuses on storage cost, our study is fundamentally different because we focus on the re-usability of stored results. We apply association rule mining technique for identifying which intermediate state results should be stored so that they can be reused for making pipelines in future.

Yuan et al. \cite{Yuan2011On-demandSystems} proposed an algorithm for determining which set of intermediate state results can be stored with the minimum cost. Our study is different because we focus on the maximum reuse of the stored results. We propose a technique on the basis of association rule mining for determining which intermediate state results from workflows should be stored for future reuse.

Koop et al. \cite{Koop2008VisComplete:Pipelines} proposed a technique for automatically completing a pipeline under progress by analyzing pipeline execution history. While their technique focuses on identifying which processing modules could be added after an anchor module in a partially completed pipeline, our study has a completely different focus. We propose a technique for identifying which of the intermediate state results from a pipeline should be stored for reusing.

Many studies \cite{Chinthaka2009CBRAssistant} \cite{Zhang2011Recommend-as-you-go:Reuse} \cite{Spjuth2015ExperiencesBioinformatics} have investigated the possibility of discovering and reusing services and workflows in a SWfMS. Our study is fundamentally different than those studies because we do not aim to help users in building pipelines. We identify which of the intermediate state results from a pipeline under progress should be stored for reuse.

From our above discussion, it is clear that none of the existing studies investigated recommending intermediate state results from a pipeline for storing so that the stored result can be reused in the future. To the best of our knowledge, our study is the first attempt towards such a recommendation. 
Our in-depth investigation on a good number of workflows indicates that our proposed technique can efficiently suggest which intermediate state result from a workflow should be stored for reusing. 

\section{CONCLUSION}
\label{conclusions}

In this chapter, we propose and investigate a novel technique (RISP) for suggesting which intermediate state result from a pipeline under progress should be considered for storing so that the result can be reused in the future. For making suggestions, our technique mines association rules from the already executed pipelines in the history and analyzes their support and confidence measures. We analyze the efficiency of our technique by applying it on 508 workflows downloaded from Galaxy public server. According to our investigation, our proposed technique can efficiently suggest which intermediate state result from a pipeline under progress should be stored for reusing. The 508 workflows that we investigated 
had 7165 intermediate state results in total. However, our recommendation technique suggests storing only 49 of these results. We find that these 49 results can be reused for building around 51\% of the entire pipelines. Moreover, the intermediate state results stored according to the suggestions from our technique exhibit a high reuse frequency.

We also apply our technique on the scientific workflow management system (SWfMS) in a plant phenotyping based image research center (P2IRC) for investigating how much gain in execution time can be achieved by reusing the stored results. From our investigation on 32 workflows that we executed on the SWfMS we find that by reusing previously stored results we can save around 74\% of the execution time that would be required if we did not reuse previously stored results. Findings from our research make us realize that our proposed recommendation technique (RISP) has the potential to significantly improve the performance of Big-Data systems by suggesting appropriate reuse of the intermediate state results from the scientific workflows. 

Although the automation of the data storing and recommendation can be achieved by the execution sequences of workflows, consideration of the states of the tools in RISP can give user proper recommendation with more user control in a SWfMS.In the next chapter, we have discussed how RISP works with the states of tools in a SWfMS.

\chapter{Optimal Storing Modes of Workflow in A Scientific Workflow Management System Considering Tool State}
\label{OptimalMode}
In a Scientific Workflow Management System (SWfMS) workflows are frequently built for processing large datasets using a sequence of processes. The processing modules through which a job passes from initiation to completion need to be configured and tuned with parameters in their composition and execution. Configuring and tuning parameters in a workflow can have a significant impact on certain types of model that can change the states of processes (i. e. modules). Different outputs might be produced for a sequence of modules with identical input datasets in different workflows for a different set of parameter configuration. Composing a lengthy workflow manually for a desired investigation where constituent processing modules might often be computationally expensive for their current states and datasets could be burdensome to SWfMS users. In this circumstance, proper management of data to reuse intermediate stage results generated by previously executed workflows would be beneficial to assist both composition and computation of a current workflow in a SWfMS. In our previous work, we propose a novel technique known as RISP(Recommending Intermediate States from Pipelines) based on association rule mining for suggesting which intermediate stage results from a workflow that a user is going to execute should be stored for reusing in the future. In this chapter, an extended version of the technique is presented to include module states of workflows for both recommending and storing phase. Using the extended version, around 40\% of the entire workflows can be created by reusing results suggested by our technique. 

\section{Motivation}
As a framework a Scientific Workflow Management System (SWfMS) offers the means to completely prepare, manage, monitor, and execute a scientific experiment in terms of sequentially organized and inter-dependent modules. Our research in this chapter deals with the sequential modules (i.e., workflows) in Scientific Workflow Management Systems. Optimizations in such SWfMSs are essential for investigating on large datasets to improve system efficiency and increase productivity. Big data analytics with a large volume of heterogeneous data and tools in a SWfMS must be considered with a cost reduction technique as a single module can generate multiple outputs, and workflow can have a large number of data and intermediate data. Workflows are frequently built and used to process large datasets, which can gradually increase data volume in a system. For this Big Data Velocity characteristics, incremental data from workflows need to be handled with an automatic recommendation technique for proper resource utilization.

In a SWfMS, users can manually build their workflows by selecting and sequentially adding processing modules from a finite set of modules available in the SWfMS for performing their desired investigations. Each workflow or pipeline works on particular input datasets provided by a user. The processing modules in a workflow are ordered in such a way that the output produced by a particular intermediate module can be used as input by the next module in the workflow. A particular module in a workflow could work in a different way than the same module in other workflows for parameter configuration and tuning. Thus, a specific module in different workflows can generate different outputs from identical inputs. The output that we obtain from the last module is considered the final output from a workflow. Users often create workflows for processing large datasets. The intermediate state results produced by the modules in such workflows can also be very large. Moreover, each of the processing modules might require a considerable amount of time for data processing. In such a situation, it would be beneficial if a user could reuse results produced by previously executed workflows when the user plans to execute a workflow on the same dataset.

In order to provide automatic support for reusing results from previously executed workflows, we need to have a mechanism for determining which of the intermediate state results obtained from a particular workflow have a high possibility of being reused in the future. In our research presented in the previous chapter \ref{optstoring}, we proposed such a mechanism (i.e., technique) which we call RISP (Recommending Intermediate States from Pipelines). RISP provides suggestions for storing intermediate state results by analyzing association rules between data and processing modules from the pipelines in history. To the best of our knowledge, RISP is the first one to investigate providing suggestions for storing intermediate state results from pipelines.

Previous studies considered storing all intermediate state results from pipelines. While storing all intermediate state results is good for provenance, it is not suitable from the perspective of reusability and resource utilization. To store all intermediate state results from all pipelines, we need a significant amount of storage space. As new pipelines will be created, the size of the stored results will continuously increase. Also, it might be seen that many of the stored intermediate state results are not being reused at all. On the other hand, if we do not store any of the intermediate state results, we might need to build and execute the same workflows again and again. This might have a significant negative impact on the efficiency when the processing modules in the workflows are time-consuming. In this situation, our proposed technique, RISP, can be useful. But matching of parameters and their values of each module while building a current workflow with previous parameters and values of modules of workflows need to be considered for the proper recommendation. RISP suggests intermediate state results for storing by analyzing their reuse possibility through mining association rules and this study presents an extended version of the RISP considering the states of modules. The following example will explain our idea.

\begin{figure}
  \includegraphics[width=1\textwidth]{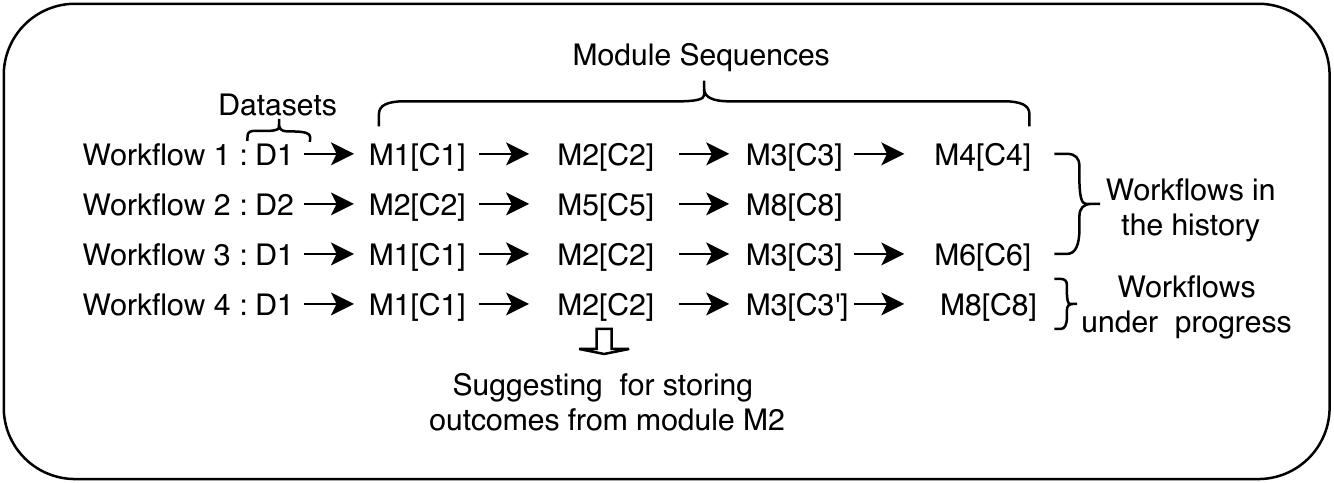}
  \caption{Automatically suggesting intermediate state results for storing from a workflow under progress.}
  \label{exampleworkflows}
\end{figure}

Let us assume that a SWfMS without any mechanism for storing intermediate state results has been used three times for executing three workflows, as shown in Fig. \ref{exampleworkflows}. A user is now going to execute the fourth workflow. In such a situation, if we integrate RISP with this SWfMS, then it will suggest storing the result that will be obtained from the module, M2, of the fourth workflow. The reason behind making this suggestion is that the dataset D1 that is going to be processed in the fourth workflow was also processed in the first and third workflows and the modules M1 and M2 were executed serially in both of these workflows. In this situation, one might also think that RISP will suggest storing the outcome of M3 as the sequence M1, M2, and M3 also executed serially in first and third workflows. But, the M3 module in the fourth workflow runs with different parameter configuration set C3', which is different from the previously executed sequence configuration sets. So, the extended version will only suggest for the outcome of M2 by satisfying the common parameters configuration. Thus, there is a high possibility that when a user attempts to create a workflow in the future using dataset D1, the user will first apply the two modules M1 and M2 serially on D1 including the same parameter configuration sets (i.e., C1 and C2). Moreover, when executing the fourth workflow, if the result from M2 is stored in the system, then in the future when a user attempts to create a workflow using dataset D1, Extended version of the RISP will notify the user about the presence of the result(s) with parameter matching information that was stored from the fourth workflow. 

We implement our extended version of RISP, as a prototype tool and apply it on hundreds of pipelines created and used by the researchers and users in two scientific workflow management systems. We have the following findings:

(1) By analyzing association rules from the previously executed pipelines and considering module state, the extended version of the RISP can automatically suggest which intermediate state results from a pipeline under progress should be stored for future reuse.

(2) The intermediate state results that can be stored according to the suggestions from our technique have a high frequency of being reused. The 534 pipelines that we investigated had 8510 possible intermediate state results. However, our technique suggests storing only 61 of these. Each of these 61 results can be reused 3 times on an average. The stored results can be reused for creating and executing around 40\% of the entire set of pipelines.


Our findings indicate that our proposed technique (RISP) can significantly improve the performance of Big-Data systems through appropriate reuse of the intermediate state results from the workflows.

The rest of the chapter is organized as follows. Section \ref{background_toolstate} defines and describes association rules, Section \ref{experimentsetuptoolstate} describes our experiment setup, Section \ref{comparisontoolstate} compares our proposed technique with three other techniques in suggesting intermediate state results to store, Finally Section \ref{conclusiontoolS} concludes the chapter.

\section{Background}
\label{background_toolstate}
Association rules are conditional statements that assist to determine the probability of connections among items in a large set. Association rules have been used in many research areas for discovering correlations and can be defined in the following way.

An association rule is consist of two parts: antecedent (i.e., if) and consequent (i.e., then). The antecedent is an item of the previous history, and consequent is an item that comes with the antecedent. 

\begin{figure}
  \includegraphics[width=1\textwidth]{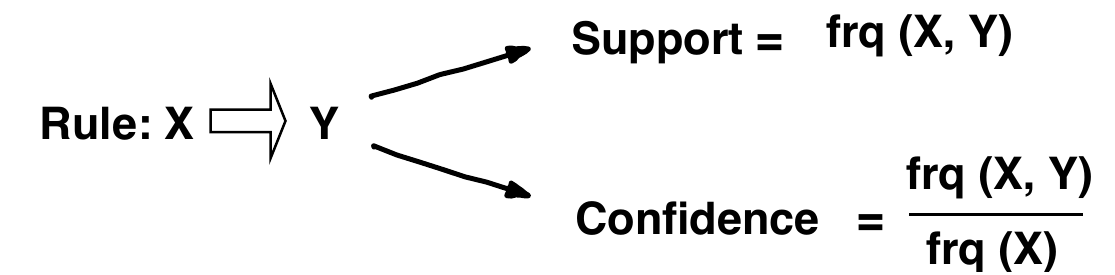}
  \label{AssociationRule}
\end{figure}

\textbf{Support} is an indication of how frequently the rules appear in the history. 

\textbf{Consequent} indicates the probability of the rules by the support of the antecedent. 

Here, $(X, Y)$ is the union of $X$ and $Y$, and so $\mathit{support}(X=> Y) = \mathit{support}(Y => X)$.
\emph{Confidence of an association rule, $X => Y$, determines the probability that $Y$ will change in a commit operation provided that $X$ changed in that commit operation}. So the confidence of $X => Y$ can be determined  in the following way.

In our technique RISP, we derive association rules between datasets and modules from pipelines and investigate those for providing suggestions regarding which intermediate state result from a workflow under progress should be stored. 

\section{Experiment Setup}
\label{experimentsetuptoolstate}
For our experimental analysis, we considered 534 workflows from Galaxy public server at \emph{usegalaxy.org} and these workflows were JSON formatted text-based files. We parsed the dataset details and module execution sequence with their tool state information from each pipeline using our own implementation. We applied our recommendation technique (adaptive RISP) on these workflows to investigate how efficiently it can recommend intermediate state results for storing so that the results can be reused in the future. Section \ref{comparisonresultstoolstate} describes the experimental results with comparison.

\section{Evaluating our proposed technique} 
\label{comparisontoolstate}
\subsection{Candidate techniques for comparison}
For evaluation, we compared our proposed adaptive RISP technique with three other candidate recommendation techniques. All recommendation techniques (including our proposed one) are listed below; these techniques are introduced in our previous chapter \ref{optstoring} with detail descriptions.

\textbf{PT (Proposed technique).}
The proposed adaptive RISP technique gradually examines each of the pipelines and recommends storing the intermediate state results based on the highest confidence value, which is calculated by the association rules. During the $n$-th pipeline execution, all possible sequences of the pipeline have been matched with previously-stored sequences of $n-1$ pipelines, then after parameter matching previously-stored results can be reused for the $n$-th pipeline that ensures to skip some module executions.

\textbf{TSAR (Technique that recommends Storing All Results).}
This technique also analyzes all possible subsequences of a workflow serially and recommends storing all the intermediate state results of each of the subsequence. While investigating the $n$-th pipeline, the technique checks previously-stored data by matching subsequences for recommending intermediate data.  If the subsequences are matched with configuration values, then the system can reuse the existing data. Otherwise, the technique recommends storing the result of the sequence after execution. Though, this technique helps to skip the highest number of processing modules because it saves all possible intermediate state results, but require huge storage space to store all of the intermediate data.

\textbf{TSPAR (Technique that recommends Storing Previously Appeared Results).}
The third technique recommends storing those intermediate state results of subsequences, which were executed at least one time in history. That means this technique considers storing the intermediate state result depends on the support values of the association rules. This is a variant of our proposed method.

\textbf{TSFR (Technique that recommends Storing the Final Result).} 
This last technique recommends storing only the ultimate outcome of a pipeline. Thus, if the same pipeline is attempted for execution in the future, then this technique can help to skip the whole execution time and reuse the output. This technique is considered only for comparison because it helps to understand how often the same pipelines get executed.

\subsection{Investigated measures}
\label{fourmeasurestoolS}
For comparing among the techniques, we take into acoount four measures such as Likeliness of Reusing from previously-stored results (LR), Percentage of Stored Results that were Reused (PSRR), Frequency of Reusing Stored Results (FRSR), Percentage of Intermediate State Results that were Stored (PISRS). The calculation mechanisms of these measures have been discussed below.

\textbf{LR Calculation.} 
To compare the performance, we measure how frequently an SWfMS can reuse intermediate state results that are stored according to the suggestions from a candidate system. By sequentially analyzing all the downloaded pipelines, we calculate the total number of pipelines that we have analyzed, and the number of pipelines for which we could reuse previously-stored intermediate state results. From these quantities, we determine LR using the following equation.

\begin{equation}
\label{measure1toolS}
LR = \frac{\mathit{\begin{split}Number~of~pipelines~for~which~we\\could~reuse~previously~stored~results\end{split}}}{Total~number~of~pipelines} \times 100
\end{equation}
     
\textbf{PSRR Calculation.} 
Here we calculate what percentages of intermediate state result user can reuse in the future based on the quantity of the stored data of a candidate recommendation technique. This measurement is essential to consider because, if a little percentage of the stored result get reused during pipeline formation, that means remain portion of the data is unused with a considerable amount of storage cost. For Calculating PSRR, we determine the total number of intermediate state results that were stored by a candidate mechanism and, the number of intermediate state results that could be reused. We then calculate PSRR by the following equation.

\begin{equation}
PSRR = \frac{No.~of~reused~results}{No.~of~stored~results} \times 100
\end{equation}

\textbf{FRSR Calculation.} This measure focuses on average how several times a saved intermediate state result was reused during workflow creation. 
If the stored intermediate state results used hardly according to the recommendation from a candidate technique, then the method should not be considered as a suitable process. For calculating FRSR, we determine the total number of intermediate state results that were stored and, the total number of times the stored results were used for creating pipelines. We then calculate this measure according to the following equation.

\begin{equation}
    \mathit{FRSR} = \frac{No.~of~times~ stored~ results~ were~ reused}{No.~ of~ stored~ results}
\end{equation}

\textbf{Calculation mechanism for PISRS Calculation.} 
This measure calculates the percentage of the intermediate state results were stored for reuse while executing the pipeline according to the suggestions from a candidate recommendation technique. If the value of this measurement is comparatively low of any technique, then that technique ensures cost-effectiveness, which is considered an efficient one.  For measuring PISRS, we examine the total number of possible intermediate states (including the final states), and the total number of intermediate state results that were stored. We determine PISRS using the following equation.

\begin{equation}
    PISRS = \frac{No.~ of~ stored~ results}{No.~ of~ intermediate~ states} \times 100
\end{equation}

\subsection{Comparing the candidate recommendation techniques on the basis of the measures}
\label{comparisonresultstoolstate}
\begin{table}[]
	\caption{Workflow information for the comparison}
	\label{workflowinfotoolS}
		\begin{tabular}{|p{90mm}|p{6mm}|p{12mm}|p{14mm}|l|} \hline
			\textbf{Description of calculated measures}&\textbf{PT}&\textbf{TSAR}&\textbf{TSPAR}&\textbf{TSFR}\\ \hline
			Number of investigated pipelines considering each candidate technique & 534& 534& 534 &534\\ \hline
			Number of pipelines for which we could reuse previously stored data &200	&247	&209	&69\\ \hline
			Number of intermediate state results that were saved according to the suggestion from a candidate technique &61	&7598	&197	&475\\
			\hline
			\multicolumn{5}{|l|}{\scriptsize  \textbf{PT} = Proposed Technique \textbf{TSAR} = Technique that Recommends Storing  All intermediate Results}\\
			\multicolumn{5}{|l|}{\scriptsize  \textbf{TSPAR} = Technique that Recommends Storing Previously Appeared Results \textbf{TSFR} = Technique that Recommends Storing the Final Result}\\
			\hline
			
		\end{tabular}
\end{table}

\begin{figure}[t]
	\pgfplotstableread{
		1	39.761433 
		2	49.10537
		3	41.55069
		4	13.717694
	}\dataset
	\begin{tikzpicture}
	\begin{axis}
	[ybar=0.12cm,
	bar width=0.45cm,
	width=\textwidth,
	height=4.5cm,
	ymin=0,
	ymax=60,        
	yticklabel style={font=\scriptsize},
	xtick=data,
	xticklabels = {	PT, TSAR, TSPAR, TSFR},
	xticklabel style={yshift=0ex, rotate=0, font=\scriptsize},
	major x tick style = {opacity=0},
	minor x tick num = 1,
	minor tick length=1ex,
	ymajorgrids = true,
	legend entries={LR (\% of pipelines that could be made by reusing previously stored intermediate state results)
	},			
	legend style={
		at={(0.5,1.02)},
		anchor=south,
		legend columns=1,
		font=\scriptsize,
		text width=2.909in,
		minimum height=0.28in,
	},
	]
	\addplot[draw=black, fill=black!70] table[x index=0,y index=1] \dataset;
	\end{axis}
	\end{tikzpicture}
	\caption{Comparing the candidate techniques according to the percentage of pipelines that could be made by reusing previously stored results.}
	\label{lrtools}
\end{figure}
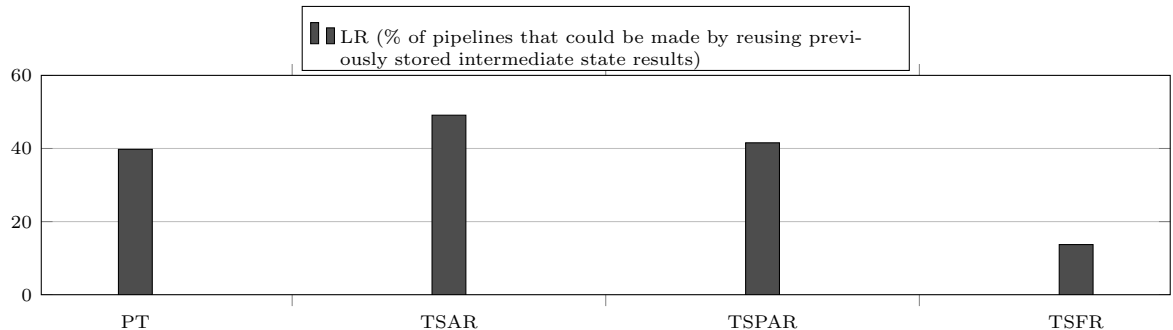

The following paragraphs compare the four candidate techniques (PT, TSAR, TSPAR, and TSFR) on the basis of the four measures (LR, PSRR, FRSR, and PISRS).

\textbf{Comparison regarding LR}

From Fig. \ref{lrtools}, we can see that the percentage of reusing from previously-stored intermediate state results of our proposed technique (PT) is around 40\%. However, the second candidate technique (TSAR) shows a higher percentage than the PT, at 49\%. According to Table \ref{workflowinfotoolS}, we can see that the PT method stores only 61 intermediate states result from 534 workflows where TSAR technique stores a high number of intermediate state results, that is 7598. For that reason, a SWfMS with TSAR can encounter a storage overhead problem.

Besides, it is seen that (Fig. \ref{lrtools}) the data usage percentage of our proposed technique is higher than the other two recommending techniques, TSPAR and TSFR. Moreover, the third and fourth techniques suggest storing 197 and 475 intermediate state results respectively (Table \ref{workflowinfotoolS}). Therefore, our technique ensures a higher likeliness of reusing by storing a considerably smaller number of intermediate state results.

\begin{figure}[t]
	\pgfplotstableread{
1	32.78688
2	1.553040
3   28.934011
4	4.842105
	}\dataset
	\begin{tikzpicture}
	\begin{axis}
	[ybar=0.12cm,
	bar width=0.45cm,
	width=\textwidth,
	height=4.5cm,
	ymin=0,
	ymax=40,        
	yticklabel style={font=\scriptsize},
	xtick=data,
	xticklabels = { PT, TSAR, TSPAR, TSFR},
	xticklabel style={yshift=0ex, rotate=0, font=\scriptsize},
	major x tick style = {opacity=0},
	minor x tick num = 1,
	minor tick length=1ex,
	ymajorgrids = true,
	legend entries={PSRR (\% of saved data that could be reused from prviously stored intermediate state results)
	},			
	legend style={
		at={(0.5,1.02)},
		anchor=south,
		legend columns=1,
		font=\scriptsize,
		text width=2.909in,
		minimum height=0.28in,
	},
	]
	\addplot[draw=black, fill=black!70] table[x index=0,y index=1] \dataset;
	\end{axis}
	\end{tikzpicture}
	\caption{Comparing the candidate techniques on the basis of the percentage of saved data that could be reused from previously stored results}
	\label{psrrtoolS}
\end{figure}
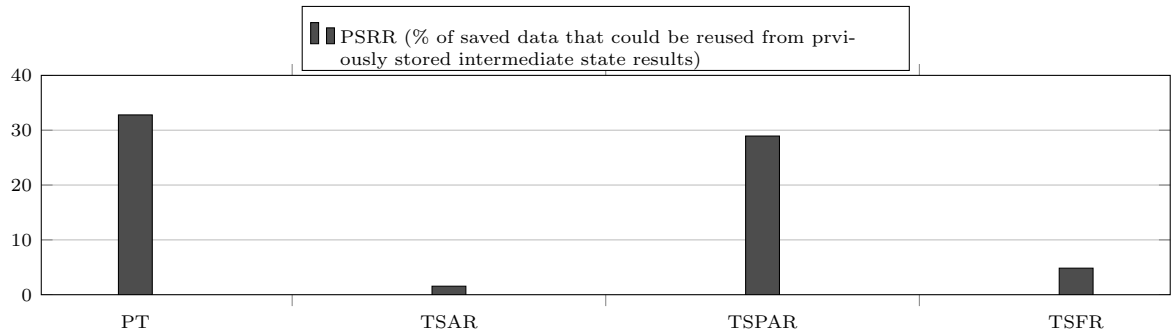

\textbf{Comparison regarding PSRR}
In Figure \ref{psrrtoolS}, our proposed technique can reuse around 32\% from the stored intermediate states results, and it is the highest rate among the four technique. Besides, all intermediate results storing process (TSAR) presents a small percentage (1.55\%) value can be reused by storing data, where most of the stored results will be unused. The other two techniques TSPAR (i.e., stores if previously appeared at once) and TSFR (i.e., keeps the outcome only) exhibit the PSRR value is 28\% and 4\% respectively. Conclusively, with PSRR calculation, among the four technique, our proposed technique ensures the highest reuse of stored results.

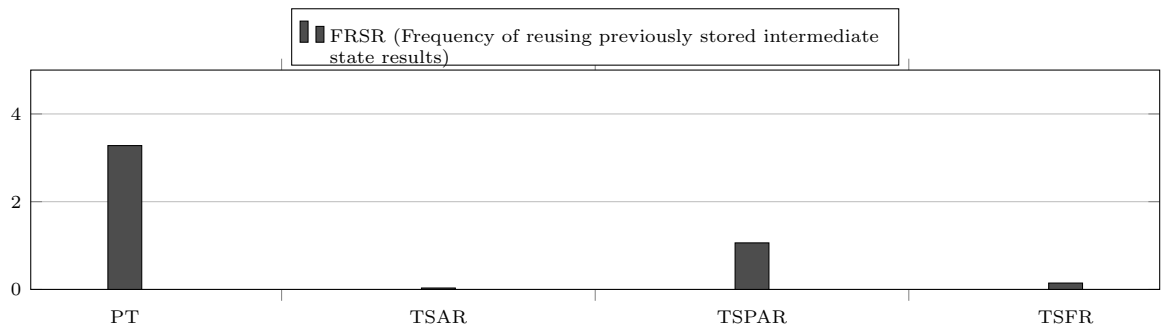
\begin{figure}[t]
	\pgfplotstableread{
1	3.2786884
2	0.0325085
3	1.0609137
4	0.1452631
	}\dataset
	\begin{tikzpicture}
	\begin{axis}
	[ybar=0.12cm,
	bar width=0.45cm,
	width=\textwidth,
	height=4.5cm,
	ymin=0,
	ymax=5,        
	yticklabel style={font=\scriptsize},
	xtick=data,
	xticklabels = { PT, TSAR,TSPAR, TSFR},
	xticklabel style={yshift=0ex, rotate=0, font=\scriptsize},
	major x tick style = {opacity=0},
	minor x tick num = 1,
	minor tick length=1ex,
	ymajorgrids = true,
	legend entries={FRSR (Frequency of reusing previously stored intermediate state results)
	},			
	legend style={
		at={(0.5,1.02)},
		anchor=south,
		legend columns=1,
		font=\scriptsize,
		text width=2.909in,
		minimum height=0.24in,
	},
	]
	\addplot[draw=black, fill=black!70] table[x index=0,y index=1] \dataset;
	\end{axis}
	\end{tikzpicture}
	\caption{Comparing the candidate techniques on the basis of the frequency of reusing previously stored intermediate state results}
	\label{frsrtoolS}
\end{figure}

\textbf{Comparison regarding FRSR}
Here we compare the value of how frequently the intermediate state results stored according to the suggestion from a candidate technique get reused during the creation of workflows.  In figure \ref{frsrtoolS}, we realize that our proposed technique (PT) exhibits the highest reuse frequency (i.e., around 3) among all four techniques.  On the other hand, the other three techniques show reuse frequency below two times. Therefore, the matrics, FRSR, establishes the proposed method is an efficient one in terms of frequent reusability.

\textbf{Comparison regarding PISRS}
In the mesaures PISRS, we calculate the percentage of intermediate state results that were stored for four techniques. According to the definition, if a technique has the lowest PISRS value, that should be the most efficient one. In this experiment in Figure \ref{pisrs}, our proposed technique suggests storing only 61 results where total intermediate state result is 8510, so the PISRS value of PT is only 0.71\%. The second technique TSAR stores total 7598 intermediate state result, so the percentage regarding candidate technique TSAR is 89\%. Furthermore, The PISRS value of the other two technique is higher compared to our proposed technique. Therefore, in the last measure (i.e., PISRS), our proposed technique also performs the best among all four candidate techniques.

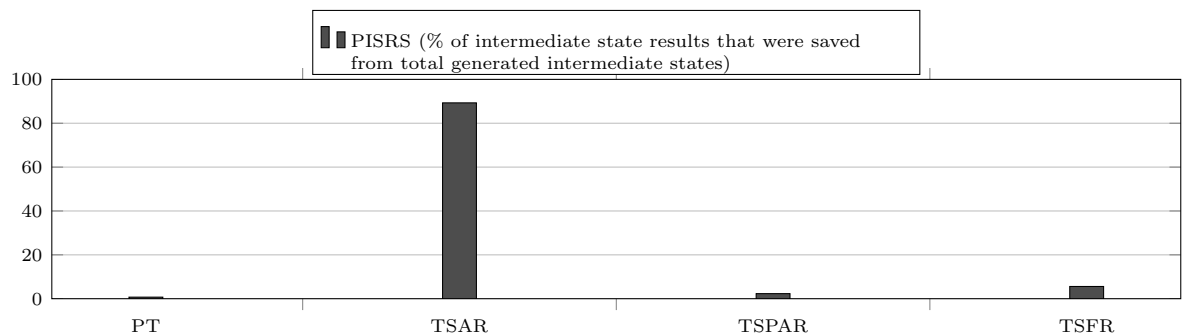
\begin{figure}[t]
	\pgfplotstableread{
1	0.7168038
2	89.283195
3	2.3149235
4	5.581669
	}\dataset
	\begin{tikzpicture}
	\begin{axis}
	[ybar=0.12cm,
	bar width=0.45cm,
	width=\textwidth,
	height=4.5cm,
	ymin=0,
	ymax=100,        
	yticklabel style={font=\scriptsize},
	xtick=data,
	xticklabels = { PT, TSAR, TSPAR, TSFR},
	xticklabel style={yshift=0ex, rotate=0, font=\scriptsize},
	major x tick style = {opacity=0},
	minor x tick num = 1,
	minor tick length=1ex,
	ymajorgrids = true,
	legend entries={PISRS (\% of intermediate state results that were saved from total generated intermediate states)
	},			
	legend style={
		at={(0.5,1.02)},
		anchor=south,
		legend columns=1,
		font=\scriptsize,
		text width=2.909in,
		minimum height=0.28in,
	},
	]
	\addplot[draw=black, fill=black!70] table[x index=0,y index=1] \dataset;
	\end{axis}
	\end{tikzpicture}
	\caption{Comparing the candidate techniques on the basis of the percentage of intermediate state results that were saved from all intermediate states}
	\label{pisrstoolS}
\end{figure}

\section{Conclusion}
\label{conclusiontoolS}
In a SWfMS, users usually create workflows manually by selecting and sequentially adding processing modules from a finite set of modules. This process of creating workflows is time-consuming, and computationally expensive modules require a considerable amount of time in a workflow execution for accomplishing a desired task. In our previous work, we proposed a recommendation technique RISP, which suggests storing the intermediate state result in a SWfMS. In this work, by emphasizing parameter configurations and tuning of modules, we consider an extended version of the technique. The technique is designed to make the recommendation compatible in a SWfMS where a tool as a module is presented with its state instance. By analyzing 534 workflows from Galaxy public server, we found that total 8510 intermediate state results recommendation technique suggests storing only 61 of these results. Besides, the proposed technique can be used to reuse data for around 40\% of the entire workflows in a SWfMS with a high frequency of reusability. Therefore, the proposed extended version of the RISP has a significant potential to improve the efficiency of data management in a SWfMS even with the tool states of modules in workflows.

To test the effectiveness and efficiency of the RISP in a SWfMS we also conducted several performance evaluations and user studies. Effectiveness and efficiency of our proposed technique RISP are discussed in the next chapter.

\chapter{Designing for Recommending Intermediate States in A Scientific Workflow Management System}
\label{DesignRISP}
To process a large amount of data sequentially and systematically,  proper management of workflow components (i.e., modules, data, configurations, associations among ports and links) in a Scientific Workflow Management System (SWfMS) is inevitable. Managing data with provenance information in a SWfMS to support reusability of workflows, modules, and data is not a simple task.  Handling such components is even more burdensome for frequently assembled and executed complex workflows for investigating large datasets with different technologies (i.e., various learning algorithms or models). However, a great many studies propose various techniques and technologies for managing and recommending services in a SWfMS, but only a very few studies consider the management of data in a SWfMS for efficient storing and facilitating workflow executions. Furthermore, there is no study to inquire about the effectiveness and efficiency of such data management in a SWfMS from a user perspective. In this chapter, we present and evaluate an GUI version of such a novel approach of intermediate data management with some use cases. The technique, we call RISP (Recommending Intermediate States from Pipelines) can facilitate execution of workflows with stored processed data and can thus reduce the computational time of some modules in SWfMS. For investigating the effectiveness of the techniques in a SWfMS, storing and recommending scenarios of the techniques are analyzed from a user viewpoint. We integrated RISP with an existing workflow management system called SciWorCS.  In SciWorCS, we also designed an interface using which users use for selecting the recommendation for intermediate states. We investigated GUI-RISP's effectiveness from users perspectives along with measuring its overhead in terms of storage and efficiency of workflow execution.

\section{Motivation}
In Big-data analytics when a large volume of heterogeneous data need to be processed with different technologies (i.e., various learning algorithms or models), proper data flow and module/service associations in a SWfMS are crucial for efficiency. Users often assemble workflows manually in SWfMSs, where processing modules are selected for the purpose of a particular task. Besides, users frequently run similar workflows by changing only a few modules for finding suitable methodologies to investigate on a particular input dataset \cite{inproceedingsBigD18}. In such frequently executed workflows, processing of large datasets eventually generates some computationally expensive modules in a workflow of a SWfMS. Investigating the same dataset with several workflows comprising of computationally expensive modules ultimately requires a long execution time. In such a situation, to experiment with new methodologies by changing only a few modules of a workflow could be a tedious task. Since modules are reusable in different workflows, similarly consideration of reusable outcomes of the modules in a SWfMS could be a solution for increasing the performance of a workflow execution.

However, improving the performance of a workflow by introducing reusability and storing all of the generated data in a SWfMS is not a feasible solution as the generated data or intermediate data are also large enough to increase the storing and storage cost dramatically. As well as storing all of the generated data from a workflow can hamper the resource utilization in a SWfMS. In a workflow assembling process, where different port and parameter configurations for different types of data and data flow are involved, a proper decision-making scheme is necessary in the runtime to store, to support reusability, to introduce efficiency and to manage storage cost of the heterogeneous data. Consequently, proper management of such data and metadata (i.e., association information of module to data, module to module and module to configuration) in a system of workflow processing is essential to facilitate the execution of workflows at low costs. 

To introduce such data management in our previous work
, we proposed a technique of automatic data recommendation for storing and retrieving intermediate data of workflows in a SWfMS. The proposed technique in the study 
, with a data management scheme is intended to facilitate workflow assembling and executions, and this study is designed to evaluate the technique in a real-world SWfMS from users perspectives. Furthermore, a GUI(Graphical User Interface) version of the technique is considered in the SciWorCS system, where users can have more control and choice to select intermediate data. Multiple intermediate data in a SWfMS can be found for different states of a corresponding tool. Such as parameter configuration or tuning in a tool can generate different intermediate states in different workflows. If we consider the tool states in the association rules of the RISP by considering parameter matching, we can give more choices to users for selecting intermediate states. More details of this GUI version have been discussed in Section \ref{technique_integration_overview}.   

Nowadays, SWfMSs are used to manage modules and data in a way that can facilitate building workflows interactively for reducing configuration hassle. Modules are usually local script or web-services and most of the SWfMSs support interactive drag and drop facilities to add such modules while building workflows from a set of available modular tools. For a particular task, port linking, dataset importing, and parameter configuring are also possible interactively to assemble sequential execution or build workflows as an acyclic graph in a SWfMS. Besides, a SWfMS also coordinates processing scripts, datasets and resource allocations in various steps of workflow execution to ensure dependencies of dataflow. For such systems, quantitative analysis of performance is hard to measure, and only organizational performance by user evaluation can be interpreted with use cases \cite{REIJERS2016126} \cite{772961}. Also, the effectiveness of a SWfMS and integration of a new methodology in the SWfMS to increase efficiency could be assessed by user evaluation. Thus, in this study, we evaluate our intermediate data recommendation technique in the SciWorCS system and assess the feasibilities in various aspects of assembling and executing workflows.

The study is designed to demonstrate the integration synopsis of the GUI-RISP in the SciWorCS and evaluate the SWfMS with the data recommendation technique from user usage perspectives by executing various types of workflows. We have the following findings:

1. By analyzing workflow assembling and executing log from our user study It can be illustrated that the recommendation technique can help users to execute pipelines efficiently (e.g., 56\% less user request, and 25\% less time in workflow execution).

2. GUI-RISP can help users more while developing long, complex workflows (e. g., 150\% more use in intermediate data than the short workflows from the available intermediate data).

3. The GUI version of RISP in the SciworCS can give users more understandably to design workflows with intermediate data to get more information and control from the web interface.

Our findings indicate that our proposed technique can significantly improve the performance and efficiency of a SWfMS by reusing the intermediate states from previously executed workflows.

The rest of the chapter is organized as follows. Section \ref{RelatedWorks} discusses the related work, Section \ref{technique_integration_overview} presents the architecture of our proposed technique , Section \ref{background} defines and describes some basic terms related to the study, Section \ref{ImplementationDetails} describes our experimental setup, Section \ref{ExperimentalStudiesandResults} presents the user evaluations, Section \ref{discussion_and_future_direction} shows some future directions, Section \ref{ThreattoValidity} describes possible threats to validity, and finally, Section \ref{ConclusionandFutureWorks} concludes the chapter by mentioning our future direction. 

\section{Related Works}
\label{RelatedWorks}
Most of the SWfMSs are implemented as a process-aware based information system where data management is mostly neglected in terms of reuse for facilitating workflow execution. In this chapter, our study is mainly focused on user evaluation of a data recommendation and management technique in a SWfMS. Existing studies on recommendation, management and new technique evaluation for both data and processes in SWfMSs are presented below in two sections to compare with our proposed technique. 
\subsection{Recommendations in SWfMSs}
A number of studies \cite{Woodman2015WorkflowProvenance} \cite{Yuan2011On-demandSystems} \cite{Koop2008VisComplete:Pipelines} \cite{Gil:2011:MYM:2063076.2063082} \cite{10.1007/978-3-540-85502-6_18} have been conducted on recommending and managing modules and data in  SWfMSs. We discuss these studies in the following paragraphs.
Woodman et al. \cite{Woodman2015WorkflowProvenance} proposed an algorithm for determining which subset of the intermediate state results from a pipeline could be stored at the lowest cost. Similarly, Yuan et al. \cite{Yuan2011On-demandSystems} proposed an algorithm for determining which set of intermediate state results could be stored at a minimum cost from a SWfMS. Koop et al. \cite{Koop2008VisComplete:Pipelines} proposed a technique for automatically completing a pipeline under progress by analyzing pipeline execution history. Gil et al. \cite{Gil:2011:MYM:2063076.2063082} used metadata in their technique to select an appropriate model for a particular dataset and to set up modules' parameters dynamically. Likewise, Leake et al. \cite{10.1007/978-3-540-85502-6_18} demonstrated the use of semantic information of provenance for suggesting services. Besides, many studies \cite{Chinthaka2009CBRAssistant} \cite{Zhang2011Recommend-as-you-go:Reuse} \cite{Spjuth2015ExperiencesBioinformatics} have investigated the possibility of discovering and reusing services and workflows in a SWfMS. While the above studies solely focus on storage cost of data or module recommendation, our study is fundamentally different because we focus on a re-usability technique by recommending the storing of the outcomes of modules and in this study, we evaluate the effectiveness of the technique.
From the above inspection, it is also clear that none of the existing studies investigated on intermediate state results for reuse in a SWfMS. In this study, we present the details architecture and user evaluation of our recommendation technique of intermediate data storing and reusing in a SWfMS. Additionally, use cases from different areas of workflow building are presented to evaluate the proposed technique.

\subsection{Case Studies, Technology Integrations and Evaluations}
A great many studies have been done on Use-case analysis of system integration and architecture implementation in the domain of Scientific Workflow Management. Some of them are discussed here to understand the effectiveness and evaluation process of a SWfMS. Wu et al. \cite{wu2012} illustrated a distributed WfMS's implementation with web services and emphasized on the quality of mapping schemes of service composing algorithms for performance enhancement. Zheng et al. \cite{Zheng:2015:ICW:2755979.2755984} analyzed the effects of Docker container for workflows to make them compatible in various infrastructures and explored the possibilities of virtual container management for low overhead in a SWfMS. Muniswamy-Reddy et al. \cite{Muniswamy-Reddy:2006:PSS:1267359.1267363} proposed a storage system by considering low overhead in transactions where provenance traces are considered as metadata to provide additional functionalities of a typical file system. Brown et al. \cite{Brown2007}  analyzed LIGO WfMS by integrating techniques and technologies from various WfMSs. In their study of LIGO system, gravitational wave data are used to explore future directions and usefulness of the techniques of different WfMSs in a single system. Wang et al. \cite{Wang:2009:KHG:1645164.1645176} demonstrated the architecture of the distributed Kepler and explored its performance and future directions for data-intensive tasks. Bahsi et al. \cite{Bahsi:2007:CWM:1377549.1377550} have analyzed different SWfMSs from the perspective of conditional-workflow-building process and stated that conditions could be managed in various ways in different SWfMSs respect to their core implementations.

Gathering the facts from the above studies on evaluating and verifying the effectiveness of SWfMSs, in this study, we also want to investigate the use cases of our technique in a SWfMS. As well as, we want to observe the effectiveness of our technique in both workflow building and execution phases. 

\section{Risp in a SWfMS}
\label{technique_integration_overview}
\begin{figure}
  \includegraphics[width=\textwidth]{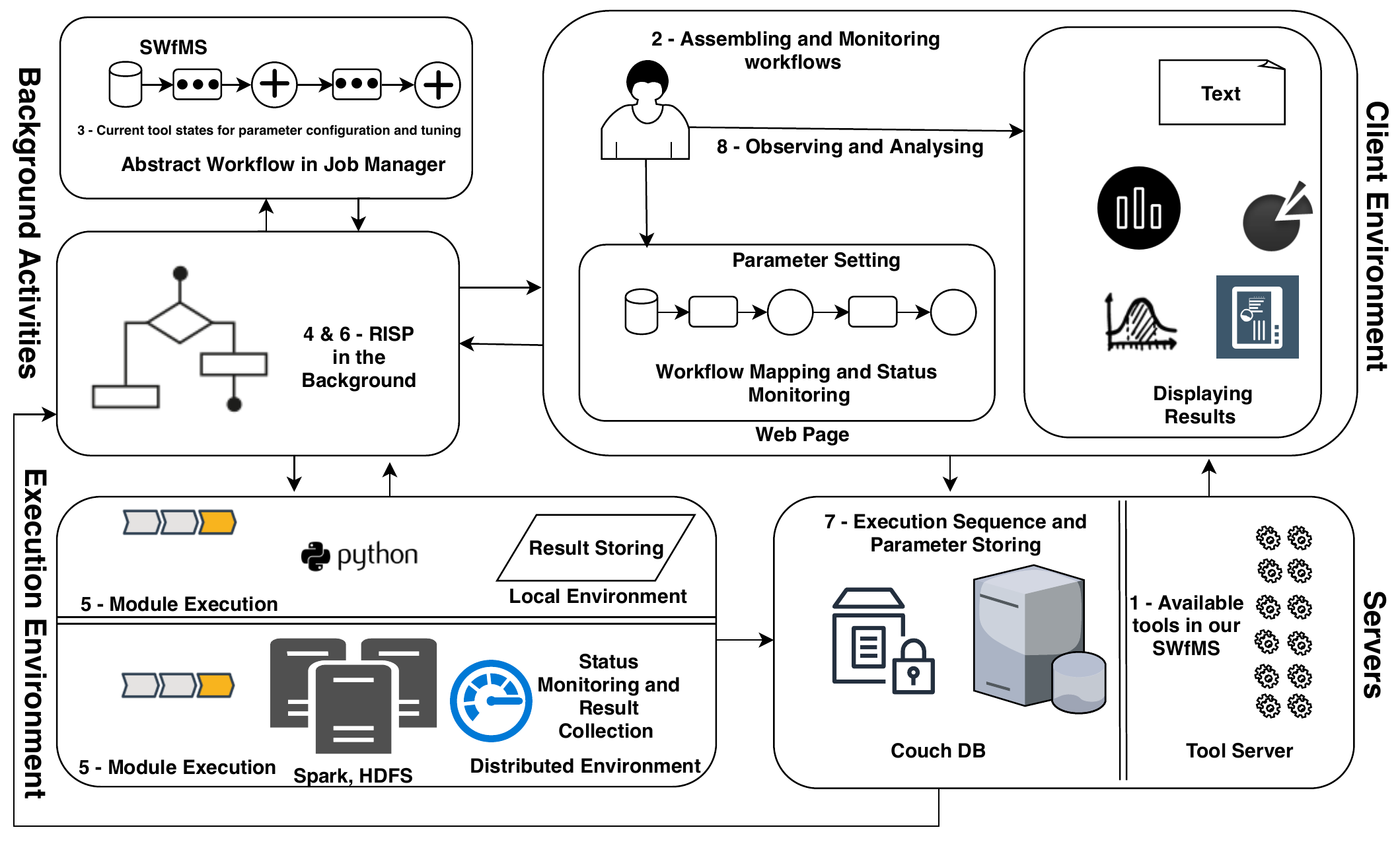}
  \caption{High level architecture of the SWfMS with RISP}
  \label{RISPArchitecture}
\end{figure}

Before going to present the experimental studies of our user usage log interpretation, in this section, we present a brief overview of SciWorCS architecture with the integrated GUI version of RISP (RISP - Our developed tool to assemble workflows with recommended reusable intermediate data). 
By Integrating RISP in the SWfMS, scientific analysis can be done by a set of reusable modules and intermediate data. Intermediate data could be preserved in the SciWorCS by different users from previously executed workflows. In SciWorCS, workflows are assembled on a web interface where modules could be executed in both distributed and local environments by their implementation. As well as, datasets could be stored in a local file system or a distributed file system (i.e., HDFS) for their purpose. RISP mainly works on background by getting some control information from the web-based GUI. Figure \ref{RISPArchitecture} illustrates the high-level overview of the architecture and the RISP integration of the SciWorCS. In the figure, major components and user activities are numbered from 1 to 8 for clarifying the data flow in the SciWorCS while building a workflow by using the GUI-RISP.

\subsection{System Overview}
\subsubsection{Reusable Module and Data}
Workflows for scientific data analysis are often considered as Directed Acyclic Graphs (DAGs) \cite{GARG2015256} \cite{Prodan:2005:DSS:1066677.1066835}. By considering intermediate data, a workflow can be presented as $W = (D, M, E, ID, O)$, where $M$ is a set of process modules with $n$ elements $m_i, (1 \leq i \leq n)$ and $E$ is set of directed edges of data flow $e_{ij} = (m_i, m_j), (1 \leq i \leq n, 1 \leq j \leq n, i \neq j)$ denoting the dependency and condition among modules and datasets. $D$ is the input data set, $O$ is the output data set, and $ID$ represents the set of outcomes of different modules (i.e., intermediate data). Intermediate data sets (IDs) from various workflows are considered to be managed with RISP for their incremental storage consumption.  A module from a set M especially performs a specific task (i. e., data preparation or feature extraction or model fitting or statistical analysis) in the SWfMS (i.e., SciWorCS 
). Computation scenario of such a module can be further customized with parameter settings of a specific configuration set $C$ -  where parameter configuration set is $P$, $c_i \epsilon C, (1 \leq  i \leq P)$.  By using a specific parameter set $P$, a module of set M might have in a different state from a state of the same module in similar or dissimilar workflows. Hence a workflow module and intermediate data can be generalized by $Definition$ \ref{def_module} and $Definition$ \ref{def_data} in the SciWorCS respectively.

\begin{remark}
Process module is defined as a tuple, $m \Rightarrow \langle id, I, O,$ $ C, S, T, Id \rangle $ where $id$ is the unique identifier of the module in a system, $I$ and $O$ supported input and output formats. $C$ is the configuration set of different parameters, $S$ is the source code, $T$ is the state of the module in workflow execution, and the $Id$ is a set of intermediate data generated from a module in workflow execution. The generalized definition can be helpful to readers for understanding the reusable modules of the SWfMS while building workflows.
\label{def_module}
\end{remark}

\begin{remark}
Intermediate dataset is defined as a tuple, $Id \Rightarrow \langle Sid, Did, S,$ $ sT, lT, m, T  \rangle $ where $Sid$ is the unique identifier of the dataset in a system, $Did$ raw dataset id. $S$ is the size of the datasets, $sT$ and $lT$ are the required saving and loading time for the dataset, $m$ is the module that generated the dataset with state $T$. The generalized definition can be helpful to readers for understanding the reusable intermediate data of the SWfMS while building workflows.
\label{def_data}
\end{remark}

\subsubsection{Workflow Composition with Intermediate Data}
\begin{figure}
  \includegraphics[width=1\textwidth]{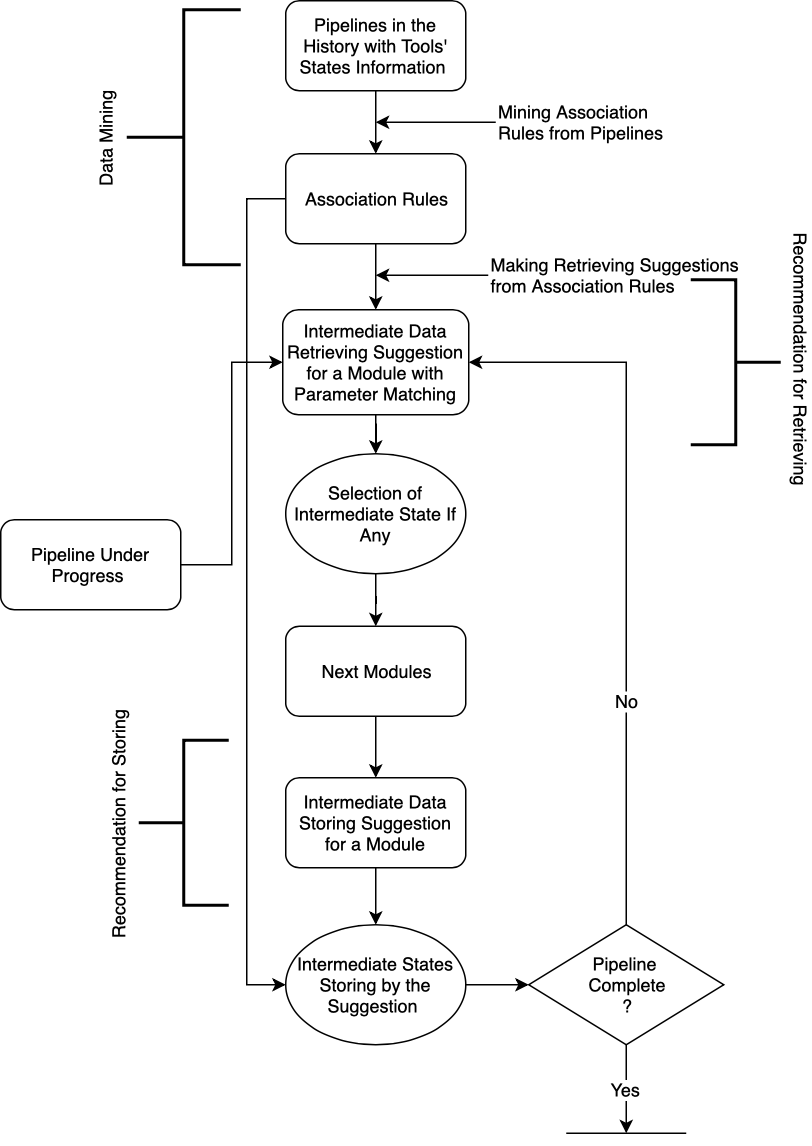}
  \caption{Workflow composition with the proposed technique }
  \label{IDRecommendationSystem}
\end{figure}

To perform data analysis with a workflow $W = (M, E)$, modules, $M$, from the toolbox of web panel need to be selected and assembled in the main composition window of the SciWorCS. Besides, ports need to be mapped with links for the set $E$ to sustain the required dependencies of the workflow. Simply executing all of the modules from the workflow in a SWfMS would cost a considerable amount of time. For example, any given instance of the workflow execution, $W_i$, there are some modules outcomes $(Ids \epsilon ID)$ and a set $Id$ passed from one module to another for a dependency - defined by a directed edge, $e_{ij} = (m_i, m_j) \epsilon E$, and each module with its state $T$ perform some operations on the outcomes of the previous module $(Id_i)$ to produce some new outcomes $(Id_j)$. Usually, to perform operations on a large dataset, a single module takes a considerable amount of time, suppose $t_i$ time to process and generate an intermediate data $(Id_j)$. In most of the cases, these generated Id sets are also large and incrementally produced in a SWfMS. As a workflow could be composed with a considerable number of modules, the execution time of the workflow might be too long. Again in a SWfMS, users generally build similar workflows frequently by changing only a few modules, i.e., $(m_k \dots m_l) \epsilon M $. In such a situation, rather than executing all modules $(m_1 \dots m_n) \epsilon M $ for a new workflow, in a SWfMS previously executed results ($IDs$) could be used to reduce execution cost. Besides, consideration of the state of a tool for derived data could increase user choice in a SWfMS. 
Hence, we adopt the GUI-RISP in  SciWorCS for each possible outcome of a module to consider the state of modules. By using the GUI version, we can recommend to reuse existing appropriate intermediate data while building new workflows in a SWfMS with proper parameter matching and data generation time information. Figure \ref{IDRecommendationSystem} demonstrates the core architecture of our data management and recommendation technique of GUI-RISP in the SciWorCS.  

\subsection{Background}
\label{background}

Association rules have been used in many research areas for discovering correlations and can be defined in the following way. 

\textbf{Association Rule.}
An association rule \cite{Agrawal:1993:MAR:170036.170072} is an expression of the form $X => Y$ where $X$ is the antecedent, and $Y$ is the consequent. Each of $X$ and $Y$ is a set of one or more program entities. 
The meaning of such a rule is that if $X$ gets changed in a particular commit operation, $Y$ also has the tendency of being changed in that commit.

\textbf{Support and Confidence.}
As defined by Zimmermann et al. \cite{Zimmermann:2004:MVH:998675.999460}, \emph{support is the number of commits in which an entity or a group of entities changed together}.
The support of an association rule is determined in the following way.

\begin{equation}
    \mathit{support}(X=>Y)=\mathit{support}(X, Y)
\end{equation}

Here, $(X, Y)$ is the union of $X$ and $Y$, and so $\mathit{support}(X=> Y) = \mathit{support}(Y => X)$.
\emph{Confidence of an association rule, $X => Y$, determines the probability that $Y$ will change in a commit operation provided that $X$ changed in that commit operation}. So the confidence of $X => Y$ can be determined  in the following way.

\begin{equation}
    \mathit{confidence}(X => Y) = \mathit{support}(X, Y) / \mathit{support}(X)
\end{equation}

In the technique RISP, we derive association rules between datasets and modules from pipelines and investigate those for providing suggestions regarding which intermediate state result from a workflow under progress should be stored.  

\textbf{Determining the supports and confidences of the  association rules obtained from the workflows.}

In the RISP, all the distinct association rules from all the previously executed workflows are determined, then the support and confidence of each of the rules are calculated in the following way. 

\textbf{Support of an association rule.} Support of an association is the number of times it can be generated from the workflows. The support of an association rule, $D \Rightarrow e_{ij}$ can be expressed formally in the following way.

\begin{equation}
support (D \Rightarrow e_{ij}) = No.\,of\,times\,edge\,e_{ij}\,found\,in\,history.
\end{equation}

Here $e_{ij}  \epsilon E = (m_i, m_j), (1 \leq i \leq n, 1 \leq j \leq n, i \neq j)$.

\textbf{Confidence of an association rule.} We determine the confidence of the association rule $D \Rightarrow e_{ij}$ in the following way from its support value:

\begin{equation}
confidence (D \Rightarrow e_{ij}) = \frac{support (D \Rightarrow e_{ij})}{support(D)}
\end{equation}

Here, support(D) is the number of times input dataset D was used in all workflows.

\subsection{RISP Interaction Modelling}

We analyzed different situations in which users will utilize RISP in the user interface of a SWfMS. In the following we describe different situations.

In this section of the study, we describe the possible workflow creation scenarios that inspired to design the user interface for our data recommendation technique. After investigating several SWfMSs and their workflows, we have figured out the following major composition scenarios.

\textbf{Composition scenarios.}
\begin{itemize}  
\item Creating a new workflow by dragging and dropping tools. In this scenario, users may compose both small and large workflows from scratch, where sub workflows are unique to the system
\item Creating a workflow by dragging and dropping tools. In this case,  sub workflows could be in the system from previous executions of different users with proper access control
\item Importing an existing workflow in the system from a saved one or a shared source. In this case, sub workflows could be in the system from previous executions of different users with proper access control.
\end{itemize}

Also, to increase the efficiency of the Gui-RISP selection process, we have considered the workflow change scenarios of workflow after composing the workflow in a SWfMS.

\textbf{Change scenarios.}
\begin{itemize} 
\item Users could change modules, tune modules, delete modules from a composed workflow. Even user may want to save the modified workflow in the system
\item Parameter configurations could be happened in a module
\item Tool state could be changed for a workflow.
\end{itemize}

By considering the above scenarios, recommendations are arranged on the SWfMS's interface.

\textbf{Recommendations.} 
\begin{itemize}
\item If a subset of modules is unique in a workflow then users will not get any recommendation to use any datasets rather a recommendation for saving datasets may appear on the interface
\item If a subset of modules can be found in previous executions, then users will get a recommendation for the subset if datasets are available otherwise a recommendation for saving datasets will appear on the interface
\item A message or recommendation list for a subsequence appears after adding a new module and connecting it to a previous one
\item Recommendation can be changed based on CRUD operations in a workflow
\item Recommendation can be updated after each parameter tuning, and configuration
\item Recommendations are based on tool states not the ids of tools.  
\item  Rather than creating a workflow from scratch if users choose to import it from existing ones, a dedicated button can be used to check the available intermediate data for subsequences
\item If the data loading time is more than the execution time for a module, users are warned with a message while using the recommendations
\item Required time to execute a module from previous execution is presented on the module for better understandability.
\end{itemize}

\subsection{System Design}
Job manager of the SciWorCS is responsible for executing the modules of an assembled workflow based on their dependencies and implementations (i.e., modules could be executed in a single server or a cluster). RISP acts alongside the job manager to suggest intermediate datasets automatically. Besides, users can manually check and load intermediate datasets of previously executed workflows while composing and executing a new workflow for reusing available datasets by the necessity of the new workflow execution using the GUI-RISP. The SciWorCS also provides real-time status monitoring, data mapping, and parameter setting mechanisms for each module individually to enable more modularized control. Job manager further performs the storing procedure of intermediate data and other necessary data by getting the recommendation from RISP. All of the execution records and configuration information of modules in a workflow are stored in a CouchDB server. RISP mainly associates the execution sequences and parameter information of previously executed workflows for recommending intermediate data while composing workflows in the SWfMS.  

\subsubsection{Implementation Details}
\label{ImplementationDetails}
The SciWorCS is built with Flask (A Python Microframework) to provide a web-based solution for the end-users. Various resource management techniques such as HDFS, CouchDB, UnixFileSystem are incorporated to enable interoperability among the used cutting edge technologies (i.e., Yarn, Spark, Python Subprocess and so on) in the SciWorCS for our data management scheme. The core architecture of the SciWorCS is implemented in Python 2, and the latest web technologies (i. e., JavaScript, GoJS, and so on) are used to provide the interactive user experience. The core implementation is hosted on a Linux server where all of the local scripts of modules are executed. A distributed Spark cluster (i.e., a Compute Canada Cluster of five-node, 40 cores and a total of 200GB RAM)  and a CouchDB server are also associated with the SWfMS to submit distributed jobs and update log information from the web server by using their respective APIs.
\subsubsection{Rules}
Figure \ref{UI} shows the interactive composition window of the SciWorCS to assemble and execute workflows. Modules of a composed workflow could be executed in the Linux server as local scripts or in the Spark cluster as distributed jobs. Both distributed and local modules can be dragged and dropped from the left side toolbox panel (labeled as L) to the main container of workflows (labeled as MCW). After placing the required modules from the panel L for a workflow, users need to connect them with links to resolve their conditions and dependencies. The right side panel (labeled as R) in Figure \ref{UI} shows the available input datasets and output produced by workflows in the SWfMS.Datasets and parameters could be set for a module by double-clicking the module on a pop-up window (labeled as MPW). Each module in a workflow has a particular status label for the recommendation technique, GUI-RISP (labeled as RISP). Initially, this status for each module is 'Not Checked' but after a certain time of assembling, checking or executing a module the status could be 'Checked Not Found', 'Checked Found', or 'Load Data'. With the status, 'Load Data', the SWfMS enables a button for loading intermediate data for an assembled workflow. The GUI version of RISP, where parameter matching and time are considered for giving multiple options to select intermediate data is placed below this status label as a List (labeled as A-RISP). By clicking an item from the list, a user can serve the same procedure of loading intermediate data for a workflow.  

\subsubsection{Suggestion}
\begin{figure*}
  \includegraphics[width=\textwidth]{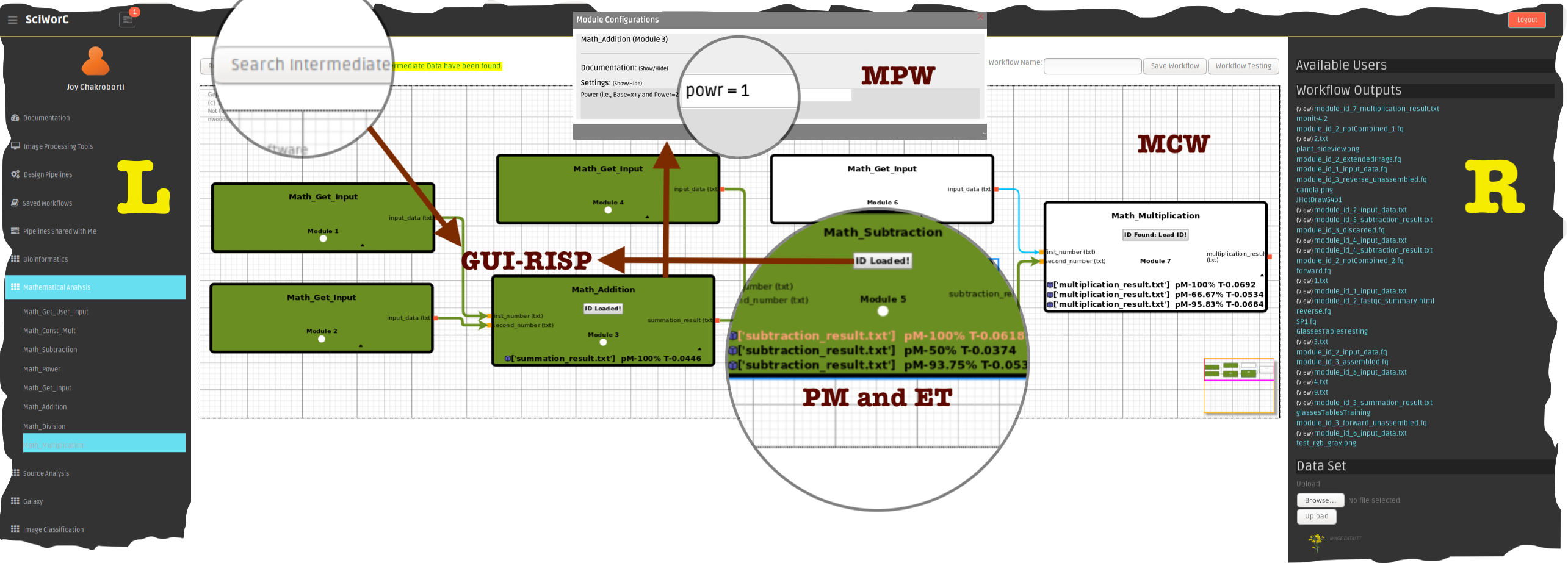}
  \caption{User interface for building workflows}
  \label{UI}
\end{figure*}

Figure \ref{UI} shows a workflow composition process in the SciWorCS with the help of our proposed data recommendation technique, RISP. Following the assembling process for a desired workflow, the 'Search' option could be used to know the existing relevant intermediate data that were stored by the recommendation of RISP. Module status label also displays the information for the possibility of loading and executing workflows with previously-stored intermediate data. In the figure, green-colored modules and links represent that the intermediate data are loaded up to the colored point, and there is no need of executions of those modules for the depicted workflow. In this case, module subtraction (labeled as m4) had the option for selecting intermediate data, and the data loaded up to this module execution point. By observing the parameter matching information from the options of the GUI-RISP and selecting an intermediate dataset also serves the same fashion to execute a workflow in the Scientific Workflow Management System.  

\section{Experimental Studies and Results}
\label{ExperimentalStudiesandResults}
For assessing the workload of the RISP in the SciWorCS, we have done the load testing, where we considered the SWfMS with the RISP and without the RISP. In our evaluation,  to avoid biases and unintentional delays from users, we used Apache Jmeter \cite{halili2008apache} and Google Analytics \cite{clifton2012advanced} for assembling workflows automatically and keeping logs with sampling. First, we recorded workflow compositions and executions of the two scenarios (i.e., the composition without RISP and with RISP) in the SWfMS using the Script Recorder of the Jmeter. Then we ran each of the scripts from Jmeter for 100 times to report its performance. Table \ref{table:overallload} compares various performance matrix of both of the scenarios for the SWfMS. In the table, we can see that using the RISP in SciWorCS does not change so much in average response time for various requests from our web application. Besides, the throughput increase with the RISP, and this is for the reason that only a very few requests and responses are needed in a workflow composition using the RISP in the SciWorCS (I.e., 56\% fewer requests).

\begin{table}
\centering
\caption{Overall load of the SWfMS}
\begin{tabular*}{\columnwidth}{ @{\extracolsep{\fill}}|r|r|r| } 
\hline
\thead{Measures} & \thead{Composition without RISP} &  \thead{Composition with RISP} \\
\hline
Requests & 743 & 321 \\ 
Average Time  & 210.3(Sec) & 200.5(Sec)\\ 
Throughput/Sec & 3.5 & 5\\ 
Received KB/sec & 7.32 & 19.88\\ 
Sent KB/sec & 4.63 & 7.77\\ 
\hline
\end{tabular*}
\label{table:overallload}
\end{table}

Before going to the extensive user evaluation, first, we inquire the overall end-user experience of the RISP and to do this, we followed the application performance monitoring(APM) strategy. By following the strategy, we mainly answered a research question to assess the overall user experience of RISP in the SciWorCS.

\textbf{RQ1:} What is the overall performance of the RISP in the system? Are there any components making overall performance poor?




To answer the question and present the overall performance we calculate the Apdex Standard \cite{sevcik2005defining}, the equation to calculate the standard is given below -

\begin{equation}
Apdext = (SatisfiedCount + (ToleratingCount/2))/TotalSamples
\end{equation}
 
We chose Apdex in our initial evaluation because the Apdex score gives overall satisfaction for a system from all users rather than the satisfaction of a few individuals from different categories \cite{sevcik2005defining}. In the scoring process of Apdex to categorize the responses from users for a system, we need to define a threshold time. In the scoring, we use the default threshold value (0.5 sec) to divide the satisfied and unsatisfied responses. To calculate the average single request response time of workflows, we execute 18 workflows in the SciWorCS with RISP and keep log of the average response time of each request (except module operation requests). 

From our Apdex scoring, with the value 0.89, we can say that near 90\% users are satisfied by the performances of the SciWorCS after RISP integration. Thus, \textbf{RQ1} can be answered with a positive statement for the RISP being used in the SWfMS. Since the average response time is not varying much after RISP integration, so components can interact normally with each other after RISP integration in the SWfMS which answers a part of \textbf{RQ1} for assessing the overall performance of the SWfMS. Furthermore, by integrating the RISP in the SciWorCS, higher number of KB/sec processing (Table \ref{table:overallload}) between a client and the server show the increased throughput for the SWfMS. So, no component is making the overall performance poor rather improving the performance of the SWfMS using the RISP (this answers the last part of \textbf{RQ1}). 
 
For the details user evaluation, we mainly cover two popular scientific workflow composition areas (i.e., Image Processing and Bio-informatics workflow compositions) by assembling and executing workflows with/without RISP in the SciWorCS . Both of the studies of the two areas are discussed below to answer some important research questions. Research questions are designed to investigate the following three hypotheses:

\textbf{H1:} RISP will not increase the overhead of the SciWorCS. We hypothesize the technique will minimize the request and responses between a client and the server of the SciWorCS. 

\textbf{H2:} User would use the intermediate data while executing workflows if the opportunity is given to them. Here, we want to explore how much intermediate data are used in the SWfMS from available intermediate data.  

\textbf{H3:} RISP will help to compose workflows efficiently. We hypothesize workflow execution time would be reduced if users use intermediate data.  
  
\subsection{Participants}
We selected participants who had at least heard the term workflow (i.e., business or scientific workflow) as this study investigates the effectiveness of a technique in a SWfMS.  We selected a total of 7 participants aged from 24 to 35 from a university, including both graduate and undergraduate students. The participants had an average age of 28, where 43\% are female, 57\% are male. 

\subsection{Study 1: Image Processing Workflow Composition}
In study 1, we want to explore the user satisfaction towards the performance of composing and executing data intensive workflows by the aided data of our technique (i.e., RISP). Image processing workflows are used in this study where differences from typical workflow composition to our proposed technique composition are investigated in the SciWorCS. Generally, work pattern and efficiency of workflow composition and execution are explored for both of the scenarios (i.e., composition with and without RISP) in the SWfMS to answer our \textbf{RQ2}.

\textbf{RQ2:} How is the RISP being used by users in the system? 

In workflows of image processing where input datasets are normally large in size, the processing time of modules can be lengthy, and their produced intermediate data can be large to manage. A reusability technique of data might be effective for such workflows, and in this study, we want to ensure the effectiveness of our technique with image processing workflows. In the case of large data analysis, dependencies and conditions in workflows need to be considered especially to solve a desired problem in minimum cost. By considering these dependencies of directed acyclic type, several recommendation techniques \cite{Koop2008VisComplete:Pipelines} \cite{Chinthaka2009CBRAssistant} \cite{Zhang2011Recommend-as-you-go:Reuse} \cite{Spjuth2015ExperiencesBioinformatics} of module or service are introduced.  However, none of these studies are presented with the usefulness of the recommended services or intermediate data for reusing in workflow design. Therefore, in this study, to explore the usability of efficient pipeline design by reusing intermediate data in a SWfMS, we use five image classification workflows of data-intensive computations. Consequently, how efficient the recommendation technique for users to perform workflow assembling and executing, how much time is required to design and execute a workflow and so on are analyzed in this study. Additional factors such as effects of a length, complexity (highest degree in a workflow) and data availability have been considered in our second study to assess the proposed technique and explore future directions in the workflow building process. Hence, in this part of our evaluation, we answer the \textbf{RQ2} using the time factors and the data usage of workflow assembling and executing.






\subsubsection{Task and Stimulus}
In the study, as mentioned above to verify the effectiveness of enhancing performance and gaining time, we consider five workflows of image classification. Modules of these five workflows perform distributed processing in the SWfMS's Spark cluster. For the classification models, we choose five popular deep neural network architectures, i.e., \textit{(InceptionV3, Xception, ResNet50, VGG16, and VGG19)} to give the option of choosing a model in our model-fitting module of the five workflows. In total, the five workflows contain ten distinct modules that are implemented in the SCiWorCS and compatible with the recommendation technique, GUI-RISP. Each of the five workflows is computationally expensive and solves a classification problem using a sequence of data preparation, feature extraction, model fitting, and analysis. In this study, to eliminate biases users had the opportunity to choose the classification models arbitrarily to perform image classification on a given dataset using a workflow from the five workflows. The dataset for the classification problem is collected from the P2IRC project of the University of Saskatchewan and has more than 3K images (2K for training, 1K for testing). In study sessions, workflow structures are printed and provided to the participants. Three of the five workflows using the five models are explained to them as model testing problems of image classification. Workflow design procedure in the SciWorCS such as dragging and dropping, port linking, dependency making of modules are described to the participants before the main study. Besides, how to check, load and use available intermediate data in the SWfMS are demonstrated using demo workflows to the participants.

\subsubsection{Experiment Procedure and Data Collection}
All of the occurred event, and their timestamps are logged in the background by defining JavaScript Tracking Snippet of Google Analytics \cite{clifton2012advanced} while building pipelines by the participants.  To capture valuable information of assembling and executing workflows from both of the scenarios (i.e., the SciWorCS with and without the RISP) each event is tracked with \textit{hitType, eventCategory, eventAction, eventValue, and eventLabel}. Similarly, timing information are tracked with \textit{hitType, timingCategory, timingVar, timingValue, and timingLabel} using the JavaScript Snippets of Google Analytics. At the end of the study,  participants were asked to fill out the \textit{NASA-TLX} questionaries for analyzing the workload of executing image processing workflows in the SWfMS.

\subsubsection{Results}
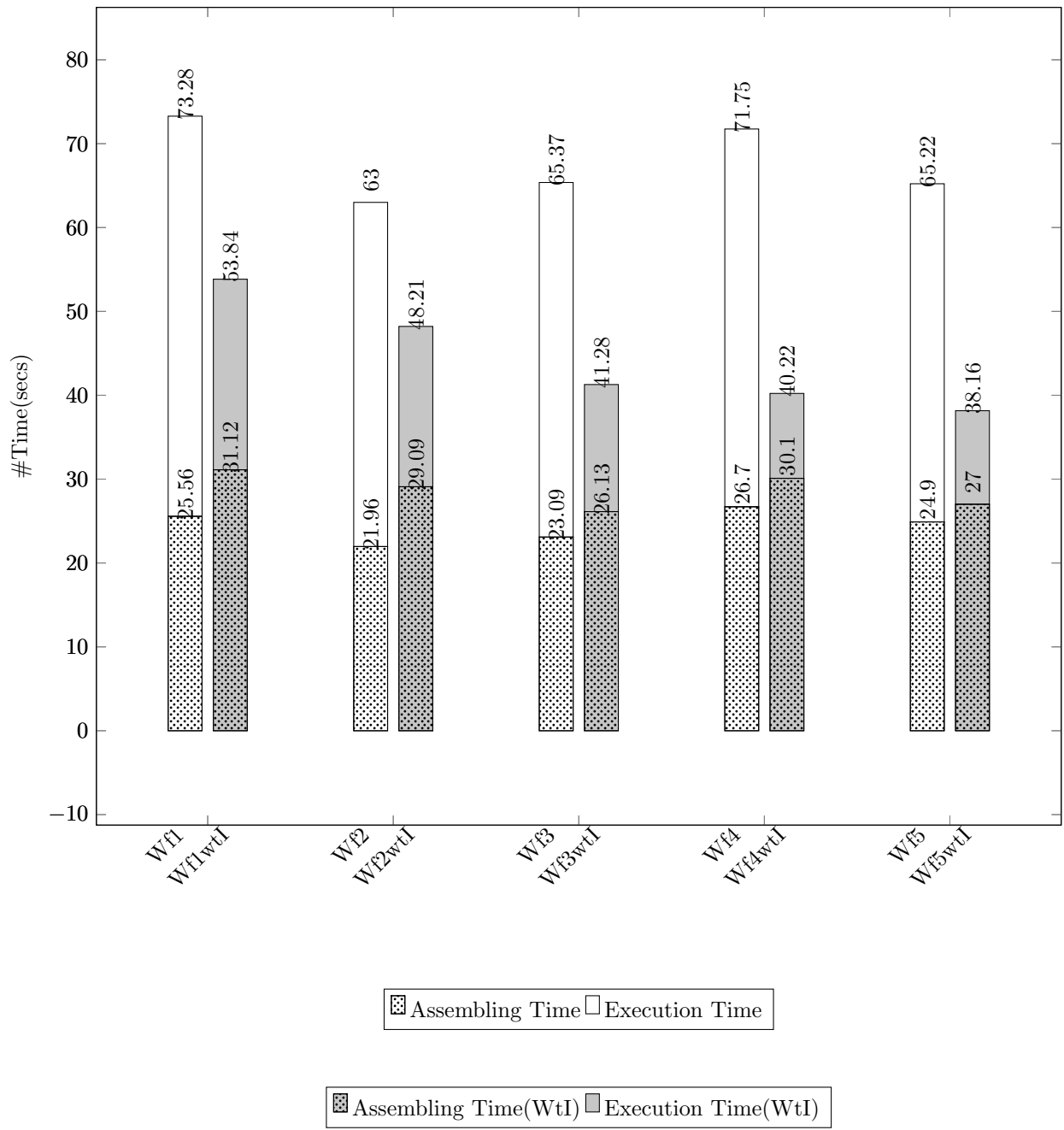
\begin{figure}[htbp]
\begin{tikzpicture}
\begin{axis}[
    ybar stacked,
    width=\textwidth,
	bar width=15pt,
	ymin=0,
	ymax=75,
	bar shift=-10pt,
	xticklabel style = {xshift=-10pt},
	nodes near coords,
	every node near coord/.append style={xshift=-3pt,yshift=10pt,rotate=90},
    enlargelimits=0.15,
    legend style={at={(0.5,-0.20)},
      anchor=north,legend columns=-1},
    ylabel={\#Time(secs) },
    symbolic x coords={Wf1, Wf2, Wf3, Wf4, Wf5},
    xtick=data,
    x tick label style={rotate=45,anchor=east},
    ]
\addplot[ybar, black, draw=black, fill=darkgray!00, postaction={pattern = crosshatch dots}] plot coordinates {
(Wf1,25.56) (Wf2,21.96) (Wf3,23.09) (Wf4,26.70) (Wf5,24.9)};
\addplot[ybar, black, draw=black, fill=darkgray!00] plot coordinates {
(Wf1,47.72) (Wf2,41.04) (Wf3,42.28) (Wf4,45.05) (Wf5,40.32)};
\legend{\strut Assembling Time,  \strut Execution Time}
\end{axis}
\begin{axis}[
    ybar stacked,
    width=\textwidth,
	bar width=15pt,
	bar shift=10pt,
	ymin=0,
	ymax=75,
	xticklabel style = {xshift=10pt},
	nodes near coords,
	every node near coord/.append style={xshift=17pt,yshift=10pt,rotate=90},
    enlargelimits=0.15,
    legend style={at={(0.5,-0.32)},
      anchor=north,legend columns=-1},
    symbolic x coords={Wf1wtI, Wf2wtI, Wf3wtI, Wf4wtI, Wf5wtI},
    xtick=data,
    x tick label style={rotate=45,anchor=east},
    ]
    \addplot+[ybar, black, draw=black, fill=darkgray!30, postaction={pattern = crosshatch dots}] plot coordinates {
    (Wf1wtI,31.12) (Wf2wtI,29.09) (Wf3wtI,26.13) (Wf4wtI,30.10) (Wf5wtI,27)};
  \addplot+[ybar, black, draw=black, fill=darkgray!30] plot coordinates {
    (Wf1wtI,22.72) (Wf2wtI,19.12) (Wf3wtI,15.15) (Wf4wtI,10.12) (Wf5wtI,11.16)};
\legend{\strut Assembling Time(WtI), \strut Execution Time(WtI)}
\end{axis}
\end{tikzpicture}
\caption{Required assembling and execution time in image processing workflows}
\label{a_e_time}
\end{figure}

In the study of image processing workflow's performances analysis, users activities of building and executing workflows are traced separately. Average assembling and executing time from all the participants for all the five workflows are plotted in Figure \ref{a_e_time}. In Figure \ref{a_e_time} gray-colored bars represent assembling and executing time of the five workflows with the proposed technique, GUI-RISP. Dots pattern in a bar denotes the assembling time, and the solid part depicts the execution time. In the figure, we can see that assembling time of workflows with the GUI-RISP is higher than the assembling time of workflows without the GUI-RISP (i.e., The dotted gray bars are higher than the dotted white bars). However, all of the five gray bars (i.e., aggregated dotted and solid) of the five workflows are lower in height than their corresponding white bar of the same workflow. That means overall performance of all of the workflows with the proposed technique GUI-RISP is better than the performances of workflows without the technique in the SciWorCS.  The reason behind this performance is that participants had the option to select recommended intermediate data by the GUI-RISP, and they used intermediate data to skip some modules operations. Though the assembling time of workflows with GUI-RISP is higher than the assembling time of workflows without GUI-RISP, the extra time in assembling for the recommendation and data configuration is negligible compared to the overall performance for all of the workflows. Thus the proposed technique can efficiently handle intermediate data to recommend for reuse in a real-life data-intensive workflow building environment. 

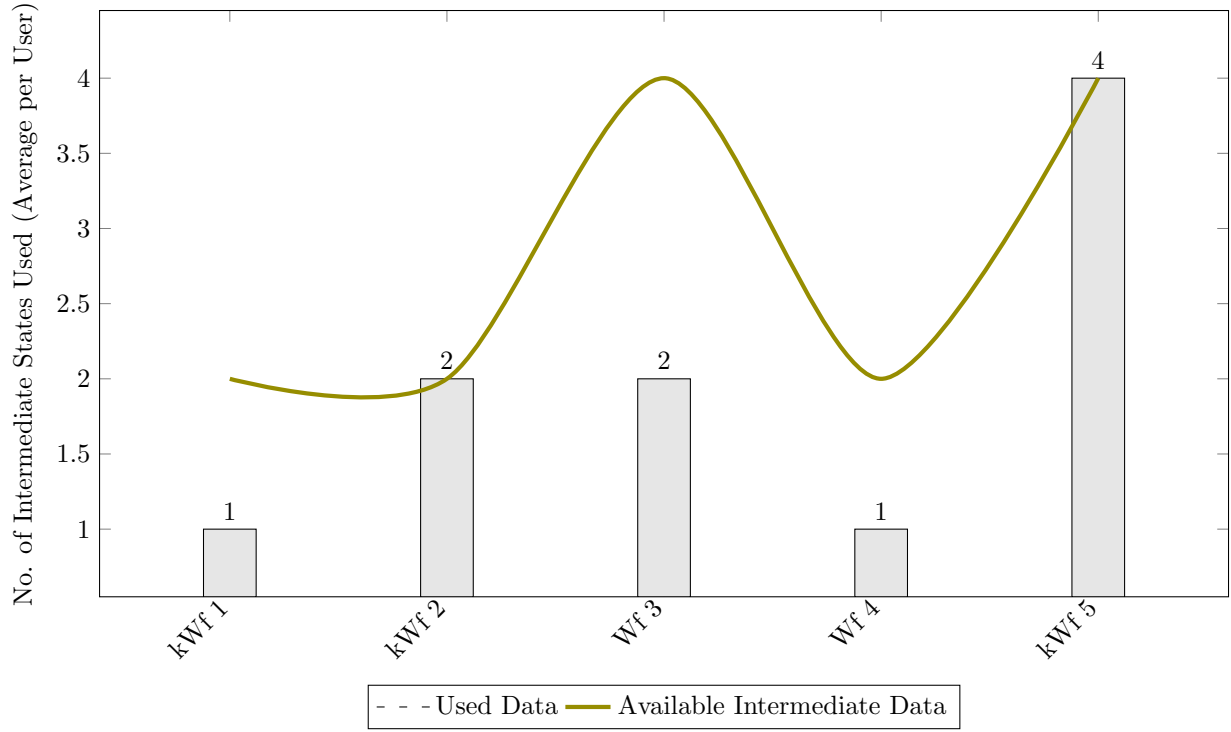
\begin{figure}[htbp]
\begin{tikzpicture}
\begin{axis}[
    width=\textwidth,
    enlargelimits=0.15,
    legend style={at={(0.5,-0.15)},
      anchor=north,legend columns=-1},
    ylabel={No. of Intermediate States Used (Average per User)},
    bar width=7mm,
    y=20mm,
    symbolic x coords={kWf 1, kWf 2, Wf 3, Wf 4, kWf 5},
    xtick=data,
    x tick label style={rotate=45,anchor=east},
    nodes near coords align={vertical},
    ]
\addplot[ybar, nodes near coords, fill=black!10] 
    coordinates {(kWf 1,1) (kWf 2,2) (Wf 3,2) (Wf 4,1) (kWf 5,4)};
\addplot[draw=olive!100,ultra thick,smooth] 
    coordinates {(kWf 1,2) (kWf 2,2) (Wf 3,4) (Wf 4,2) (kWf 5,4)};
\legend{ Used Data, Available Intermediate Data}    
\end{axis}
\end{tikzpicture}
\caption{Use of intermediate states in image procesing workflows}
\label{u_o_intermediate}
\end{figure}

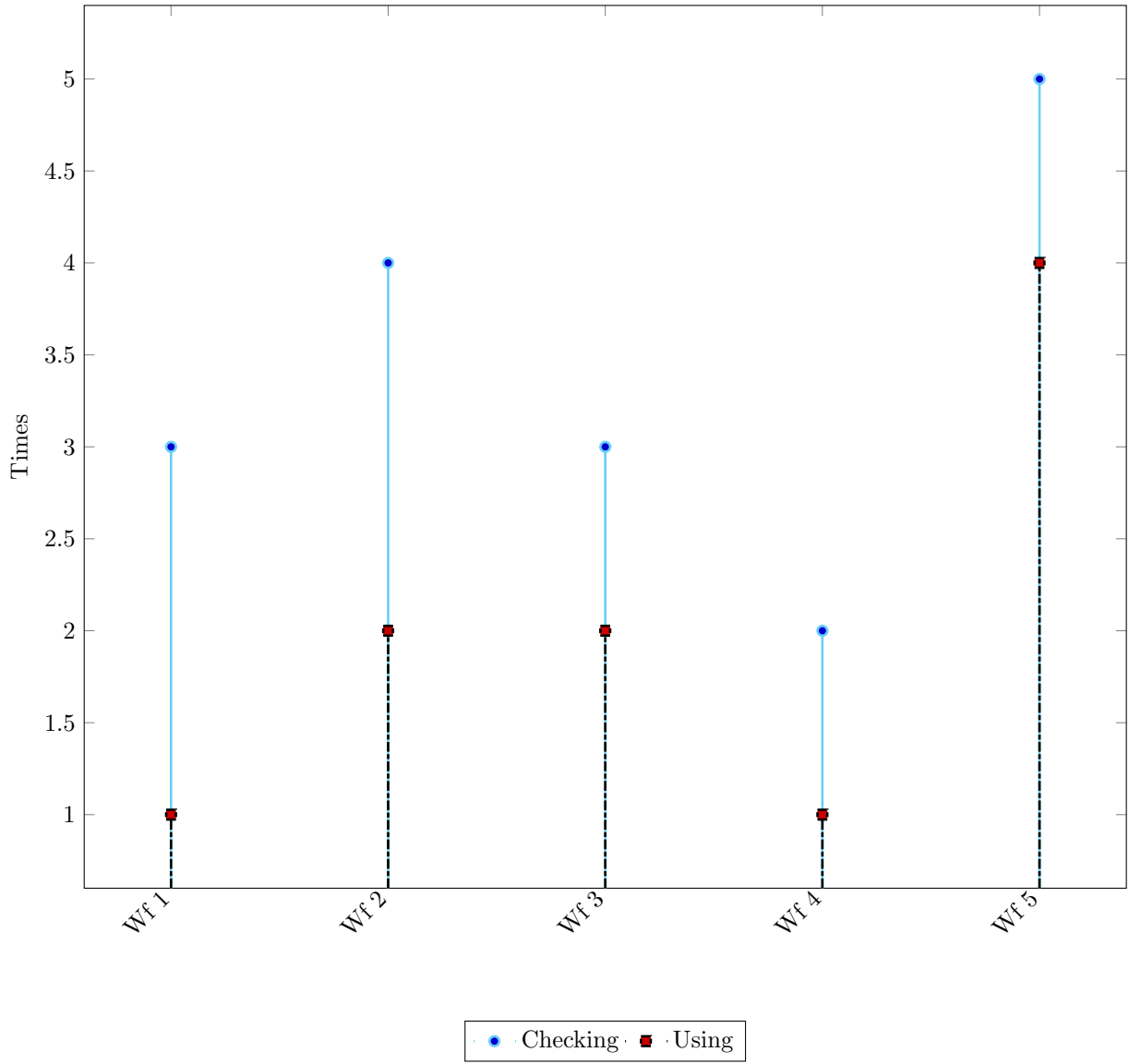
\begin{figure}[htbp]
\begin{tikzpicture} 
\begin{axis}[
    ylabel={Times},
    width=\textwidth,
    symbolic x coords={Wf 1, Wf 2, Wf 3, Wf 4, Wf 5},
    xtick=data,
    x tick label style={rotate=45,anchor=east},
    legend style={at={(0.5,-0.15)}, anchor=north,legend columns=-1},
]
    \addplot+ [line width=1pt, draw=cyan!50,
        ycomb,
    ] coordinates {
        (Wf 1,3) (Wf 2,4) (Wf 3,3) (Wf 4,2) (Wf 5,5)
    };
    \addplot+ [line width=1pt,dash pattern=on 4pt off 1pt on 2pt off 1pt, draw=black!100,
        ycomb,
    ] coordinates {
        (Wf 1,1) (Wf 2,2) (Wf 3,2) (Wf 4,1) (Wf 5,4)
    };
\legend{ Checking, Using}  
\end{axis}
\end{tikzpicture}
\caption{Checking and Using Intermediate States in Image Processing Workflows}
\label{c_u_intermediate}
\end{figure}

Figure \ref{u_o_intermediate} depicts the data usage pattern among known and unknown workflows. Known workflow to a user means the same workflow is executed by the same user previously. In our case, workflow 1, 2, and 5 (i.e., \textit{InceptionV3, Xception, and VGG19}) are executed by the participants prior to the main study while showing the demo of workflow building. We found that participants tend to use more intermediate data for known workflows rather than unknown workflows. In Figure \ref{u_o_intermediate}, around 87\% of available intermediate data are used for the known workflows, which are higher than the 50\% of two unknown workflows (N.B. all workflows had at least one intermediate data from previous executions). In addition, we found participants checked and used more intermediate data for computationally expensive modules (e.g., in our case, participant checked and used for the model fitting modules) of workflows rather than inexpensive modules. Traces of checking and using intermediate data in the workflows are plotted in Figure \ref{c_u_intermediate}. Figure \ref{c_u_intermediate} also represents the possibility of using intermediate data by checking the data from the users, and the results are more significant in all three known workflows (i.e., Workflow 1, 2 and 5) than the unknown 3 and 4. In this study, from the provided raw data and generated intermediate data, in total intermediate data and raw data showed the split of  83 \% and 17 \% usage while building pipelines by the all participants. This information also showed that if we use the RISP, we can increase the reusability of a SWfMS. All of the findings in this study can be used to answer the \textbf{RQ2} partially, and up to this point, results show the GUI-RISP could be used efficiently for data-intensive workflows.

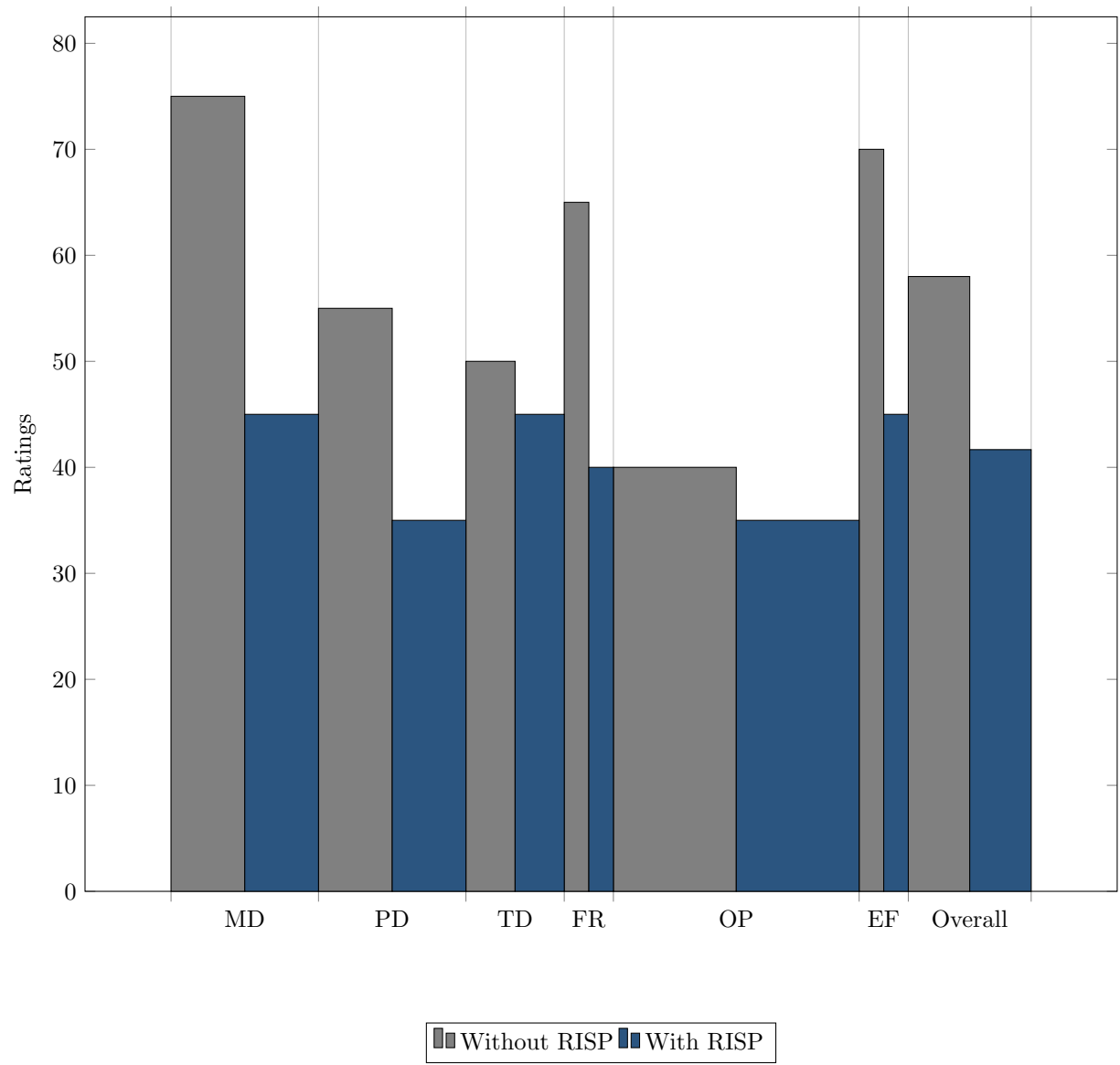
\begin{figure}[htbp]
\pgfplotstableread{
  Scale    Mean  Weight
  MD      75    3
  PD      55    3
  TD      50    2
  FR      65    1
  OP      40    5
  EF      70    1
  Overall 58   2.5
  Dummy    0       0
}{\tlxweightedratings}
\pgfplotstableread{
  Scale    Mean2  Weight2
  MD      45    3
  PD      35    3
  TD      45    2
  FR      40    1
  OP      35    5
  EF      45    1
  Overall 41.67   2.5
  Dummy    0       0
}{\tlxweightedratingsInt}


\begin{tikzpicture}
  \begin{axis}[
    ymin=0,
    width = \textwidth,
    ybar interval, xtick=data,
    ylabel={Ratings},
    legend style={at={(0.5,-0.15)},
      anchor=north,legend columns=-1},
    xticklabels={MD, PD, TD, FR, OP, EF, Overall},
  ]
    \addplot+[black,fill=black!50] table[x=Interval, y=Mean,
      create on use/Interval/.style={
        create col/expr={\pgfmathaccuma + \prevrow{Weight}}},
    ] {\tlxweightedratings};
        \addplot+[black, fill={rgb:red,1;green,2;blue,3}] table[x=Interval, y=Mean2,
      create on use/Interval/.style={
        create col/expr={\pgfmathaccuma + \prevrow{Weight2}}},
    ] {\tlxweightedratingsInt};
\legend{Without RISP, With RISP}    
  \end{axis}
\end{tikzpicture}
\caption{NASA-TLX responses for image processing workflows}
\label{nasatlx_image}
\end{figure}

NASA-TLX responses from all of the participants for this study are plotted in Figure \ref{nasatlx_image}. In the figure, dark blue bars represent the average scores of all the participants for the SciWorCS with the RISP. The gray bars represent the average scores of all the participants for the SciWorCS without RISP. All of the subscales, such as  Mental Demand, Physical Demand, Temporal Demand, Frustration Level, Overall Performance, Effort of NASA-TLX are considered in this study with an overall score.
From Figure \ref{nasatlx_image}, it can be interpreted that on average, users require a low level of demands with intermediate data in all subclasses (i.e., the SWfMS with the GUI-RISP). Though assembling and executing time of workflows vary in the two systems. One reason could be, assembling workflow with the proposed technique in the SWfMS requires more demands as the participants needed to know more control of the data management. However, executing workflow with the RISP, users needed to wait less time, which could reduce the subclasses demands in the SWfMS. Overall, the aggregated effects of the two phases exhibit the low level of demands for the GUI-RISP system. 

\subsection{Study 2: Bioinformatics Workflow Composition}
Scientific analysis of bioinformatics data is another popular domain that relies on SWfMSs for composing heterogeneous tools. In this study, we incorporate the bioinformatics workflow composition using the GUI-RISP and compare the composition with traditional workflow composition. In the previous study, we considered the same length workflows of heterogeneous technologies and large datasets. In this study, we consider composing micro-services or simple modules of bioinformatics (i.e., workflows are categorized in simple smaller, and complex longer) with comparatively smaller datasets. For the active bioinformatics communities, various micro-services and services are available and ready to use. To analyzing bioinformatics data in a SWfMS, both simple and complex workflows are important for the availability of micro-services and services. Thus in this study, we conduct our experiment for both smaller and longer bioinformatics workflows to complete the answer of our \textbf{RQ2}.





\subsubsection{Task and Stimulus}

In the previous study, we chose image processing workflows of distributed modules to process a massive amount of data. Contrast to that, here we choose four bioinformatics workflows of local-script modules and comparatively smaller datasets. First, two workflows are smaller in size and relatively easy to follow. The last two workflows are long in size and contain the first two workflows in its structure. In total, the four workflows contain four distinct modules such as \textit{Get\_input,  Pear, Fastqc, and Flash}. All of these modules are imported from Galaxy \cite{Afgan2016TheUpdate} to integrate into the SciWorCS and compatible with the GUI-RISP. Each of the four workflows is not computationally expensive comparate to the previous study but complex in their structure for so many port mappings. The datasets for the bioinformatics problems are collected from the P2IRC project of the University of Saskatchewan. The dataset has mainly two files, one is Forward FastQC, and another is Reverse FastQC. 
In study sessions, workflow structures are printed and provided to the participants. All of the four workflows with some modules are explained to them as genome sequencing problems of bioinformatics. Likewise, previous study, workflow design procedure in the SWfMS such as dragging and dropping, port linking, dependency making of modules are described to the participants before the main study. Besides, how to check, load and use available intermediate data in the SciWorCS are again demonstrated using demo bioinformatics workflows to the participants. 

\subsubsection{Experiment Procedure and Data Collection}
Same as before all of the activities of the participants are logged and timestamped using JavaScript Tracking Snippet of Google Analytics while building workflows in the study season. Likewise before, we also capture the same information of assembling and executing workflows for both of the scenarios (i.e.,  SciWorCS with and without the GUI-RISP). Such as every event is tracked with \textit{hitType, eventCategory, eventAction, eventValue, and eventLabel} for all of the four workflows. Similarly, timing information is tracked with \textit{hitType, timingCategory, timingVar, timingValue and timingLabel} using the JavaScript Snippets of Google Analytics for all of the workflows. At the end of the main study, participant filled out the NASA-TLX questionnaire to evaluate the workload of executing bioinformatics workflows in the SWfMS.

\subsubsection{Results}
In this study of bioinformatics workflows, unlike the previous study assembling and executing time of each workflow is traced together. Average completion times of all of the four workflows are plotted in Figure \ref{s_c_time_bio}. The first box of each group in the box plot represents the total completion time of a workflow with the GUI-RISP (i.e., with intermediate data). The second box of each group represents workflow completion time of a workflow without the RISP (i.e., without intermediate data). First two groups in the figure depict completion time for the two simple workflows and last two depict completion time for the two complex workflows. From Figure \ref{s_c_time_bio}, it can be stated that for the complex workflows user can gain more time for the availability of intermediate data for a long sequence of modules than the simple/short workflows using the GUI-RISP. Availability of intermediate data and their usage can be clearly illustrated from Figure \ref{u_o_intermediateS2}. In the Figure, preferences of data usage while building workflows with GUI-RISP by the participants for both long and short workflows are represented. Similar to the previous study, this study also has workflows with at least one intermediate data from previous executions. In the figure, we can see that for long workflows user used more intermediate data than the short workflows. In Figure \ref{s_c_useKvU}, we show the intermediate data usage for known and unknown modules. Known module means operation and relation of the modules in each workflow are explained to the participants prior to the study. For this study, known modules are Galaxy\_Pear and Galaxy\_Flash. From the figure, we can see that participants used more intermediate data to skip module operations for known modules than the unknown modules. In Figure \ref{s_c_useKvU}, 64\% of available intermediate data are used for the known modules, which are greater than the 36\% of unknown modules. This usage pattern could be from the reason that the participant felt comfortable to skip the operation of those modules by knowing that what they are skipping. So better understandability can improve the use of intermediate data, and in future, we will shift the SWfMS towards data-centric awareness system. In this study, from the provided raw data and generated intermediate data, in total intermediate data and raw data showed the split of  63 \% and 37 \% usage while building pipelines by all the participants. Experimental results of this study show that if we use the GUI-RISP, we can increase the reusability of a SWfMS for all types of workflows. The increased reusability and efficiency in the various types of workflows of this study complete the answer to our \textbf{RQ2}. 

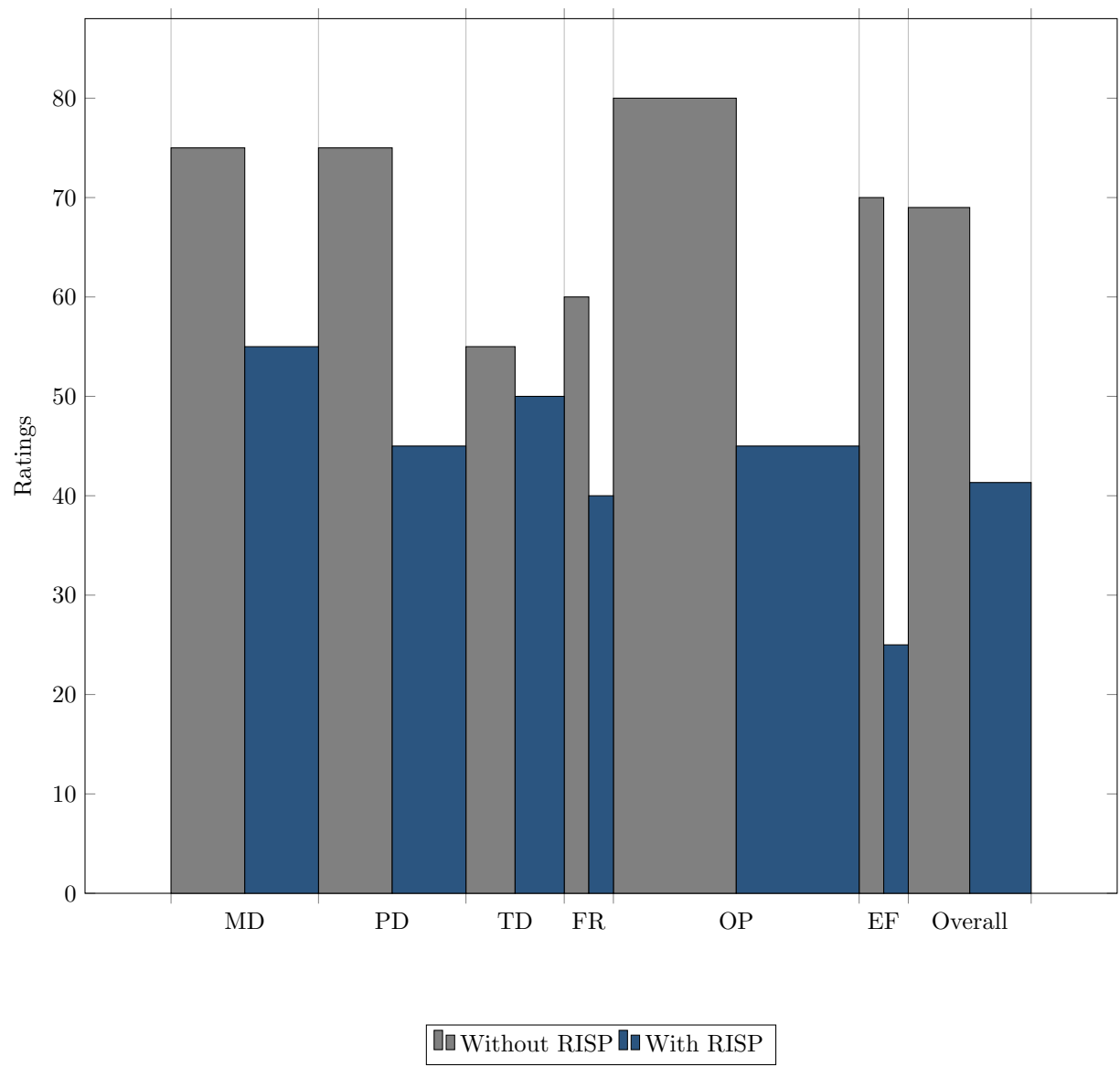
\begin{figure}[htbp]
\pgfplotstableread{
  Scale    Mean  Weight
  MD      75    3
  PD      75    3
  TD      55    2
  FR      60    1
  OP      80    5
  EF      70    1
  Overall 69   2.5
  Dummy    0       0
}{\tlxweightedratings}
\pgfplotstableread{
  Scale    Mean2  Weight2
  MD      55    3
  PD      45    3
  TD      50    2
  FR      40    1
  OP      45    5
  EF      25    1
  Overall 41.33   2.5
  Dummy    0       0
}{\tlxweightedratingsInt}


\begin{tikzpicture}
  \begin{axis}[
    ymin=0,
    width = \textwidth,
    ybar interval, xtick=data,
    ylabel={Ratings},
    legend style={at={(0.5,-0.15)},
      anchor=north,legend columns=-1},
    xticklabels={MD, PD, TD, FR, OP, EF, Overall},
  ]
    \addplot+[black,fill=black!50] table[x=Interval, y=Mean,
      create on use/Interval/.style={
        create col/expr={\pgfmathaccuma + \prevrow{Weight}}},
    ] {\tlxweightedratings};
        \addplot+[black, fill={rgb:red,1;green,2;blue,3}] table[x=Interval, y=Mean2,
      create on use/Interval/.style={
        create col/expr={\pgfmathaccuma + \prevrow{Weight2}}},
    ] {\tlxweightedratingsInt};
\legend{Without RISP, With RISP}    
  \end{axis}
\end{tikzpicture}
\caption{NASA-TLX responses for bioinformatics workflows}
\label{nasatlx_bio}
\end{figure}

NASA-TLX responses from all of the participants for this study are also collected and plotted in Figure \ref{nasatlx_bio}. Same as the previous study, dark blue bars represent the average scores of all the participants for the SciWorCS with the GUI-RISP. The gray bars represent the average scores of all the participants for the SWfMS without GUI-RISP. All of the subscales, such as  Mental Demand, Physical Demand, Temporal Demand, Frustration Level, Overall Performance, and Effort of NASA-TLX are also considered in this study with an overall score. From Figure \ref{nasatlx_bio}, it can be interpreted that on average, a user requires the low levels of subclass demands in GUI-RISP system. This may be the reason that the completion time of workflows varies for their complexity and workflow completion with the proposed technique in the SWfMS for long workflows shows a higher gain in execution than the of short workflows.So, on average users are satisfied with the proposed technique in a SWfMS for its efficiency. Aggregated results of our NASA-TLX responses of both of the studies can be used for answering the \textbf{RQ3}.

\textbf{RQ3:} Whether GUI-RISP is helping to compose workflows more efficiently?

With the response values from different subclasses of NASA-TLX in both of the studies, it can be interpreted that RISP does not increase workloads while composing workflows in the SciWorCS. This result satisfies the \textbf{RQ3} of our experimental studies.   

\begin{figure}[htbp]

\begin{tikzpicture}[scale=\textwidth/22cm]
\begin{axis}[
boxplot/draw direction=y,
ylabel={Assembling Time + Execution Time (secs)},
boxplot={
    %
    draw position={1/3 + floor(\plotnumofactualtype/2) + 1/3*mod(\plotnumofactualtype,2)},
    %
    box extend=0.3,
},
x=5cm,
xtick={0,1,2,...,10},
x tick label as interval,
xticklabels={%
    {Simple Workflow 1\\{\tiny Without/With}},%
    {Simple Workflow 2\\{\tiny Without/With}},%
    {Complex Workflow 1\\{\tiny Without/With}},%
    {Complex Workflow 2\\{\tiny Without/With}},%
},
    x tick label style={
        text width=2.5cm,
        align=center
    },
]
\addplot[draw=black,fill={rgb:red,4;green,2;yellow,1}]
table[row sep=\\,y index=0] {data\\46\\54\\75\\78\\80\\};
\addplot[draw=black,fill={rgb:red,3;green,1;yellow,2}]
table[row sep=\\,y index=0] {data\\41\\44\\55\\70\\73\\};
\addplot[draw=black,fill=green!50!red]
table[row sep=\\,y index=0] {data\\49\\56\\60\\67\\86\\};
\addplot[draw=black,fill=green!50!red!50]
table[row sep=\\,y index=0] {data\\45\\50\\52\\60\\65\\};
\addplot[draw=black, fill=black!50]
table[row sep=\\,y index=0] {data\\68\\69\\73\\85\\91\\};
\addplot[draw=black, fill=black!20]
table[row sep=\\,y index=0] {data\\60\\62\\64\\75\\79\\};
\addplot[draw=black, fill={rgb:red,1;green,2;blue,3}]
table[row sep=\\,y index=0] {data\\85\\88\\91\\96\\100\\};
\addplot[draw=black, fill={rgb:red,0;green,0.5;blue,1}]
table[row sep=\\,y index=0] {data\\77\\81\\82\\91\\95\\};
\end{axis}
\end{tikzpicture}
\caption{Time differences among various types of bioinformatics workflow}
\label{s_c_time_bio}
\end{figure}
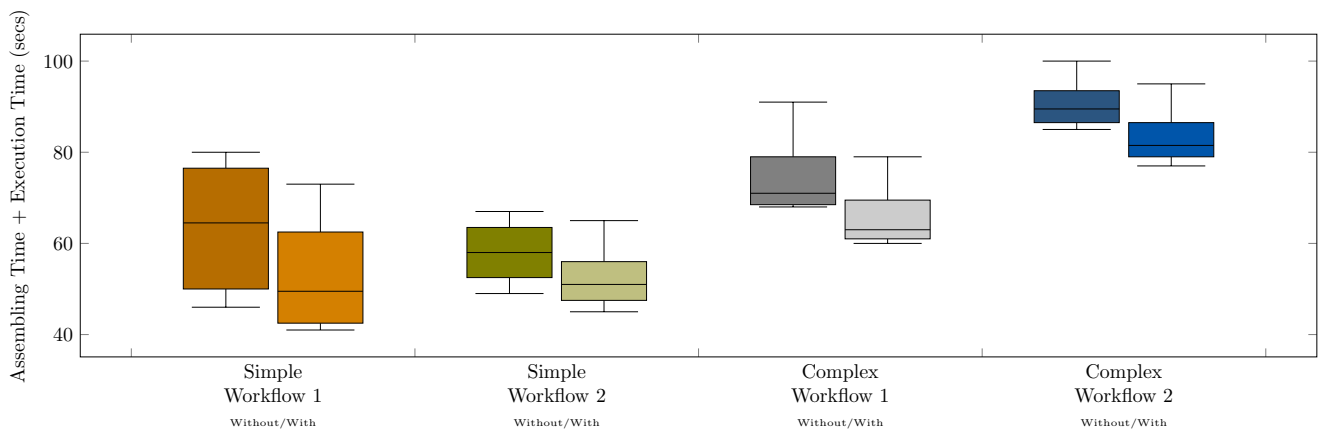

\begin{figure}[htbp]
\begin{tikzpicture}
\begin{axis}[
    width = \textwidth,
    enlargelimits=0.15,
    legend style={at={(0.5,-0.15)},
      anchor=north,legend columns=-1},
    ylabel={No. of Intermediate States Used (Average per User)},
    bar width=7mm,
    y=20mm,
    symbolic x coords={sWf 1, sWf 2, lWf 1, lWf 2},
    xtick=data,
    x tick label style={rotate=45,anchor=east},
    nodes near coords align={vertical},
    ]
\addplot[ybar, nodes near coords, fill=black!10] 
    coordinates {(sWf 1,1) (sWf 2,1) (lWf 1,2) (lWf 2,3)};
\addplot[draw=olive!100,ultra thick,smooth] 
    coordinates {(sWf 1,1) (sWf 2,3) (lWf 1,2) (lWf 2,3)};
\legend{ Used Data, Available Intermediate Data}    
\end{axis}
\end{tikzpicture}
\caption{Use of intermediate states in bioinformatics workflows}
\label{u_o_intermediateS2}
\end{figure}

\begin{figure}[htbp]
\begin{tikzpicture}
\begin{axis}[
    ybar,
    width = \textwidth,
    enlargelimits=0.15,
    legend style={at={(0.5,-0.15)},
      anchor=north,legend columns=-1},
    ylabel={\#No of Modules},
    symbolic x coords={Wf1,Wf2,Wf3,Wf4},
    xtick=data,
    nodes near coords,
    nodes near coords align={vertical},
    ]
\addplot[fill={rgb:red,4;green,2;yellow,1}] coordinates {(Wf1,2) (Wf2,2) (Wf3,3) (Wf4,4)};
\addplot[fill=green!50!red] coordinates {(Wf1,0) (Wf2,0) (Wf3,2) (Wf4,2)};
\addplot[fill=black!50] coordinates {(Wf1,1) (Wf2,2) (Wf3,2) (Wf4,2)};
\addplot[fill={rgb:red,1;green,2;blue,3}] coordinates {(Wf1,1) (Wf2,1) (Wf3,2) (Wf4,3)};
\legend{UnKnown, UnKnown Used, Known, Known Used}
\end{axis}
\end{tikzpicture}
\caption{Use of intermediate states in bioinformatics workflows}
\label{s_c_useKvU}
\end{figure}

\subsection{Study 3: User Behavior in the Interface}
\label{study3}
In this study of user behavior analysis, we consider the time-spent and gaze-point matrics to explore the eye-tracking data of the participants from the workflow composition and execution. One of the main reasons to chose the eye-tracking technology for this study is to validate the user involvement on the intermediate data recommendation GUI (i.e., validation of the users' interest in intermediate data while composing workflows in the SciWorCS). All of the workflows from the previous two studies are used to track user gaze-point and time duration by specifying different areas on the composing interface. Particularly, where participants are looking mostly and how much time they are spending in those areas are examined by Areas of Interest of Tobii Studio to answer the \textbf{RQ4}.

\textbf{RQ4.} What are the major areas where participants are mostly involved while assembling and executing workflows with the GUI-RISP?

Assembling and executing workflows in a SWfMS requires significant attention from users as there are some manual operations from users for both assembling and configuring the workflows. Thus, on an interface of workflow composition, we need to analyze where the gaze points are and where users are spending most of the time while composing workflows with the proposed management technique. Besides, we need to know whether the participant really interested in intermediate data or is the GUI-RISP really helping them to compose workflows - consideration of these facts with our selected metrics should be adequate to answer the \textbf{RQ4}.

\subsubsection{Task and Stimulus} In this study of behavioral analysis, we did not consider any additional workflows. We collect both gaze data and time spent data (with Heat Map) for all of the workflows that are composed by the participant in Study 1 and 2.

\subsubsection{Experimental Procedure and Data Collection}
Both the user behavioral data (i.e., gaze data and time spent data) are collected using a Tobii Eye-tracker while building workflows by the participants. The tracker was calibrated for each participant, and the participants maintained the recommended distance (i.e., between 18 to 40 inches) from the computer monitor. Important areas such as suggestion and composition areas are divided into separate sections as 'area of interest' for tracing eye-tracking data. Tobii Pro software solution is used for analyzing and representing the tracked data for the comparison. 

\begin{table}[]
\centering
\caption{Total Visit Duration}
\label{TotaVisitDuration}
\resizebox{\columnwidth}{!}{%
\begin{tabular}{|c|c|c|c|c|c|c|}
\hline
\multicolumn{7}{|c|}{p2irc-shipi.usask.ca:5000/cvs}                                                                                                                                                                                                                                                                                                                                 \\ \hline
\multirow{2}{*}{Summary Only} & \multicolumn{3}{c|}{Composition Area}                                                                                                                                    & \multicolumn{3}{c|}{Suggestion Area}                                                                                                                                     \\ \cline{2-7} 
                              & \begin{tabular}[c]{@{}c@{}}N\\ (Count)\end{tabular} & \begin{tabular}[c]{@{}c@{}}Mean\\ (Seconds)\end{tabular} & \begin{tabular}[c]{@{}c@{}}Sum\\ (Seconds)\end{tabular} & \begin{tabular}[c]{@{}c@{}}N\\ (Count)\end{tabular} & \begin{tabular}[c]{@{}c@{}}Mean\\ (Seconds)\end{tabular} & \begin{tabular}[c]{@{}c@{}}Sum\\ (Seconds)\end{tabular} \\ \hline
All Recordings                & \multicolumn{1}{r|}{7}                             & \multicolumn{1}{r|}{1539}                               & \multicolumn{1}{r|}{10773}                              & \multicolumn{1}{r|}{7}                             & \multicolumn{1}{r|}{596.30}                                & \multicolumn{1}{r|}{4174.1}                                \\ \hline
\end{tabular}%
}
\end{table}

\subsubsection{Results} 

\begin{figure}
  \includegraphics[width=\textwidth]{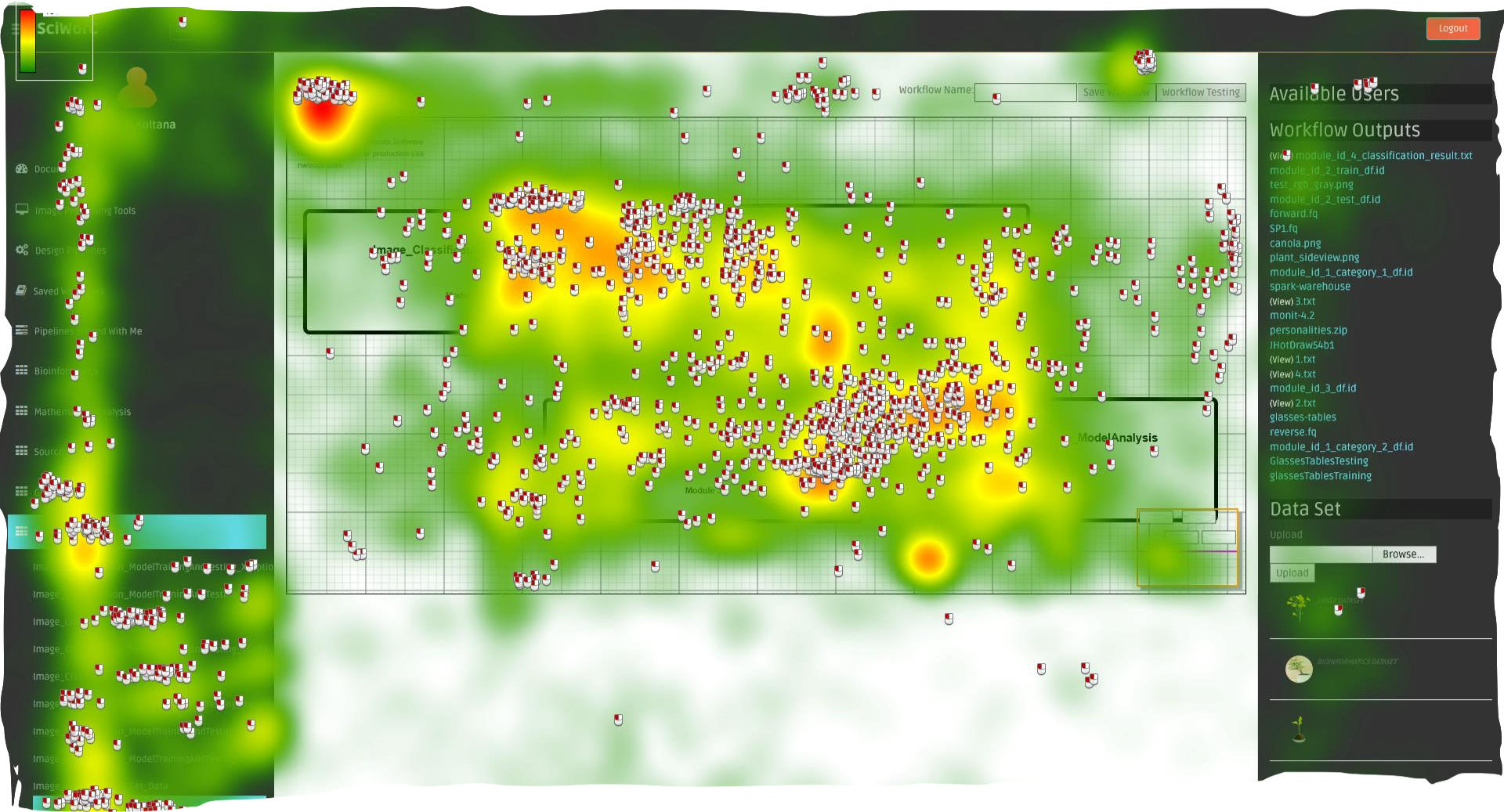}
  \caption{Heatmap of the interface after building workflows}
  \label{heatmap}
\end{figure}

Figure \ref{heatmap} is representing the heat map of image processing workflows.  In the heat map, we can see that users spent a considerable amount of time on modules after dragging and dropping them in the composition areas. Mostly on module 2 and 3, where users had the option to select intermediate data. So from the heat map, it can be interpreted that users are interested in the suggestion of the recommendation technique.
Areas of our interest are divided into two regions, i.e., composition area and suggestion area. Table \ref{TotaVisitDuration} represents the average time spent data in these two areas separately. From the table, it can be seen that around 39\% time of total composition area time is spent on suggestion areas. This means with the facilities of using intermediate data with parameter and time information user tried to understand the suggested data with interest to use them. Result of this study answers our research question 4 (\textbf{RQ4}) and explore the potentiality of suggested data on the interface.


\section{Discussion and Future Directions}
\label{discussion_and_future_direction}
The results of the three studies indicate that users generally prefer intermediate data for reusing while executing workflows in a SWfMS. In the study of computationally expensive image processing workflows, we found that the participants prefer mostly to use intermediate data for computationally expensive modules. In future, we have a plan to work on prioritizing the expensive modules in the SciWorCS for storing their outcomes with the technique, so that we can increase the reusability respect to users' usage pattern. Other significant findings that explored some research directions are also considered in this study for further improvement of our recommendation system. Especially the finding of users' usage preferences on intermediate data usage for known modules and known workflows of the two studies could be used to design a new architecture of the SWfMS where data associations with module and workflows will be expedited for efficiency. For instance, in real-life complex workflows are formed from a different domain, and many collaborators of different areas are involved in the workflow design process. In such a case, intermediate data management can help a collaborator to design and execute a known part of a workflow, and other collaborators could reuse this part for the rest of the execution. This type of system is only possible if we can increase the data awareness in a SWfMS, and we are currently working on it to use intermediate data in a collaborative environment.  In our second study of bioinformatics workflows, we see that while assembling long workflows users prefer to use more intermediate data. Since in real life workflow building process, sometimes 100 or more modules are involved so RISP can have a positive impact on SWfMS. Furthermore, each of the overall scores of NASA-TLX responses from both of the studies separately shows that after integrating the GUI-RISP with the SciWorCS participants experience less overhead. Moreover, the subclasses scores are less after integrating the GUI-RISP in the SWfMS. Hence we believe the technique of recommending intermediate data can increase the efficiency in a SWfMS without any shortcomings.

\section{Threat to Validity}
\label{ThreattoValidity}
We considered workflows of two areas (Image Processing and Bioinformatics)  in our user study for analyzing the effectiveness of the GUI-RISP in the SciWorCS. While more areas could make our findings more generalized, we see that the GUI-RISP shows effectiveness on workflows from both of the areas.  Thus, we believe that our findings cannot be attributed to chance. The proposed technique RISP can be considered an efficient technique for managing workflows.
\section{Conclusion and Future Works}
\label{ConclusionandFutureWorks}
To implement an efficient system for processing a large amount of data with numerous tools, proper data management in a system of workflow management is crucial. We proposed a technique of data recommendation for both storing and retrieving data while building workflows in a SWfMS and this study is intended to investigate on case studies of an GUI version of the technique for comprehending the behaviors and expectations of users in real-world workflow building. In our two case studies, from the provided raw data and generated intermediate data, in total intermediate data and raw data showed the split of 73 \% and 27 \% usage while building workflows by the all participants. This usage trend and preference on intermediate data imply that the technique can fulfill the expectation of users in most of the cases of efficient workflow execution. Additionally, we found that the technique is more useful for the long-running and complex workflows, so we believe our technique (i.e., RISP) has the potential to contribute to composing workflows with big data and heterogeneous tools. Moreover, using the NASA-TLX study, we found that demand in each subscale of NASA-TLX does not produce any negative impact while assembling and executing workflows in the SWfMS with the technique. Therefore, we believe the technique can be used in any SWfMS without any additional burden for introducing reusability and building workflow efficiently.

\chapter{Conclusion}
\label{main_conclusion}

This chapter is presented to conclude the thesis. We present a summary of all of our work in Section \ref{conclusion_summary}, and Section \ref{conclusion_future} indicates possibilities of future research.

\section{Summary}
\label{conclusion_summary}
A Scientific Workflow Management System (SWfMS) is a software solution to execute scientific workflows and manage scientific data sets in various computing environments \cite{Liu:2015:SDS:2884709.2884750} \cite{10.1007/978-3-540-68111-3_78}. In such a SWfMS, an automated technique and provenance tracking commit advantages for both composition and execution \cite{Davidson:2008:PSW:1376616.1376772} \cite{4404805}. Furthermore, a SWfMS needs to be reliable and efficient to coordinate and automate data flow and task execution in various environment such as in distributed computational, clusters, cyberinfrastructures, commercial and academic clouds environments \cite{Deelman:2015:PWM:2775768.2776457}. In scientific workflows, scientific data processing activities are assembled and automated to reduce the execution time while executing the activities \cite{Liu:2015:SDS:2884709.2884750}. Consequently, such systems face big data explosion issues with massive velocity and variety characteristics for the heterogeneous large amount of data from different domains. One of the main reasons for this explosion is the frequent execution of workflows in a SWfMS that generates a huge amount of data and characteristics of such data are always incremental. Therefore a large amount of heterogeneous data need to be managed in a Scientific Workflow Management System (SWfMS) with a proper decision mechanism. 

Previous studies also emphasized on data management with metadata and awareness of data in a scientific workflow management system for their complexity, dimension, and volume \cite{Simmhan:2005:SDP:1084805.1084812} \cite{4534284} \cite{Gray:2005:SDM:1107499.1107503}. Besides, reproducibility and reuse of essential components of workflows from SWfMSs in scientific communities is crucial \cite{Davidson:2008:PSW:1376616.1376772}  \cite{10.1007/11890850_2}. Although few studies mentioned that automation and reproducibility support could accelerate and transform workflow processes \cite{4404805}, but there is no explicit technique to do that.  Even the existing data management techniques do not support data reusability directly, most of the system is designed by emphasizing rules and cost \cite{article34636}. By emphasizing the reusability in a SWfMS, in this thesis, we propose an intermediate data management scheme. Furthermore, for a concrete decision mechanism, RISP (Recommending Intermediate States in Pipelines) is introduced that applies mining association rules to recommend intermediate data for storing and reusing while building workflows in a SWfMS. We have done several experiments for analyzing the performance and exploring the effectiveness of the technique in a SWfMS. All of our major contributions towards data management for a SWfMS is summarized below in four sections. 

\subsubsection{Management Scheme for Micro-Level Modular Programs} 
Although some of the previous works \cite{Big_Data_Usage} \cite{statistical_models} \cite{Data_Multiverse} \cite{Software_Architecture} \cite{Signal_Processing} were discussed with the fault-tolerant concept by managing intermediate states, none of them discussed the possibility and necessity of managing intermediate data for efficient scientific analysis. In this work, we proposed a novel data management scheme for intermediate data to support reusability in a SWfMS. Using the proposed management technique error recovery and efficient processing can be ensured in a SWfMS. We conducted several experiments with the proposed techniques for both the local file system and distributed file system to validate the efficiency of the technique. Our experimental studies show that the proposed technique can execute workflows efficiently by increasing performance up to 87 \%. 

\subsubsection{Optimized Storing of Workflow  Outputs}
A number of studies focused on recommending modules or services in  the workflow building process in a SWfMS  \cite{Koop2008VisComplete:Pipelines} \cite{Chinthaka2009CBRAssistant} \cite{Zhang2011Recommend-as-you-go:Reuse} \cite{Spjuth2015ExperiencesBioinformatics}. Although the recommendation or discovery of modules or services in the composition phase of a workflow can ensure the reusability of processes but not the outcomes of modules which users may need to store in a SWfMS. Besides, the data or the intermediate data are needed to be managed systematically and automatically is also emphasized by some studies \cite{Woodman2015WorkflowProvenance} \cite{Yuan2011On-demandSystems}. Where the previous studies on intermediate data are presented in terms of cost management and techniques are introduced to minimize the cost of data storing. In this work, by experimenting on different workflows from well known SWfMSs, we emphasized on reusability of data and proposed an automatic technique for both storing and retrieving data. Using the technique reusability can be increased up to 51\% and time gain can be reached up to 74\% in a SWfMS. 

\subsubsection{Storing Modes of Workflow}
Associations of  parameter configuration sets  with modules and management of parameter configuration sets in a  SWfMS are counted in some studies \cite{Altintas:2006:PCS:2165554.2165572} \cite{10.1007/11890850_2} \cite{Deelman2008DataWorkflows} \cite{4782949} \cite{SEFFINO1999105} to support the efficient re-execution of some parts of a workflow. Configured parameters of modules as an important part of metadata management and provenance are considered in many studies and common in practice. Despite dynamic parameter tuning is considered by Dias et al. \cite{Dias:2011:SDP:2110497.2110502} to improve the performance of workflow execution. Consequently, in our SWfMS we want to manage the parameters in a systematic way to use them properly for our data management scheme. Parameter set as a part of a configuration set to provide the tool state of a specific tool is considered in our system. In this study, we want to explore the usefulness of our technique, RISP considering the tool state concepts. To do this, we have analyzed 534 workflows from Galaxy, and our experimental results show that the technique can work with the tool state concept and reports around 40\% reusability in a SWfMS. 

\subsubsection{Usability Study}
A number of studies \cite{wu2012} \cite{Zheng:2015:ICW:2755979.2755984} \cite{Muniswamy-Reddy:2006:PSS:1267359.1267363} \cite{Brown2007} \cite{Wang:2009:KHG:1645164.1645176} \cite{Bahsi:2007:CWM:1377549.1377550} have considered the use case analysis and user study to reveal the insight of real-world workflows composition in a SWfMS. Though the studies are considered to evaluate SWfMSs from the user perspective, but there is a lack of inquiring a data management technique in a SWfMS to see the efficiency and effectiveness in the real world. One main reason for this lacking is that the previous studies of data management considered only the cost management in a SWfMS. In this work, we have considered reusability of data while storing them in a SWfMS, and with the concept of reusability user behavior and style of work are explored for a data management technique. We conducted several user studies of our SWfMS with the technique, RISP for various domain. Our studies reveal that the technique is very useful for composing workflows as well as illustrate some pieces of evidence that can be useful for future improvement.

\section{Future Directions}
\label{conclusion_future}

In future, we will consider the following actions in a technique of data management for a SWfMS.

\subsection{Integration of the technique in a fully functional provenance model}
Provenance model become a crucial part of a SWfMS that can be seen from a number of studies \cite{bhuyan2018quality} \cite{Davidson:2008:PSW:1376616.1376772} \cite{Gil:2011:MYM:2063076.2063082} \cite{Altintas:2006:PCS:2165554.2165572} \cite{Woodman2011AchievingVersioning} \cite{Missier2016ProvenanceAnalysis}. The lineage of scientific data, such as the origin, derivation, and contexts of datasets are tracked in a provenance model \cite{Bose:2005:LRS:1057977.1057978}. Recent studies \cite{Gil:2011:MYM:2063076.2063082} \cite{Altintas:2006:PCS:2165554.2165572} \cite{Woodman2011AchievingVersioning}  investigated on regeneration process of workflows using provenance tracing in a scientific management system. Integration of our technique in a scientific provenance model could reduce the cost of derived outcomes storing and increase the reusability of workflow components using our recommendation. In the future, we have a plan to work on such a provenance model where configuration sets as metadata with the proposed technique can be useful to reconstruct a full workflow at a low cost.  

\subsection{Prioritizing the expensive modules}
In our user study of the technique in a SWfMS, we found users had tendencies to use intermediate data for expensive modules. Chapter \ref{DesignRISP} represents the study of user usage pattern for intermediate states.  However, for the limitation of the user study and for the number of workflows, we can not generalize this. In the future, we will do more investigation on data types and their usage. A priority-based model could be useful for users to get more relevant recommendations of intermediate data in a SWfMS based on their usages.

\subsection{Genetic algorithm to recommend data}
Genetic algorithms are already popular in SWfMS community for maintaining workflow scheduling. In our SWfMS, we have a plan to introduce such technologies for data management to execute workflows in a low budget for a particular environment. In this case, we need to design a fitness function by maintaining the trade-off between data transition time and processing time. However, to design such a function in a SWfMS, the SWfMS has to be matured enough for the validation of the function. Our SWfMS is just evolving now, and in future, we will investigate on genetic algorithms based data management techniques in our Scientific Workflow Management System. 
 
\subsection{Data and domain awareness}
Heterogeneity of tools and data in a SWfMS urges the collaboration from different domains in the SWfMS. In the collaboration process, where multiple domain experts are involved, a domain-specific suggestion mechanism would help participants to understand more about the data usability. With the help of metadata, our recommendation process could increase the domain awareness in a SWfMS, and in future, we will work on a technique of data management of a SWfMS where data awareness will be emphasized for various domains.

%
% ...
%




\uofsbibliography{main}

\uofsappendix

\begin{appendices}

\chapter{Data Management Scheme}

\section{Code fragment of a modularized job}
\lstset{language=Python, caption={An example code of modularized programs},label=clusteringcode}
\begin{lstlisting}
def hdfsClustering():
    server = sys.argv[1]
    uname = sys.argv[2]
    upass = sys.argv[3]
    data_path = sys.argv[4]
    save_path = sys.argv[5]
    img_type = sys.argv[6]
    k = int(sys.argv[7])
    iterations = int(sys.argv[8])


    pipes = ImgCluster(server, uname, upass)
    pipes.setLoadAndSavePath(data_path, save_path)
    processing_start_time = time()
    rdd = loadExecuteHDFS(sc, data_path, pipes)
    rdd = common_transform(pipes, rdd, (0))
    rdd = common_estimate(pipes, rdd, ("sift"))
    model = cluster_model(pipes, rdd, (k, iterations))
    rdd = cluster_analysis(pipes, rdd, model)
    processing_end_time = time() - processing_start_time

    pipes.saveClusterResult(rdd.collect())
    print "SUCCESS: Images procesed in {} seconds".format(round(processing_end_time, 3))

    url = 'http://XXX.XX.XXX.XX:5984/plantphenotype/2bbdc24f3d74d6859423e4311409b7be/'
    payload = {
   "_id": "2bbdc24f3d74d6859423e4311409b7be",
   "_rev": "3-6372072da016fa257640892ea85e06c7",
   "inputdataset": "hdfs://hadoop-master:54310/user/hadoop/leaves_images",
   "execution_time": processing_end_time,
   "the_doc_type": "module",
   "special_weight": "XX",
   "number_of_time_called": X,
   "caller_workflow": "2bbdc24f3d74d6859423e4311409aefa",
   "outputdataset": "hdfs://hadoop-master:54310/user/hadoop/leaves_result_new",
   "is_intermediate_states": False,
   "module_id": "module_id_1"
    }
    headers = {}
    res = requests.post(url, data=payload, headers=headers)


if(__name__=="__main__"):
spark = SparkSession.builder.appName("Clustering on Server").getOrCreate()
sc = spark.sparkContext
sqlContext = SQLContext(sc)
npartitions = XX
# loger = sc._jvm.org.apache.log4j
# logger = loger.LogManager.getRootLogger()
# describeLeaves()
# matchLeaves()
# hdfsImgSeg()
hdfsClustering()
sc.stop()
\end{lstlisting}

\chapter{RISP}

\section{Code fragment of RISP analyzer}
\lstset{language=Java, caption={An example code of RISP evaluation},label=RISPAnalyzer}
\begin{lstlisting}
public class Associationrules {


    public static void main(String[] args) {     
        
        
        System.out.println ("\n\nProposed System.");
        Pipelines p = new Pipelines ();
        p.readPipelines();
        p.analyzePipelines();
        p.getResult();
        
        System.out.println ("\n\nSystem that always saves.");
        Pipelines p2 = new Pipelines ();
        p2.readPipelines();
        p2.analyzePipelinesAlwaysSave ();
        p2.getResult();
        
        System.out.println ("\n\nSystem that saves those results that were previously used at least once.");
        Pipelines p3 = new Pipelines ();
        p3.readPipelines();
        p3.analyzePipelinesSupportOne ();
        p3.getResult();

        System.out.println ("\n\nSystem that saves final.");
        Pipelines p4 = new Pipelines ();
        p4.readPipelines();
        p4.analyzePipelinesFinalResult ();
        p4.getResult();                         
    }
    
}
\end{lstlisting}

\chapter{Adaptive RISP}

\section{JSON structure of a workflow}
\lstset{language=json, caption={An example workflow of stranded and prokaryotic RNAseq data analysis},label=DetrprokV4Frag}
\begin{lstlisting}
{
    "a_galaxy_workflow": "true", 
    "annotation": "", 
    "format-version": "1.1", 
    "name": "Detrprok", 
    "steps": {
        "0": {
            "annotation": "Select here the GFF annotation file containing coding genes, tRNAs, rRNAs and known ncRNAs. You can download it from GJI (http://genome.jgi.doe.gov) or NCBI (http://ftp.ncbi.nih.gov/genomes/Bacteria). Caution: 1srt column should have exactly the same identifier as the genome sequence used for mapping.", 
            "id": 0, 
            "input_connections": {}, 
            "inputs": [
                {
                    "description": "Select here the GFF annotation file containing coding genes, tRNAs, rRNAs and known ncRNAs. You can download it from GJI (http://genome.jgi.doe.gov) or NCBI (http://ftp.ncbi.nih.gov/genomes/Bacteria). Caution: 1srt column should have exactly the same identifier as the genome sequence used for mapping.", 
                    "name": "genome annotations"
                }
            ], 
            "label": null, 
            "name": "Input dataset", 
            "outputs": [], 
            "position": {
                "left": 206, 
                "top": 152
            }, 
            "tool_errors": null, 
            "tool_id": null, 
            "tool_state": "{\"name\": \"genome annotations\"}", 
            "tool_version": null, 
            "type": "data_input", 
            "user_outputs": [], 
            "uuid": "8b700d1d-acf8-423e-92e3-d125a2a96526"
        }, 
        "1": {
            "annotation": "Select here the bam file resulting from the mapping of RNA-seq reads on the genome sequence. You may use Bowtie or another mapper (use appropriate mapping option to retain only uniquely mapping reads).", 
            "id": 1, 
            "input_connections": {}, 
            "inputs": [
                {
                    "description": "Select here the bam file resulting from the mapping of RNA-seq reads on the genome sequence. You may use Bowtie or another mapper (use appropriate mapping option to retain only uniquely mapping reads).", 
                    "name": "read alignments on genome"
                }
            ], 
            "label": null, 
            "name": "Input dataset", 
            "outputs": [], 
            "position": {
                "left": 200, 
                "top": 403
            }, 
            "tool_errors": null, 
            "tool_id": null, 
            "tool_state": "{\"name\": \"read alignments on genome\"}", 
            "tool_version": null, 
            "type": "data_input", 
            "user_outputs": [], 
            "uuid": "00d8d4c6-aca2-4eed-8844-cfa3cf9aa12c"
        }, 
        .
        .
        .
        .
        .
        .
        .
        .
        .
        .
        .
        .
        .
        "42": {
            "annotation": "Uses an RGB color tag for visualization of long 5'UTR in the Artemis environment.", 
            "id": 42, 
            "input_connections": {
                "referenciesFile": {
                    "id": 41, 
                    "output_name": "outputFile"
                }
            }, 
            "inputs": [], 
            "label": null, 
            "name": "colorGff", 
            "outputs": [
                {
                    "name": "outputFile", 
                    "type": "gff"
                }
            ], 
            "position": {
                "left": 3158, 
                "top": 641
            }, 
            "post_job_actions": {
                "RenameDatasetActionoutputFile": {
                    "action_arguments": {
                        "newname": "long 5_UTR list"
                    }, 
                    "action_type": "RenameDatasetAction", 
                    "output_name": "outputFile"
                }
            }, 
            "tool_errors": null, 
            "tool_id": "toolshed.g2.bx.psu.edu/repos/clairetn/detrprok_scripts/colorGff/1.0.0", 
            "tool_state": "{\"__page__\": 0, \"__rerun_remap_job_id__\": null, \"RGBcolor\": \"\\\"250 105 180\\\"\", \"referenciesFile\": \"null\"}", 
            "tool_version": "1.0.0", 
            "type": "tool", 
            "user_outputs": [], 
            "uuid": "78bb45b9-1d16-498a-bea8-40bf3447a4bd"
        }
    }, 
    "uuid": "d190d09c-2654-4217-884e-a0ef7dbe05fc"
}
\end{lstlisting}

\chapter{GUI-RISP: User Study}

\section{Code fragment of a distributed job}
\lstset{language=Python,caption={Spark implementation of InceptionV3 model},label=InceptionV3code}
\begin{lstlisting}
# spark-submit --packages databricks:spark-deep-learning:1.2.0-spark2.3-s_2.11 /home/joy/PycharmProjects/ToolsforSwfMS/ClassificationModelTrainingAndTesting.py
import pickle
from pyspark.ml.image import ImageSchema
import os
os.environ["CUDA_VISIBLE_DEVICES"]="-1"
from pyspark.ml.classification import LogisticRegression
from pyspark.sql.functions import lit
from sparkdl import DeepImageFeaturizer
from pyspark.ml import Pipeline
from pyspark import SparkContext, SparkConf, SQLContext
from pyspark.sql.types import *
from pyspark.sql import SparkSession

import requests
from time import time
import sys
import ast

CouchDBServer = 'XXX.XXX.X.XX'
HDFS_Dir = "/user/dec657/"

conf = SparkConf().setAppName('ClassificationModelTrainingAndTestingInceptionV3')
sc = SparkContext(conf=conf)
spark = SparkSession.builder.getOrCreate()


dict = {}
dict = ast.literal_eval(sys.argv[-1])
workflowdocid = dict['workflowdocid']
modulerev = dict['modulerev']
module_id = dict['module_id']
input_dataset = dict['input'].split(",")
output_dataset = dict['output'].split(",")


processing_start_time = time()
train_df = spark.read.load(HDFS_Dir + input_dataset[0].strip())
train_df.show()
test_df = spark.read.load(HDFS_Dir + input_dataset[1].strip())
test_df.show()
featurizer = DeepImageFeaturizer(inputCol="image", outputCol="features", modelName="InceptionV3")
lr = LogisticRegression(maxIter=20, regParam=0.05, elasticNetParam=0.3, labelCol="label")
p = Pipeline(stages=[featurizer, lr])
p_model = p.fit(train_df)
df = p_model.transform(test_df)
df.write.mode('overwrite').save(output_dataset[0].strip())

processing_end_time = (time() - processing_start_time)
print "SUCCESS: Images processed in {} seconds".format(round(processing_end_time, 3))


url = 'http://'+CouchDBServer+':5984/plantphenotype_inter_v1/'+module_id+'/'
payload = {
    "_id": ""+module_id,
    "_rev": "" + modulerev,
    "inputdataset": input_dataset,
    "execution_time": processing_end_time,
    "the_doc_type": "module",
    "special_weight": 5,
    "number_of_time_called": 0,
    "caller_workflow": ""+workflowdocid,
    "outputdataset": output_dataset,
    "is_intermediate_states": 1,
    "module_id": "Image_Classification_ModelTrainingAndTesting_InceptionV3"
}
headers = {}
res = requests.put(url, json=payload)
\end{lstlisting}

\section{Distributed job submission}
\lstset{language=Python, caption={Example code for submitting a job},label=JobSubmission}
\begin{lstlisting}
import paramiko
import time
from io import BytesIO as StringIO
import requests
import ast
import time

SERVER = app.config['DISTRIBUTEED_MAIN_NODE']

USER ='dec657'
PASSWORD = 'XXXXXXXXXXXXXXXXXXX'

ssh = paramiko.SSHClient()
ssh.set_missing_host_key_policy(paramiko.AutoAddPolicy())
ssh.connect(SERVER, username=USER, password=PASSWORD)
client = ssh.invoke_shell()
stdout = client.makefile('rb')
stdin = client.makefile('wb')
job_name = app.config['LOCAL_HOME']+'ClassificationModelTrainingAndTestingInceptionV3.py'

rawcommand = ''' 
spark-submit --packages databricks:spark-deep-learning:1.2.0-spark2.3-s_2.11 {job_name} "{plugin_params}"
'''
cmmnd = rawcommand.format(job_name=str(job_name), plugin_params=str({record_ids}))

stdin.write(cmmnd)

timeout = 60
import time
endtime = time.time() + timeout
while not stdout.channel.eof_received:
     time.sleep(1)
     if time.time() > endtime:
         stdout.channel.close()
         break
print(stdout.read())

stdin.close()
ssh.close()

job_id = "job_id_X"

url = app.config['COUCHDB_SERVER']+'/'+app.config['COUCHDB_DATABASE']+'/'+'module_id_for_stutas/'
print (url)
payload = {}
headers = {}
res = requests.get(url)
res_dict = ast.literal_eval(res.text)

if res_dict['outputdataset'] != "":
    if type(res_dict['outputdataset']) is str:
        res_dict['outputdataset'] = ast.literal_eval(res_dict['outputdataset'])
    for datasetname in res_dict['outputdataset']:
        if (datasetname.strip() != ''):
            file = open("app_collaborative_sci_workflow/workflow_outputs/test_workflow/"+job_id+"_"+datasetname.strip(), "w")
            file.close()

\end{lstlisting}

\section{Source Code Snippets}
\lstset{language=Python, caption={JavaScript Tracking Snippet of Google Analytics for User Log},label=JavaScriptTrackingSnippet}
\begin{lstlisting}
<!-- Google Analytics -->
<script>
window.ga=window.ga||function(){(ga.q=ga.q||[]).push(arguments)};ga.l=+new Date;
ga('create', 'UA-XXXXXXXXX-Y', 'auto', {'sampleRate': 100, 'siteSpeedSampleRate': 100});
ga('send', 'pageview');
</script>
<script async src='https://www.google-analytics.com/analytics.js'></script>
<!-- End Google Analytics -->


ga('send', {
hitType: 'event',
eventCategory: 'IntermediateData',
eventAction: 'CheckedMultiple',
eventValue: parseInt(clickedNumber),
eventLabel: '#'+user_email+'#'+currentAssembledModule
});

ga('send', {
hitType: 'timing',
timingCategory: 'CanExecute'+is_intermediate_used,
timingVar: 'load_to_run',
timingValue: timeSincePageLoad,
timingLabel: '#'+user_email+'#'+currentAssembledModule
});

ga('send', {
hitType: 'event',
eventCategory: 'IntermediateData',
eventAction: 'AModule',
eventValue: ModNames_for_log.length,
eventLabel: '#'+user_email+'#'+currentAssembledModule
});

ga('send', {
hitType: 'event',
eventCategory: 'IntermediateData',
eventAction: 'UsedI',
eventValue: ModNames.length,
eventLabel: '#'+user_email+'#'+currentAssembledModule
});

ga('send', {
hitType: 'event',
eventCategory: 'IntermediateData',
eventAction: 'AvailableI',
eventValue: ModNames.length,
eventLabel: '#'+user_email+'#'+currentAssembledModule
});

ga('send', {
hitType: 'event',
eventCategory: 'IntermediateData',
eventAction: user_email,
eventValue: ModNames.length,
eventLabel: currentAssembledModule
});

ga('send', {
hitType: 'event',
eventCategory: 'IntermediateData',
eventAction: 'LoadedSeqKnown',
eventValue: ModNames.length,
eventLabel: '#'+user_email+'#'+currentAssembledModule
});

ga('send', {
hitType: 'event',
eventCategory: 'IntermediateData',
eventAction: 'LoadedSeqUnKnown',
eventValue: ModNames.length,
eventLabel: '#'+user_email+'#'+currentAssembledModule
});

ga('send', {
hitType: 'event',
eventCategory: 'IntermediateData',
eventAction: 'CheckedSingle',
eventValue:id_found_position,
eventLabel: '#'+user_email+'#'+currentAssembledModule
});

ga('send', {
hitType: 'timing',
timingCategory: 'CanAssemble'+is_intermediate_used,
timingVar: 'load_to_add',
timingValue: timeSincePageLoad,
timingLabel: '#'+user_email+'#'+currentAssembledModule
});
\end{lstlisting}

\section{Recorded JMeter script fragment}
\lstset{language=XML, caption={An example Jmeter recoding script},label=JmeterRecoding}
\begin{lstlisting}
<?xml version="1.0" encoding="UTF-8"?>
<jmeterTestPlan version="1.2" properties="5.0" jmeter="5.1 r1853635">
  <hashTree>
    <TestPlan guiclass="TestPlanGui" testclass="TestPlan" testname="Test Plan" enabled="true">
      <stringProp name="TestPlan.comments"></stringProp>
      <boolProp name="TestPlan.functional_mode">false</boolProp>
      <boolProp name="TestPlan.serialize_threadgroups">false</boolProp>
      <elementProp name="TestPlan.user_defined_variables" elementType="Arguments" guiclass="ArgumentsPanel" testclass="Arguments" testname="User Defined Variables" enabled="true">
        <collectionProp name="Arguments.arguments"/>
      </elementProp>
      <stringProp name="TestPlan.user_define_classpath"></stringProp>
    </TestPlan>
    <hashTree>
      <Arguments guiclass="ArgumentsPanel" testclass="Arguments" testname="User Defined Variables" enabled="true">
        <collectionProp name="Arguments.arguments">
          <elementProp name="host" elementType="Argument">
            <stringProp name="Argument.name">host</stringProp>
            <stringProp name="Argument.value">p2irc-shipi.usask.ca</stringProp>
            <stringProp name="Argument.metadata">=</stringProp>
          </elementProp>
          <elementProp name="scheme" elementType="Argument">
            <stringProp name="Argument.name">scheme</stringProp>
            <stringProp name="Argument.value">https</stringProp>
            <stringProp name="Argument.metadata">=</stringProp>
          </elementProp>
        </collectionProp>
      </Arguments>
      <hashTree/>
      <ConfigTestElement guiclass="HttpDefaultsGui" testclass="ConfigTestElement" testname="HTTP Request Defaults" enabled="true">
        <elementProp name="HTTPsampler.Arguments" elementType="Arguments" guiclass="HTTPArgumentsPanel" testclass="Arguments" testname="User Defined Variables" enabled="true">
          <collectionProp name="Arguments.arguments"/>
        </elementProp>
        <stringProp name="HTTPSampler.domain">p2irc-shipi.usask.ca</stringProp>
        <stringProp name="HTTPSampler.port"></stringProp>
        <stringProp name="HTTPSampler.protocol">https</stringProp>
        <stringProp name="HTTPSampler.contentEncoding"></stringProp>
        <stringProp name="HTTPSampler.path"></stringProp>
        <stringProp name="HTTPSampler.concurrentPool">6</stringProp>
        <stringProp name="HTTPSampler.connect_timeout"></stringProp>
        <stringProp name="HTTPSampler.response_timeout"></stringProp>
      </ConfigTestElement>
.
.
.
.
.
.
.
.
      <ResultCollector guiclass="SummaryReport" testclass="ResultCollector" testname="Summary Report" enabled="true">
        <boolProp name="ResultCollector.error_logging">false</boolProp>
        <objProp>
          <name>saveConfig</name>
          <value class="SampleSaveConfiguration">
            <time>true</time>
            <latency>true</latency>
            <timestamp>true</timestamp>
            <success>true</success>
            <label>true</label>
            <code>true</code>
            <message>true</message>
            <threadName>true</threadName>
            <dataType>true</dataType>
            <encoding>false</encoding>
            <assertions>true</assertions>
            <subresults>true</subresults>
            <responseData>false</responseData>
            <samplerData>false</samplerData>
            <xml>false</xml>
            <fieldNames>true</fieldNames>
            <responseHeaders>false</responseHeaders>
            <requestHeaders>false</requestHeaders>
            <responseDataOnError>false</responseDataOnError>
            <saveAssertionResultsFailureMessage>true</saveAssertionResultsFailureMessage>
            <assertionsResultsToSave>0</assertionsResultsToSave>
            <bytes>true</bytes>
            <sentBytes>true</sentBytes>
            <url>true</url>
            <threadCounts>true</threadCounts>
            <idleTime>true</idleTime>
            <connectTime>true</connectTime>
          </value>
        </objProp>
        <stringProp name="filename"></stringProp>
      </ResultCollector>
      <hashTree/>
      <ResultCollector guiclass="TableVisualizer" testclass="ResultCollector" testname="View Results in Table" enabled="true">
        <boolProp name="ResultCollector.error_logging">false</boolProp>
        <objProp>
          <name>saveConfig</name>
          <value class="SampleSaveConfiguration">
            <time>true</time>
            <latency>true</latency>
            <timestamp>true</timestamp>
            <success>true</success>
            <label>true</label>
            <code>true</code>
            <message>true</message>
            <threadName>true</threadName>
            <dataType>true</dataType>
            <encoding>false</encoding>
            <assertions>true</assertions>
            <subresults>true</subresults>
            <responseData>false</responseData>
            <samplerData>false</samplerData>
            <xml>false</xml>
            <fieldNames>true</fieldNames>
            <responseHeaders>false</responseHeaders>
            <requestHeaders>false</requestHeaders>
            <responseDataOnError>false</responseDataOnError>
            <saveAssertionResultsFailureMessage>true</saveAssertionResultsFailureMessage>
            <assertionsResultsToSave>0</assertionsResultsToSave>
            <bytes>true</bytes>
            <sentBytes>true</sentBytes>
            <url>true</url>
            <threadCounts>true</threadCounts>
            <idleTime>true</idleTime>
            <connectTime>true</connectTime>
          </value>
        </objProp>
        <stringProp name="filename"></stringProp>
      </ResultCollector>
      <hashTree/>
    </hashTree>
  </hashTree>
</jmeterTestPlan>
\end{lstlisting}

\section{Database Mapping Classes and View}
\lstset{language=Python, caption={Example codes for metadata},label=DBMapping}
\begin{lstlisting}
from flaskext.couchdb import Document, TextField, FloatField, DictField, Mapping, ListField, IntegerField, BooleanField
from couchdb.design import ViewDefinition

class WorkFlow(Document):
    the_doc_type = TextField(default='workflowintermediate')
    workflow_id = ListField(DictField)
    workflow_name = TextField(default='testName')
    user_email = TextField()

class Module(Document):
    the_doc_type = TextField(default='module')
    module_id = TextField()
    storing_time = FloatField(default=0.0)
    execution_time = FloatField(default=0.0)
    retrieving_time = FloatField(default=0.0)
    inputdataset = TextField()
    outputdataset = TextField()
    is_intermediate_states = FloatField(default=0.0)
    special_weight = TextField(default='0')
    number_of_time_called = TextField(default='0')
    result = TextField(default ='0')
    caller_workflow = TextField()
    parameters = ListField(DictField)

class Dataset(Document):
    the_doc_type = TextField(default='dataset')
    dataset_id = TextField()
    size = TextField(default='00')
    saving_time = FloatField(default=0.0)
    loading_time = FloatField(default=0.0)
    caller_module = TextField()

class IntermediateDataset(Document):
    the_doc_type = TextField(default='intermediatedataset')
    dataset_id = TextField()
    size = TextField(default='00')
    module_id = TextField()
    saving_time = FloatField(default=0.0)
    loading_time = FloatField(default=0.0)
    generator_module = TextField()
    from_dataset = TextField()

class Cost(Document):
    the_doc_type = TextField(default='cost')
    cost_id = TextField()
    storage_cost = FloatField(default=0.0)
    time_cost = FloatField(default=0.0)

class WorkFlowUser(Document):
    the_doc_type = TextField(default='workflowuser')
    workflow_id = TextField()
    user_name = TextField()
    user_email = TextField()


views_by_caller_workflow_for_module = ViewDefinition('intermediate_workflow', 'module', '''
    function (doc) {
         if (doc.the_doc_type && doc.the_doc_type == 'module') {
            emit(doc._id, doc.caller_workflow)
        };
    }
    ''')

views_by_workflowintermediate = ViewDefinition('intermediate_workflow', 'workflowintermediate', '''
    function (doc) {
         if (doc.the_doc_type && doc.the_doc_type == 'workflowintermediate') {
            emit(doc.workflow_id, doc.user_email)
        };
    }
    ''')

views_by_module = ViewDefinition('intermediate_workflow', 'module', '''
    function (doc) {
         if (doc.the_doc_type && doc.the_doc_type == 'module') {
            emit(doc.module_id, doc.caller_workflow)
        };        
    }
    ''')

views_by_module_for_outputdataset = ViewDefinition('intermediate_workflow', 'module', '''
    function (doc) {
         if (doc.the_doc_type && doc.the_doc_type == 'module') {
            emit(doc.module_id, doc)
        };
    }
    ''')

views_by_dataset_caller_module = ViewDefinition('intermediate_workflow', 'dataset', '''
    function (doc) {
         if (doc.the_doc_type && doc.the_doc_type == 'dataset') {
            emit(doc.caller_module, doc._id)
        };
    }
    ''')
\end{lstlisting}

\chapter{User Study: User Manual}

\section{Evaluation of RISP}
RISP (Recommending Intermediate States from Pipelines) can execute pipelines with previously processed data for reducing computational time of some modules in a SWfMS. This document will guide you through the evaluation of the usability of the technique in our SWfMS. We highly appreciate your participation.

\section{RISP in SWfMS}
In this study, you need to execute only seven workflows. You do not need to provide any feedback or answer any questions for building the workflows. All of your activities will be logged and timestamped for further analysis while assembling and executing workflows. At the end of each section (there are only two sections), you need to provide only workload responses in the NASA-TLX app on an iPad. Below is the link of the website to compose and execute the seven workflows. Workflows of this study are presented graphically in the next two sections by their types.  


L - Left side toolbox panel 
R - Right side dataset  panel
MCW - Main container of workflows
MPW - Module pop up window
R - RISP recommendation
AR - Adaptive RISP recommendation
RISP - The recommendation technique
A-RISP - Adaptive version of the technique

\begin{figure}[htbp]
\centerline{\includegraphics[width=\textwidth]{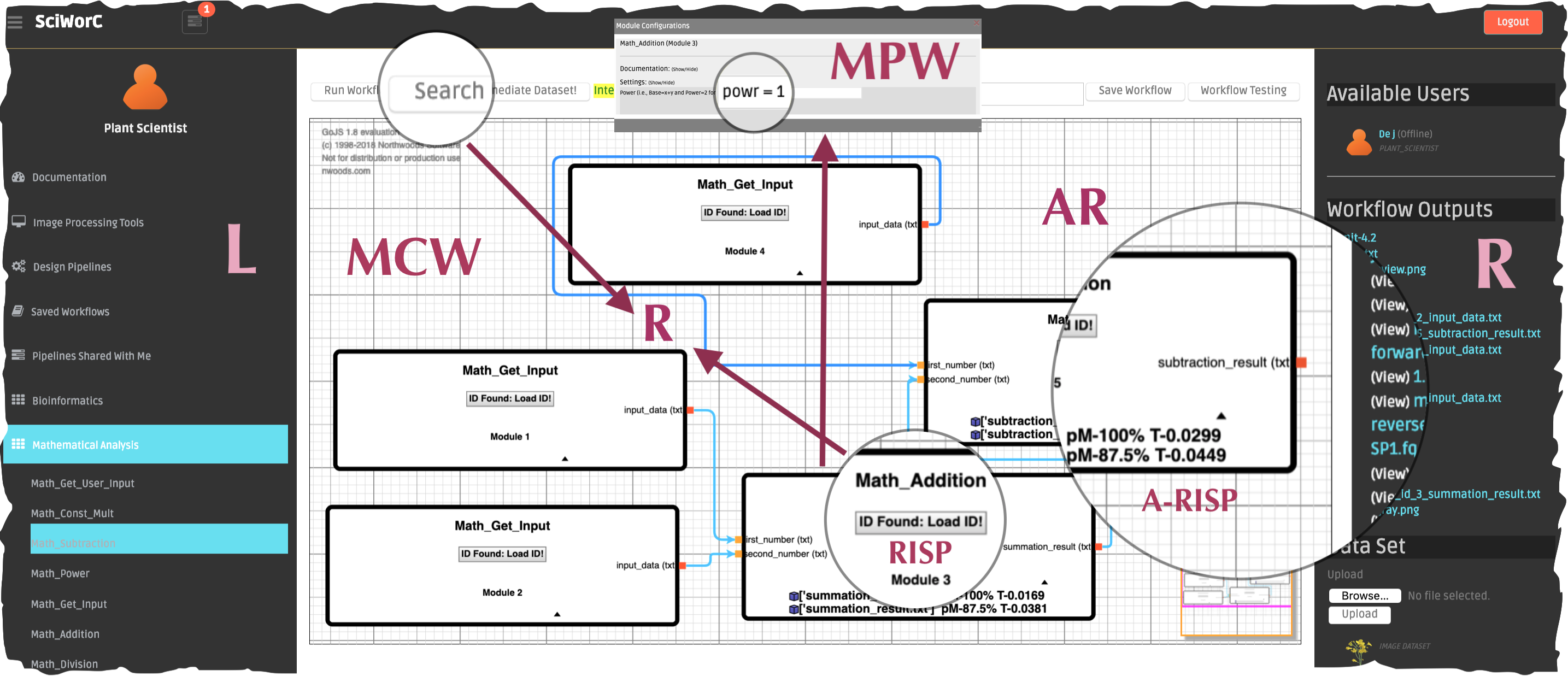}}
\caption{ User interface for building workflows.}
\label{fig_USUI}
\end{figure}

\section{Evaluation of the technique (Study 1)}
Image Processing Workflow Composition:
There are total five workflows in this study. Try to execute only two of them (same or different workflows with and without the RISP), i.e., first, you have to build a workflow without intermediate states and then you will have the choice to build another workflow with intermediate states.   

\begin{figure}[htbp]
\centerline{\includegraphics[width=\textwidth]{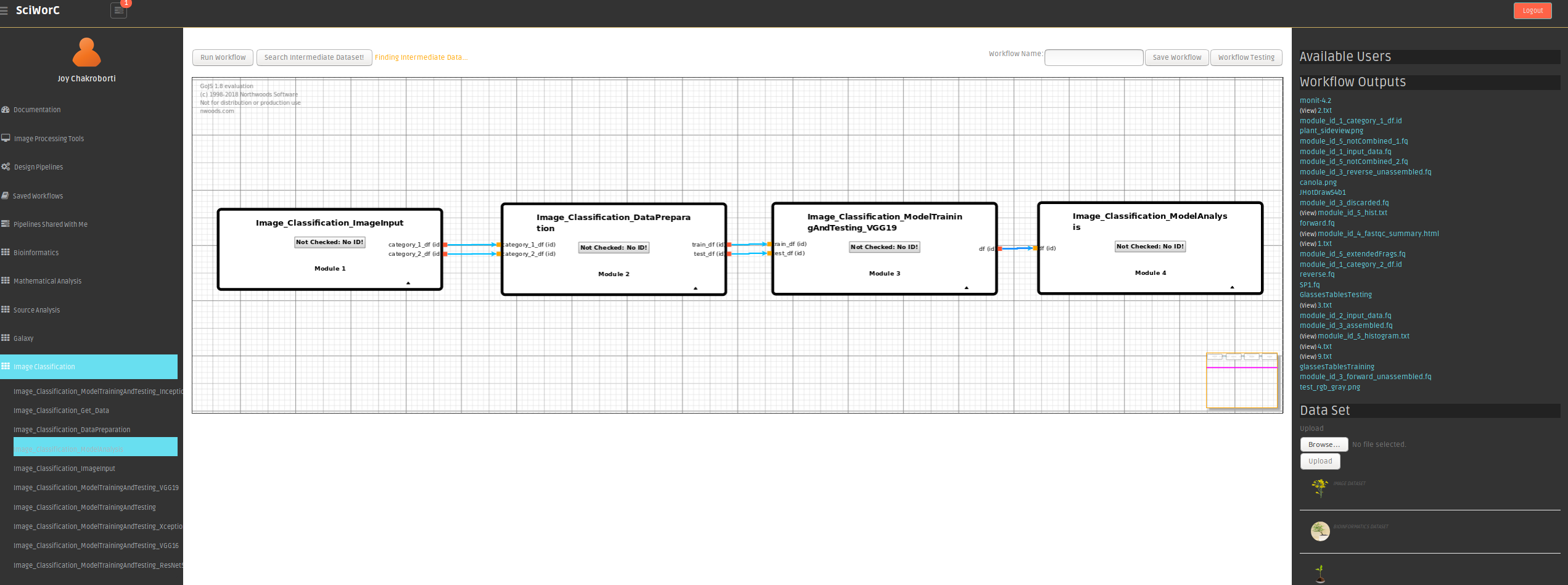}}
\caption{ Image Processing Workflow 1.}
\label{fig_IP1}
\end{figure}

\begin{figure}[htbp]
\centerline{\includegraphics[width=\textwidth]{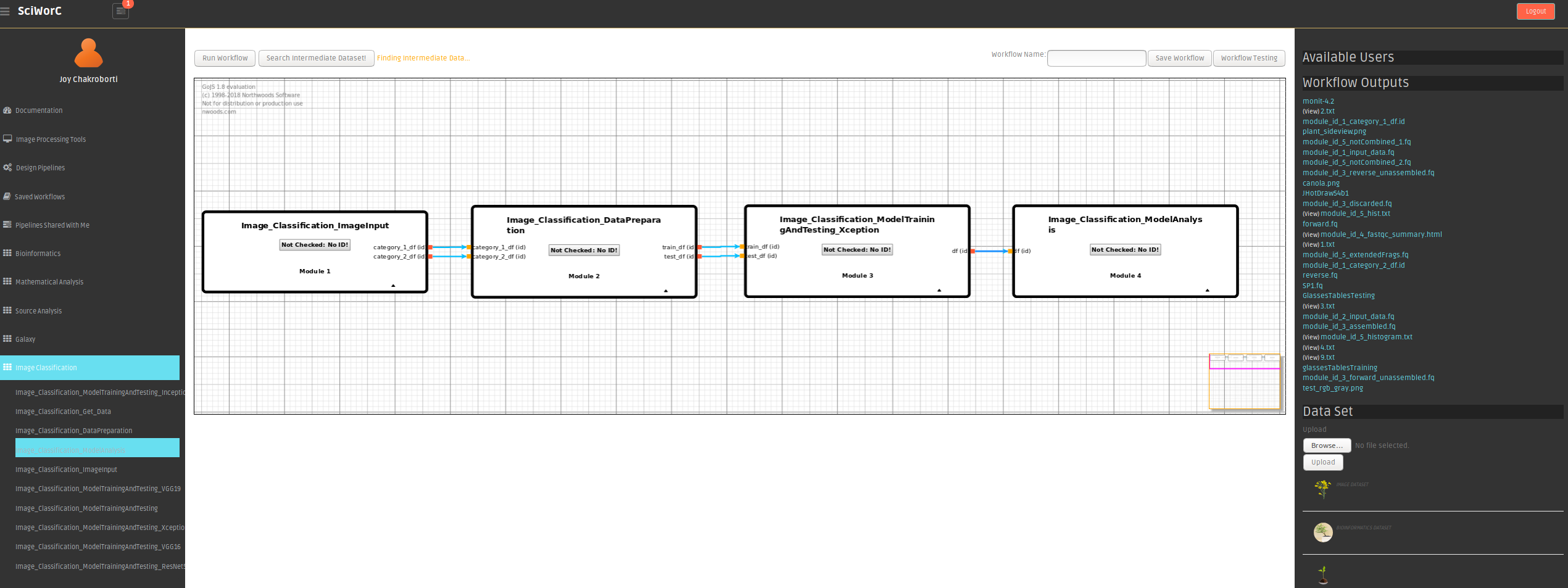}}
\caption{ Image Processing Workflow 2.}
\label{fig_IP2}
\end{figure}

\begin{figure}[htbp]
\centerline{\includegraphics[width=\textwidth]{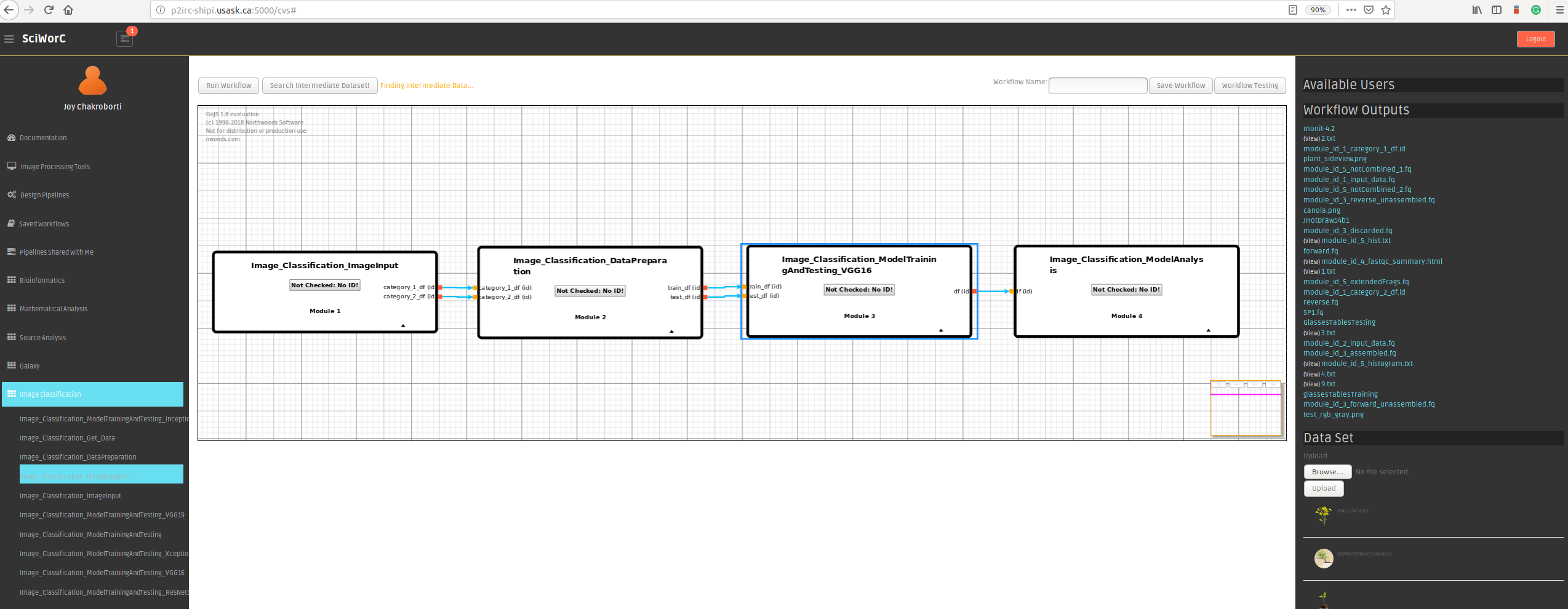}}
\caption{ Image Processing Workflow 3.}
\label{fig_IP3}
\end{figure}

\begin{figure}[htbp]
\centerline{\includegraphics[width=\textwidth]{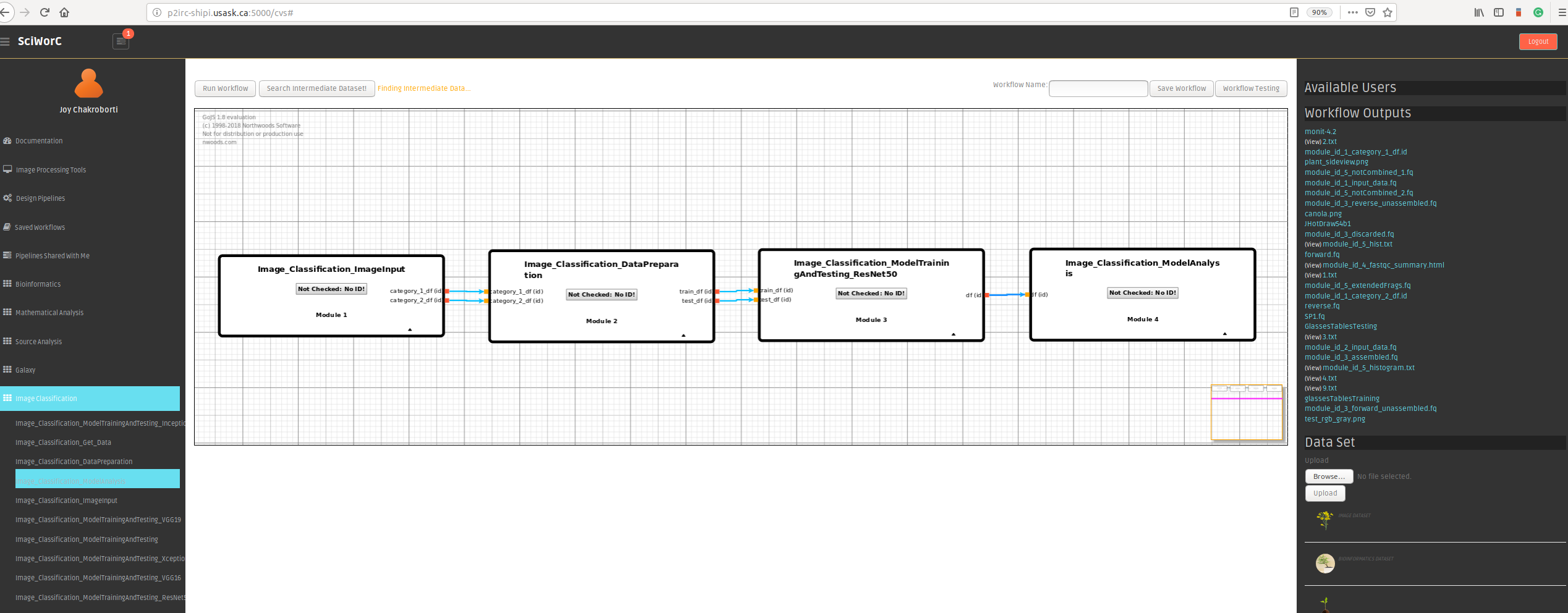}}
\caption{ Image Processing Workflow 4.}
\label{fig_IP4}
\end{figure}

\begin{figure}[htbp]
\centerline{\includegraphics[width=\textwidth]{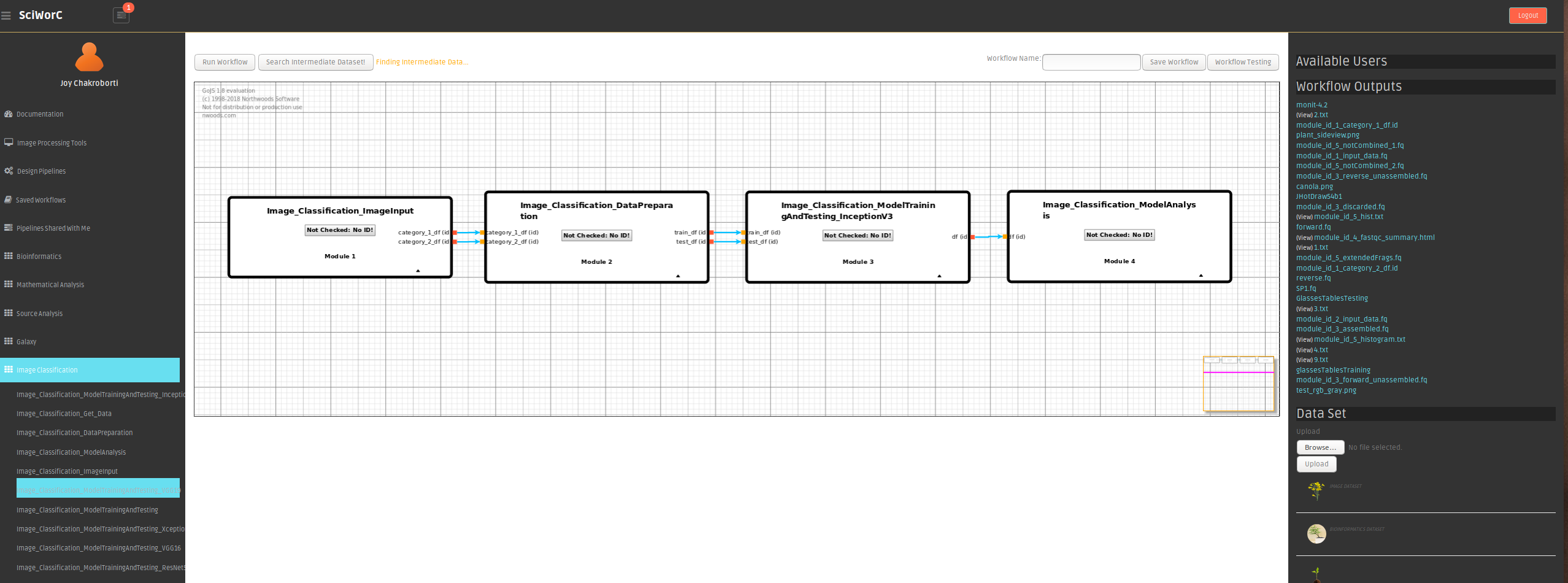}}
\caption{ Image Processing Workflow 5.}
\label{fig_IP5}
\end{figure}

\section{Evaluation of the technique (Study 2)}

Bioinformatics Workflow Composition:

In the bioinformatics workflow composition, you have to complete all of the four workflows presented below. In this study, you have the freedom to chose intermediate states by their availability. The complexity of the workflows increases gradually. 

\begin{figure}[htbp]
\centerline{\includegraphics[width=\textwidth]{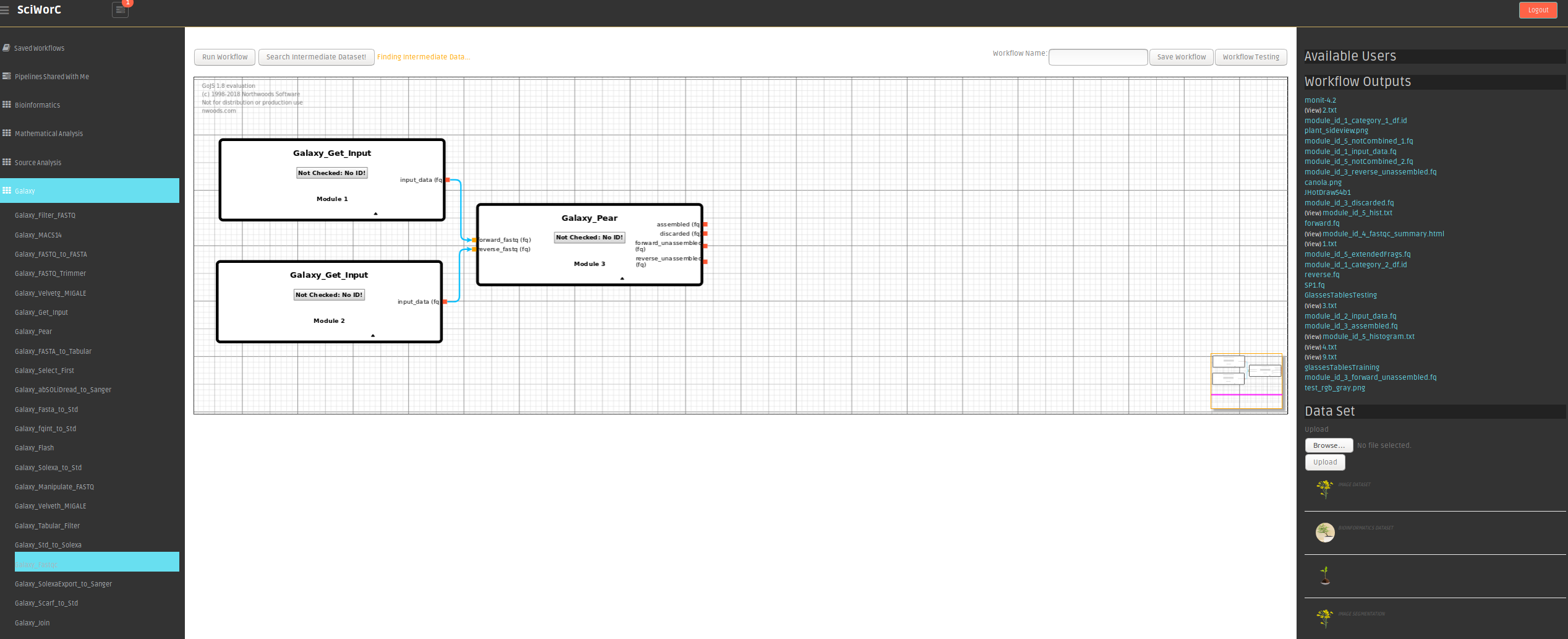}}
\caption{Bioinformatics Workflow 1.}
\label{fig_BI1}
\end{figure}

\begin{figure}[htbp]
\centerline{\includegraphics[width=\textwidth]{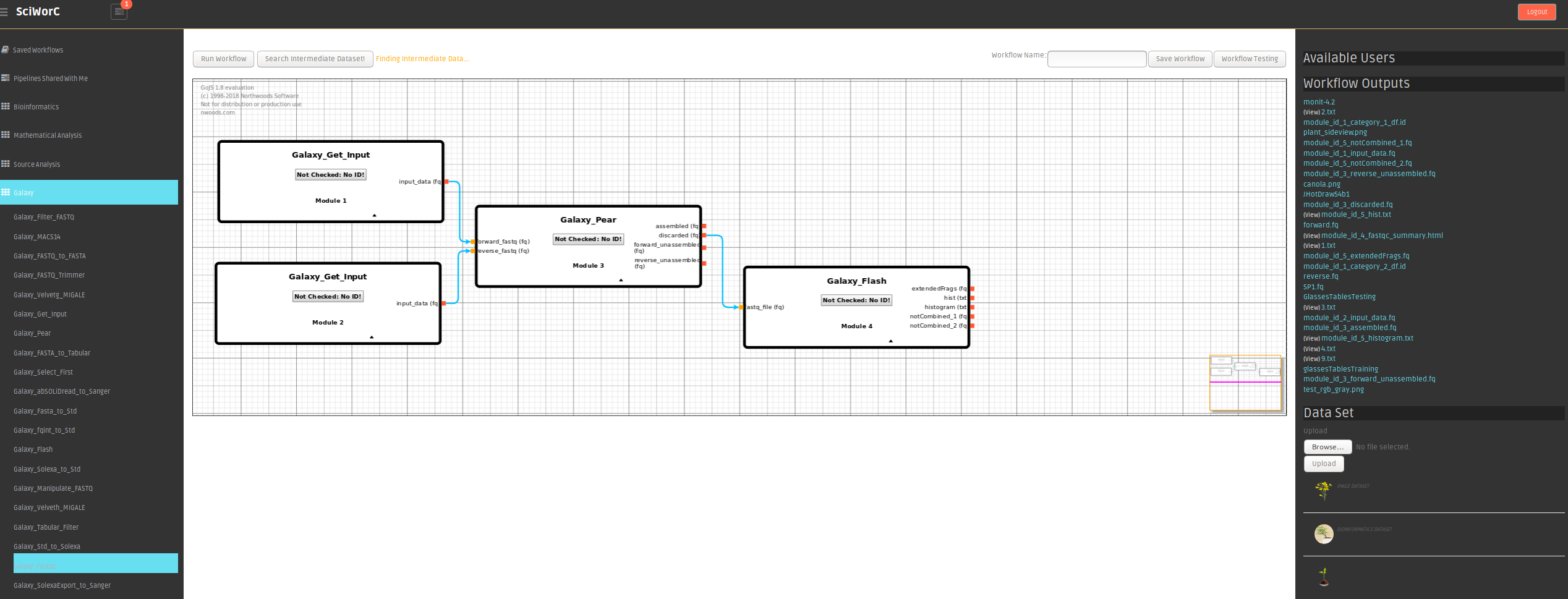}}
\caption{Bioinformatics Workflow 2.}
\label{fig_BI2}
\end{figure}

\begin{figure}[htbp]
\centerline{\includegraphics[width=\textwidth]{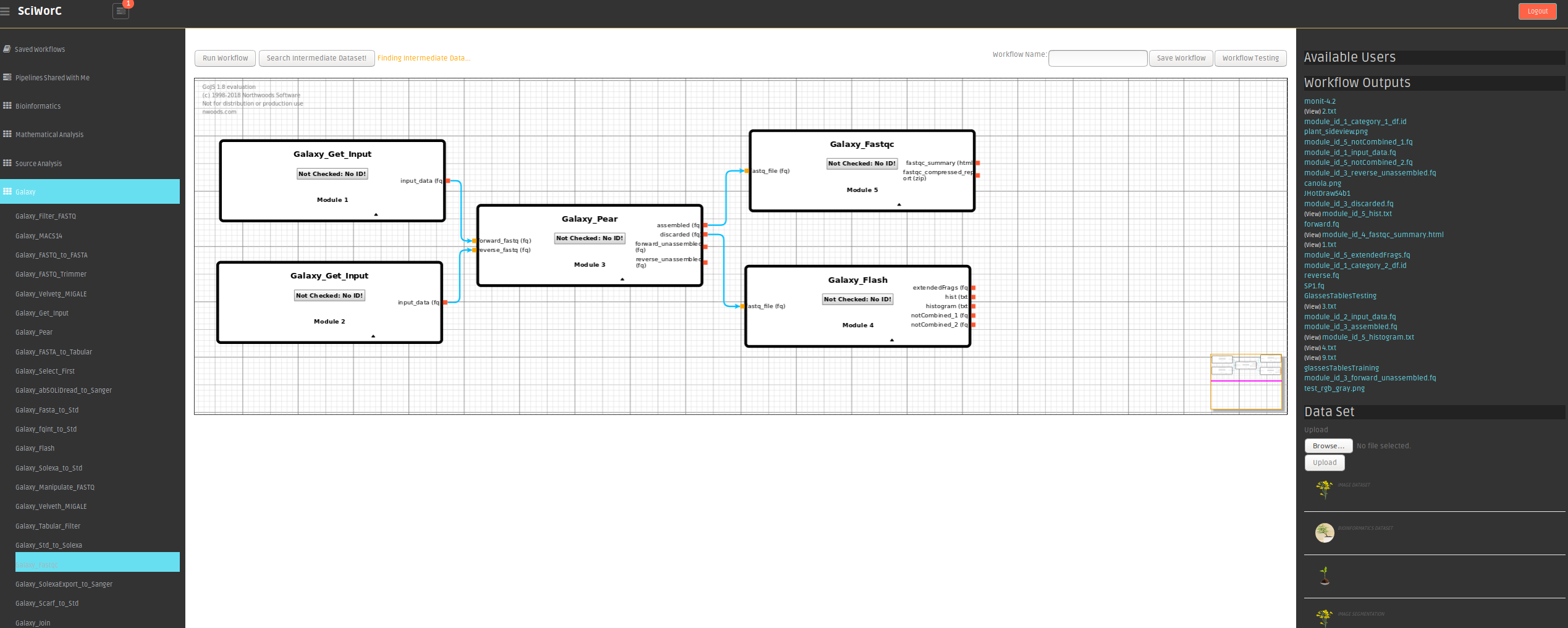}}
\caption{Bioinformatics Workflow 3.}
\label{fig_BI3}
\end{figure}

\begin{figure}[htbp]
\centerline{\includegraphics[width=\textwidth]{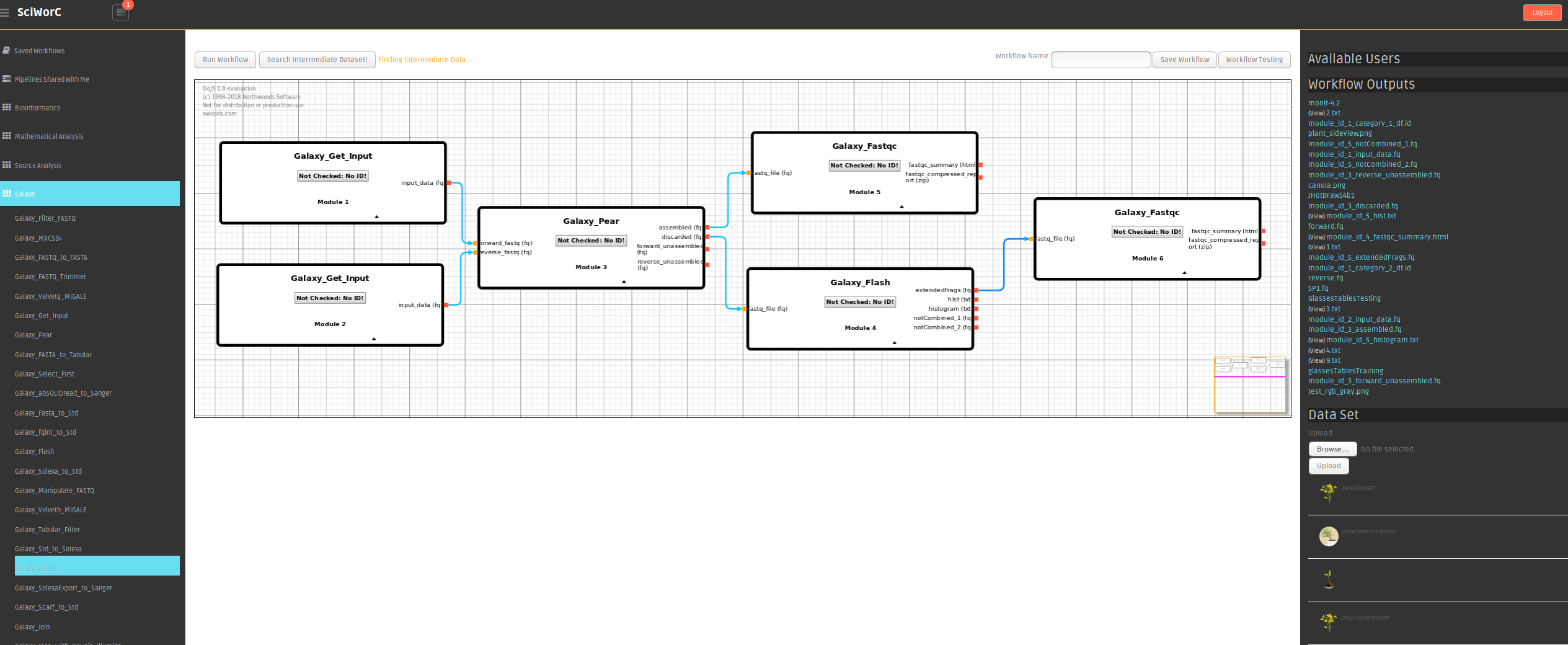}}
\caption{Bioinformatics Workflow 4.}
\label{fig_BI4}
\end{figure}

\end{appendices}


\begin{thebibliography}{100}

\bibitem{Afgan2016TheUpdate}
Enis Afgan, Dannon Baker, Bérénice Batut, Marius van den Beek, Dave Bouvier, Martin Čech, John Chilton, Dave Clements, Nate Coraor, Björn~A Grüning, Aysam Guerler, Jennifer Hillman-Jackson, Saskia Hiltemann, Vahid Jalili, Helena Rasche, Nicola Soranzo, Jeremy Goecks, James Taylor, Anton Nekrutenko, and Daniel Blankenberg.
\newblock The galaxy platform for accessible, reproducible and collaborative biomedical analyses: 2018 update.
\newblock {\em Nucleic Acids Research}, 46:W537--W544, 2018.

\bibitem{agrawal1993mining}
Rakesh Agrawal, Tomasz Imieli{\'n}ski, and Arun Swami.
\newblock Mining association rules between sets of items in large databases.
\newblock In {\em Acm sigmod record}, volume~22, pages 207--216. ACM, 1993.

\bibitem{Agrawal:1993:MAR:170036.170072}
Rakesh Agrawal, Tomasz Imieli\'{n}ski, and Arun Swami.
\newblock Mining association rules between sets of items in large databases.
\newblock {\em SIGMOD Rec.}, 22:207--216, 1993.

\bibitem{Parallel_Processing}
S.~K.~Prasad al.
\newblock Parallel processing over spatial-temporal datasets from geo, bio, climate and social science communities: A research roadmap.
\newblock {\em IEEE International Congress on Big Data (BigData Congress)}, 1:232--250, 2017.

\bibitem{Altintas:2006:PCS:2165554.2165572}
Ilkay Altintas, Oscar Barney, and Efrat Jaeger-Frank.
\newblock Provenance collection support in the kepler scientific workflow system.
\newblock In {\em Proceedings of the 2006 International Conference on Provenance and Annotation of Data}, IPAW'06, pages 118--132, Berlin, Heidelberg, 2006. Springer-Verlag.

\bibitem{Bahsi:2007:CWM:1377549.1377550}
Emir~M. Bahsi, Emrah Ceyhan, and Tevfik Kosar.
\newblock Conditional workflow management: A survey and analysis.
\newblock {\em Sci. Program.}, 15(4):283--297, December 2007.

\bibitem{10.1007/978-3-540-68111-3_78}
Adam Barker and Jano van Hemert.
\newblock Scientific workflow: A survey and research directions.
\newblock In Roman Wyrzykowski, Jack Dongarra, Konrad Karczewski, and Jerzy Wasniewski, editors, {\em Parallel Processing and Applied Mathematics}, pages 746--753, Berlin, Heidelberg, 2008. Springer Berlin Heidelberg.

\bibitem{Big_Data_Usage}
T.~Becker, J.~Cavanillas, E.~Curry, and W.~Wahlster.
\newblock {\em Big Data Usage, New Horizons for a Data-Driven Economy}.
\newblock Springer, 2016.

\bibitem{belhajjame2008metadata}
Khalid Belhajjame, Katy Wolstencroft, Oscar Corcho, Tom Oinn, Franck Tanoh, Alan William, and Carole Goble.
\newblock Metadata management in the taverna workflow system.
\newblock In {\em 2008 Eighth IEEE International Symposium on Cluster Computing and the Grid (CCGRID)}, pages 651--656. IEEE, 2008.

\bibitem{ROSLab}
N.~Bezzo, J.~Park, A.~King, P.~Gebhard, R.~Ivanov, and I.~Lee.
\newblock Demo abstract: Roslab --- a modular programming environment for robotic applications.
\newblock In Acm/ieee International, editor, {\em Conference on Cyber-Physical Systems (ICCPS)}, pages 214--214, Berlin, 2014.

\bibitem{bhuyan2018quality}
Fahima Bhuyan, Shiyong Lu, Robert Reynolds, Ishtiaq Ahmed, and Jia Zhang.
\newblock Quality analysis for scientific workflow provenance access control policies.
\newblock In {\em 2018 IEEE International Conference on Services Computing (SCC)}, pages 261--264. IEEE, 2018.

\bibitem{bioinformatics2011fastqc}
Babraham Bioinformatics.
\newblock Fastqc: a quality control tool for high throughput sequence data.
\newblock {\em Cambridge, UK: Babraham Institute}, 2011.

\bibitem{Experiences_on}
J.~Blomer.
\newblock Experiences on file systems: Which is the best file system for you?
\newblock {\em Journal of Physics: Conference Series}, 664:4, 2015.

\bibitem{Bose:2005:LRS:1057977.1057978}
Rajendra Bose and James Frew.
\newblock Lineage retrieval for scientific data processing: A survey.
\newblock {\em ACM Comput. Surv.}, 37(1):1--28, March 2005.

\bibitem{Brown2007}
Duncan~A. Brown, Patrick~R. Brady, Alexander Dietz, Junwei Cao, Ben Johnson, and John McNabb.
\newblock {\em A Case Study on the Use of Workflow Technologies for Scientific Analysis: Gravitational Wave Data Analysis}, pages 39--59.
\newblock Springer London, London, 2007.

\bibitem{callahan2006vistrails}
Steven~P Callahan, Juliana Freire, Emanuele Santos, Carlos~E Scheidegger, Cl{\'a}udio~T Silva, and Huy~T Vo.
\newblock Vistrails: visualization meets data management.
\newblock In {\em Proceedings of the 2006 ACM SIGMOD international conference on Management of data}, pages 745--747. ACM, 2006.

\bibitem{inproceedingsBigD18}
Debasish Chakroborti, Manishankar Mondal, Banani Roy, Chanchal Roy, and Kevin A.~Schneider.
\newblock Optimized storing of workflow outputs through mining association rules.
\newblock pages 508--515, 12 2018.

\bibitem{Chinthaka2009CBRAssistant}
Eran Chinthaka, Jaliya Ekanayake, David Leake, and Beth Plale.
\newblock Cbr based workflow composition assistant.
\newblock In {\em Proc. of World Congress on Services}, pages 352 -- 355, 2009.

\bibitem{chollet2018keras}
Fran{\c{c}}ois Chollet et~al.
\newblock Keras: The python deep learning library.
\newblock {\em Astrophysics Source Code Library}, 2018.

\bibitem{clifton2012advanced}
Brian Clifton.
\newblock {\em Advanced web metrics with Google Analytics}.
\newblock John Wiley \& Sons, 2012.

\bibitem{Unlocking_the}
F.~Coppens, N.~Wuyts, D.~Inz{\'e}, and S.~Dhondt.
\newblock Unlocking the potential of plant phenotyping data through integration and data-driven approaches.
\newblock {\em Current Opinion in Systems Biology}, 1:58--63, 2017.

\bibitem{crawl2008provenance}
Daniel Crawl and Ilkay Altintas.
\newblock A provenance-based fault tolerance mechanism for scientific workflows.
\newblock In {\em International Provenance and Annotation Workshop}, pages 152--159. Springer, 2008.

\bibitem{Davidson:2008:PSW:1376616.1376772}
Susan~B. Davidson and Juliana Freire.
\newblock Provenance and scientific workflows: Challenges and opportunities.
\newblock In {\em Proceedings of the 2008 ACM SIGMOD International Conference on Management of Data}, SIGMOD '08, pages 1345--1350, New York, NY, USA, 2008. ACM.

\bibitem{4534284}
E.~{Deelman} and A.~{Chervenak}.
\newblock Data management challenges of data-intensive scientific workflows.
\newblock In {\em 2008 Eighth IEEE International Symposium on Cluster Computing and the Grid (CCGRID)}, pages 687--692, May 2008.

\bibitem{Deelman2008DataWorkflows}
E.~Deelman and A.~Chervenak.
\newblock Data management challenges of data-intensive scientific workflows.
\newblock In {\em Proc. of CCGRID}, pages 687--692, 2008.

\bibitem{Deelman2008TheExample}
Ewa Deelman, Gurmeet Singh, Miron Livny, Bruce Berriman, and John Good.
\newblock The cost of doing science on the cloud: The montage example.
\newblock In {\em Proc. of SC}, pages 1--12, 2008.

\bibitem{Deelman:2015:PWM:2775768.2776457}
Ewa Deelman, Karan Vahi, Gideon Juve, Mats Rynge, Scott Callaghan, Philip~J. Maechling, Rajiv Mayani, Weiwei Chen, Rafael Ferreira~da Silva, Miron Livny, and Kent Wenger.
\newblock Pegasus, a workflow management system for science automation.
\newblock {\em Future Gener. Comput. Syst.}, 46(C):17--35, May 2015.

\bibitem{Analysis_of_Six}
B.~Depardon, G.~L. Mahec, and C.~S´eguin.
\newblock Analysis of six distributed file systems.
\newblock {\em [Research Report]}, 1, 2013.

\bibitem{Euro-Par}
F.~Desprez and P.~F. Dutot.
\newblock Euro-par 2016: Parallel processing workshops.
\newblock {\em Euro-Par}, 2016:24--26, August 2016.

\bibitem{Dias:2011:SDP:2110497.2110502}
Jonas Dias, Eduardo Ogasawara, Daniel de~Oliveira, Fabio Porto, Alvaro~L.G.A. Coutinho, and Marta Mattoso.
\newblock Supporting dynamic parameter sweep in adaptive and user-steered workflow.
\newblock In {\em Proceedings of the 6th Workshop on Workflows in Support of Large-scale Science}, WORKS '11, pages 31--36, New York, NY, USA, 2011. ACM.

\bibitem{Testing_of_several}
G.~Donvito, G.~Marzulli, and D.~Diacono.
\newblock Testing of several distributed file-systems (hdfs, ceph and glusterfs) for supporting the hep experiments analysis.
\newblock {\em Journal of Physics: Conference Series}, 513:4, 2014.

\bibitem{10.1007/11890850_2}
Juliana Freire, Cl{\'a}udio~T. Silva, Steven~P. Callahan, Emanuele Santos, Carlos~E. Scheidegger, and Huy~T. Vo.
\newblock Managing rapidly-evolving scientific workflows.
\newblock In Luc Moreau and Ian Foster, editors, {\em Provenance and Annotation of Data}, pages 10--18, Berlin, Heidelberg, 2006. Springer Berlin Heidelberg.

\bibitem{Friedman:2002:CWS:584490.584508}
Roy Friedman.
\newblock Caching web services in mobile ad-hoc networks: Opportunities and challenges.
\newblock In {\em Proceedings of the Second ACM International Workshop on Principles of Mobile Computing}, POMC '02, pages 90--96, New York, NY, USA, 2002. ACM.

\bibitem{GARG2015256}
Ritu Garg and Awadhesh~Kumar Singh.
\newblock Adaptive workflow scheduling in grid computing based on dynamic resource availability.
\newblock {\em Engineering Science and Technology, an International Journal}, 18(2):256 -- 269, 2015.

\bibitem{georgakopoulos1995overview}
Diimitrios Georgakopoulos, Mark Hornick, and Amit Sheth.
\newblock An overview of workflow management: From process modeling to workflow automation infrastructure.
\newblock {\em Distributed and parallel Databases}, 3(2):119--153, 1995.

\bibitem{4404805}
Y.~{Gil}, E.~{Deelman}, M.~{Ellisman}, T.~{Fahringer}, G.~{Fox}, D.~{Gannon}, C.~{Goble}, M.~{Livny}, L.~{Moreau}, and J.~{Myers}.
\newblock Examining the challenges of scientific workflows.
\newblock {\em Computer}, 40(12):24--32, Dec 2007.

\bibitem{Gil:2011:MYM:2063076.2063082}
Yolanda Gil, Pedro Szekely, Sandra Villamizar, Thomas~C. Harmon, Varun Ratnakar, Shubham Gupta, Maria Muslea, Fabio Silva, and Craig~A. Knoblock.
\newblock Mind your metadata: Exploiting semantics for configuration, adaptation, and provenance in scientific workflows.
\newblock In {\em Proceedings of the 10th International Conference on The Semantic Web - Volume Part II}, ISWC'11, pages 65--80, Berlin, Heidelberg, 2011. Springer-Verlag.

\bibitem{goble2010myexperiment}
Carole~A Goble, Jiten Bhagat, Sergejs Aleksejevs, Don Cruickshank, Danius Michaelides, David Newman, Mark Borkum, Sean Bechhofer, Marco Roos, Peter Li, et~al.
\newblock myexperiment: a repository and social network for the sharing of bioinformatics workflows.
\newblock {\em Nucleic acids research}, 38(suppl\_2):W677--W682, 2010.

\bibitem{Goodman2014TenData}
Alyssa Goodman, Alberto Pepe, Alexander~W. Blocker, Christine~L. Borgman, Kyle Cranmer, Merce Crosas, Rosanne Di~Stefano, Yolanda Gil, Paul Groth, Margaret Hedstrom, David~W. Hogg, Vinay Kashyap, Ashish Mahabal, Aneta Siemiginowska, and Aleksandra Slavkovic.
\newblock Ten simple rules for the care and feeding of scientific data.
\newblock {\em PLOS Computational Biology}, 10:1--5, 2014.

\bibitem{Gray:2005:SDM:1107499.1107503}
Jim Gray, David~T. Liu, Maria Nieto-Santisteban, Alex Szalay, David~J. DeWitt, and Gerd Heber.
\newblock Scientific data management in the coming decade.
\newblock {\em SIGMOD Rec.}, 34(4):34--41, December 2005.

\bibitem{halili2008apache}
Emily~H Halili.
\newblock {\em Apache JMeter: A practical beginner's guide to automated testing and performance measurement for your websites}.
\newblock Packt Publishing Ltd, 2008.

\bibitem{Hampton2013BigEcology}
Stephanie Hampton, Carly Strasser, Joshua Tewksbury, Wendy Gram, Amber Budden, Archer Batcheller, Clifford Duke, and John H~Porter.
\newblock Big data and the future of ecology.
\newblock {\em Frontiers in Ecology and the Environment}, 11:156--162, 2013.

\bibitem{Signal_Processing}
Z.~Han and M.~Hong.
\newblock {\em Signal Processing and Networking for Big Data Applications}.
\newblock Cambridge University Press, Apr 27, 2017.

\bibitem{statistical_models}
J.~Heit, J.~Liu, and M.~Shah.
\newblock An architecture for the deployment of statistical models for the big data era.
\newblock In Ieee International, editor, {\em Conference on Big Data}, pages 1377--1384, 2016.

\bibitem{hollingsworth1995workflow}
David Hollingsworth and UK~Hampshire.
\newblock Workflow management coalition: The workflow reference model.
\newblock {\em Document Number TC00-1003}, 19:16, 1995.

\bibitem{Cloud_Resource_Management}
A.~S. Kaseb, A.~Mohan, and Y.~H. Lu.
\newblock Cloud resource management for image and video analysis of big data from network cameras.
\newblock In {\em JInternational Conference on Cloud Computing and Big Data (CCBD)}, pages 287--294, 2015.

\bibitem{Development_of}
M.~Kim, J.~Choi, and J.~Yoon.
\newblock Development of the big data management system on national virtual power plant.
\newblock In {\em 10th International Conference on P2P}, pages 100--107, Grid, Cloud and Internet Computing (3PGCIC)., 2015. Parallel.

\bibitem{Koop2008VisComplete:Pipelines}
D.~Koop, C.~E. Scheidegger, S.~P. Callahan, J.~Freire, and C.~T. Silva.
\newblock Viscomplete: Automating suggestions for visualization pipelines.
\newblock {\em IEEE Transactions on Visualization and Computer Graphics}, 14:1691--1698, 2008.

\bibitem{10.1007/978-3-540-85502-6_18}
David Leake and Joseph Kendall-Morwick.
\newblock Towards case-based support for e-science workflow generation by mining provenance.
\newblock In Klaus-Dieter Althoff, Ralph Bergmann, Mirjam Minor, and Alexandre Hanft, editors, {\em Advances in Case-Based Reasoning}, pages 269--283, Berlin, Heidelberg, 2008. Springer Berlin Heidelberg.

\bibitem{lennon2009introduction}
Joe Lennon.
\newblock Introduction to couchdb.
\newblock In {\em Beginning CouchDB}, pages 3--9. Springer, 2009.

\bibitem{Distributed_metadata}
B.~Li, Y.~He, and K.~Xu.
\newblock Distributed metadata management scheme in cloud computing.
\newblock In {\em 6th International Conference on Pervasive Computing and Applications}, pages 32--38. Port Elizabeth, 2011.

\bibitem{4782949}
C.~{Lin}, S.~{Lu}, X.~{Fei}, A.~{Chebotko}, D.~{Pai}, Z.~{Lai}, F.~{Fotouhi}, and J.~{Hua}.
\newblock A reference architecture for scientific workflow management systems and the view soa solution.
\newblock {\em IEEE Transactions on Services Computing}, 2(1):79--92, Jan 2009.

\bibitem{Liu:2015:SDS:2884709.2884750}
Ji~Liu, Esther Pacitti, Patrick Valduriez, and Marta Mattoso.
\newblock A survey of data-intensive scientific workflow management.
\newblock {\em J. Grid Comput.}, 13(4):457--493, December 2015.

\bibitem{liu2007efficient}
Xin Liu and Ralph Deters.
\newblock An efficient dual caching strategy for web service-enabled pdas.
\newblock In {\em Proceedings of the 2007 ACM symposium on Applied computing}, pages 788--794. ACM, 2007.

\bibitem{lu2009collaborative}
Shiyong Lu and Jia Zhang.
\newblock Collaborative scientific workflows.
\newblock In {\em 2009 IEEE International Conference on Web Services}, pages 527--534. IEEE, 2009.

\bibitem{ludascher2006scientific}
Bertram Lud{\"a}scher, Ilkay Altintas, Chad Berkley, Dan Higgins, Efrat Jaeger, Matthew Jones, Edward~A Lee, Jing Tao, and Yang Zhao.
\newblock Scientific workflow management and the kepler system.
\newblock {\em Concurrency and Computation: Practice and Experience}, 18(10):1039--1065, 2006.

\bibitem{Development_of_a_knowledge}
L.~N. Luyen, A.~Tireau, A.~Venkatesan, P.~Neveu, and P.~Larmande.
\newblock Development of a knowledge system for big data: Case study to plant phenotyping data.
\newblock In {\em Proceedings of the 6th International Conference on Web Intelligence. ACM, , NY, USA, Article 27, 9 pages}, 2016.

\bibitem{magovc2011flash}
Tanja Mago{\v{c}} and Steven~L Salzberg.
\newblock Flash: fast length adjustment of short reads to improve genome assemblies.
\newblock {\em Bioinformatics}, 27(21):2957--2963, 2011.

\bibitem{meng2016mllib}
Xiangrui Meng, Joseph Bradley, Burak Yavuz, Evan Sparks, Shivaram Venkataraman, Davies Liu, Jeremy Freeman, DB~Tsai, Manish Amde, Sean Owen, et~al.
\newblock Mllib: Machine learning in apache spark.
\newblock {\em The Journal of Machine Learning Research}, 17(1):1235--1241, 2016.

\bibitem{Bottleneck}
M.~Minervini, H.~Scharr, and S.~Tsaftaris.
\newblock Image analysis: The new bottleneck in plant phenotyping [applications corner].
\newblock {\em IEEE Signal Processing Magazine}, 32(4):126--131, 2015.

\bibitem{Application-aware}
M.~Minervini and S.~A. Tsaftaris.
\newblock Application-aware image compression for low cost and distributed plant phenotyping.
\newblock In {\em 18th International Conference on Digital Signal Processing (DSP)}, pages 1--6, 2013.

\bibitem{Missier:2016:PDD:2915589.2915594}
Paolo Missier, Simon Woodman, Hugo Hiden, and Paul Watson.
\newblock Provenance and data differencing for workflow reproducibility analysis.
\newblock {\em Concurr. Comput. : Pract. Exper.}, 28:995--1015, 2016.

\bibitem{Missier2016ProvenanceAnalysis}
Paolo Missier, Simon Woodman, Hugo Hiden, and Paul Watson.
\newblock {Provenance and data differencing for workflow reproducibility analysis}.
\newblock In {\em Concurrency Computation}, 2016.

\bibitem{Software_Architecture}
I.~Mistrik and R.~Bahsoon.
\newblock {\em Software Architecture for Big Data and the Cloud}.
\newblock Jun 12, 2017.

\bibitem{Micro-level}
A.~K. Mondal, B.~Roy C.~K. Roy, and K.~A. Schneider.
\newblock Micro-level modularity of computaion-intensive programs in big data platforms: A case study with image data.
\newblock Technical report, Technical Report, University of Saskatchewan. Canada.

\bibitem{mouallem2010fault}
Pierre Mouallem, Daniel Crawl, Ilkay Altintas, Mladen Vouk, and Ustun Yildiz.
\newblock A fault-tolerance architecture for kepler-based distributed scientific workflows.
\newblock In {\em International Conference on Scientific and Statistical Database Management}, pages 452--460. Springer, 2010.

\bibitem{Muniswamy-Reddy:2006:PSS:1267359.1267363}
Kiran-Kumar Muniswamy-Reddy, David~A. Holland, Uri Braun, and Margo Seltzer.
\newblock Provenance-aware storage systems.
\newblock In {\em Proceedings of the Annual Conference on USENIX '06 Annual Technical Conference}, ATEC '06, pages 4--4, Berkeley, CA, USA, 2006. USENIX Association.

\bibitem{nedelcu2010nginx}
Cl{\'e}ment Nedelcu.
\newblock {\em Nginx HTTP Server: Adopt Nginx for Your Web Applications to Make the Most of Your Infrastructure and Serve Pages Faster Than Ever}.
\newblock Packt Publishing Ltd, 2010.

\bibitem{oinn2004taverna}
Tom Oinn, Matthew Addis, Justin Ferris, Darren Marvin, Martin Senger, Mark Greenwood, Tim Carver, Kevin Glover, Matthew~R Pocock, Anil Wipat, et~al.
\newblock Taverna: a tool for the composition and enactment of bioinformatics workflows.
\newblock {\em Bioinformatics}, 20(17):3045--3054, 2004.

\bibitem{oliphant2007python}
Travis~E Oliphant.
\newblock Python for scientific computing.
\newblock {\em Computing in Science \& Engineering}, 9(3):10--20, 2007.

\bibitem{Towards_Multi-site}
L.~Pineda-Morales, A.~Costan, and G.~Antoniu.
\newblock Towards multi-site metadata management for geographically distributed cloud workflows.
\newblock In Ieee International, editor, {\em Conference on Cluster Computing}, pages 294--303, IL, 2015.

\bibitem{Managing_hot_metadata}
L.~Pineda-Morales, J.~Liu, A.~Costan, E.~Pacitti, G.~Antoniu, P.~Valduriez, and M.~Mattoso.
\newblock Managing hot metadata for scientific workflows on multisite clouds.
\newblock In Ieee International, editor, {\em Conference on Big Data (Big Data)}, pages 390--397, 2016.

\bibitem{Prodan:2005:DSS:1066677.1066835}
Radu Prodan and Thomas Fahringer.
\newblock Dynamic scheduling of scientific workflow applications on the grid: A case study.
\newblock In {\em Proceedings of the 2005 ACM Symposium on Applied Computing}, SAC '05, pages 687--694, New York, NY, USA, 2005. ACM.

\bibitem{article34636}
Arcot Rajasekar, Mike Wan, Reagan Moore, and Wayne Schroeder.
\newblock A prototype rule-based distributed data management system.
\newblock 01 2006.

\bibitem{REIJERS2016126}
H.A. Reijers, I.~Vanderfeesten, and W.M.P. van~der Aalst.
\newblock The effectiveness of workflow management systems: A longitudinal study.
\newblock {\em International Journal of Information Management}, 36(1):126 -- 141, 2016.

\bibitem{Towards_a_Reference_Architecture}
B.~Roy, A.~K. Mondal, C.~K. Roy, K.~A. Schneider, and K.~Wazed.
\newblock Towards a reference architecture for cloud-based plant genotyping and phenotyping analysis frameworks.
\newblock In Ieee International, editor, {\em Conference on Software Architecture (ICSA), Gothenburg}, pages 41--50, 2017.

\bibitem{scheidegger2008querying}
Carlos~E Scheidegger, Huy~T Vo, David Koop, Juliana Freire, and Claudio~T Silva.
\newblock Querying and re-using workflows with vstrails.
\newblock In {\em Proceedings of the 2008 ACM SIGMOD international conference on Management of data}, pages 1251--1254. ACM, 2008.

\bibitem{SEFFINO1999105}
Laura~A Seffino, Claudia~Bauzer Medeiros, Jansle~V Rocha, and Bei Yi.
\newblock woodss — a spatial decision support system based on workflows.
\newblock {\em Decision Support Systems}, 27(1):105 -- 123, 1999.

\bibitem{sevcik2005defining}
Peter Sevcik.
\newblock Defining the application performance index.
\newblock {\em Business Communications Review}, 20, 2005.

\bibitem{SHAHZAD2016100}
Farrukh Shahzad, Tarek~R. Sheltami, Elhadi~M. Shakshuki, and Omar Shaikh.
\newblock A review of latest web tools and libraries for state-of-the-art visualization.
\newblock {\em Procedia Computer Science}, 98:100 -- 106, 2016.
\newblock The 7th International Conference on Emerging Ubiquitous Systems and Pervasive Networks (EUSPN 2016)/The 6th International Conference on Current and Future Trends of Information and Communication Technologies in Healthcare (ICTH-2016)/Affiliated Workshops.

\bibitem{shvachko2010hadoop}
Konstantin Shvachko, Hairong Kuang, Sanjay Radia, Robert Chansler, et~al.
\newblock The hadoop distributed file system.
\newblock In {\em MSST}, volume~10, pages 1--10, 2010.

\bibitem{Simmhan:2005:SDP:1084805.1084812}
Yogesh~L. Simmhan, Beth Plale, and Dennis Gannon.
\newblock A survey of data provenance in e-science.
\newblock {\em SIGMOD Rec.}, 34(3):31--36, September 2005.

\bibitem{sindrilaru2010fault}
Elvin Sindrilaru, Alexandru Costan, and Valentin Cristea.
\newblock Fault tolerance and recovery in grid workflow management systems.
\newblock In {\em 2010 international conference on complex, intelligent and software intensive systems}, pages 475--480. IEEE, 2010.

\bibitem{Machine_Learning}
A.~Singh, B.~Ganapathysubramanian, A.~K. Singh, and S.~Sarkar.
\newblock Machine learning for high-throughput stress phenotyping in plants.
\newblock {\em Trends in Plant Science}, 21(2):1360--1385, 2016.

\bibitem{iPlant_atmosphere}
E.~Skidmore, S.~Kim, S.~Kuchimanchi, S.~Singaram, N.~Merchant, and D.~Stanzione.
\newblock {\em iPlant atmosphere: a gateway to cloud infrastructure for the plant sciences}.
\newblock In Proceedings of the 2011 ACM workshop on Gateway computing environments. ACM, New York, NY, USA, 59-64, 2011.

\bibitem{Big_Metadata}
K.~Smith, L.~Seligman, A.~Rosenthal, C.~Kurcz, M.~Greer, C.~Macheret, M.~Sexton, and A.~Eckstein.
\newblock {\em "Big Metadata": The Need for Principled Metadata Management in Big Data Ecosystems}.
\newblock In Proceedings of Workshop on Data analytics in the Cloud. ACM, New York, NY, USA, Article 13, 4 pages, 2014.

\bibitem{Spjuth2015ExperiencesBioinformatics}
Ola Spjuth, Erik Bongcam-Rudloff, Guillermo~Carrasco Hern{\'a}ndez, Lukas Forer, Mario Giovacchini, Roman~Valls Guimera, Aleksi Kallio, Eija Korpelainen, Maciej~M. Ka{\'{n}}du{\l}a, Milko Krachunov, David~P. Kreil, Ognyan Kulev, Pawe{\l}~P. {\L}abaj, Samuel Lampa, Luca Pireddu, Sebastian Sch{\"o}nherr, Alexey Siretskiy, and Dimitar Vassilev.
\newblock Experiences with workflows for automating data-intensive bioinformatics.
\newblock {\em Biology Direct}, 10:43, 2015.

\bibitem{USING_MODULAR_PROGRAMMING}
W.~Sun, X.~Wang, and X.~Sun.
\newblock Ac 2012-3155: using modular programming strategy to prac- tice computer programming: a case study.
\newblock {\em American Society for Engineering Education}, 1, 2012.

\bibitem{talia2013workflow}
Domenico Talia.
\newblock Workflow systems for science: Concepts and tools.
\newblock {\em ISRN Software Engineering}, 2013, 2013.

\bibitem{terry2003caching}
Douglas~B Terry and Venugopalan Ramasubramanian.
\newblock Caching xml web services for mobility.
\newblock {\em ACM Queue}, 1(3):70--78, 2003.

\bibitem{truty2006systems}
Gregory~Louis Truty.
\newblock Systems and methods for adjusting caching policies for web service requests, August~15 2006.
\newblock US Patent 7,093,073.

\bibitem{Data_Multiverse}
R.~Tudoran, B.~Nicolae, and G.~Brasche.
\newblock {\em Data Multiverse: The Uncertainty Challenge of Future Big Data Analytics}, volume 10151.
\newblock Semantic Keyword-Based Search on Structured Data Sources. Lecture Notes in Computer Science, Springer, Cham, 2017.

\bibitem{van2004workflow}
Wil Van Der~Aalst, Kees~Max Van~Hee, and Kees van Hee.
\newblock {\em Workflow management: models, methods, and systems}.
\newblock MIT press, 2004.

\bibitem{vavilapalli2013apache}
Vinod~Kumar Vavilapalli, Arun~C Murthy, Chris Douglas, Sharad Agarwal, Mahadev Konar, Robert Evans, Thomas Graves, Jason Lowe, Hitesh Shah, Siddharth Seth, et~al.
\newblock Apache hadoop yarn: Yet another resource negotiator.
\newblock In {\em Proceedings of the 4th annual Symposium on Cloud Computing}, page~5. ACM, 2013.

\bibitem{Plant_phenotyping}
A.~Walter, F.~Liebisch, and A.~Hund.
\newblock Plant phenotyping: from bean weighing to image analysis.
\newblock {\em Plant Methods}, 11:1, 2015.

\bibitem{Hadoop_high}
F.~Wang, J.~Qiu, J.~Yang, B.~Dong, X.~Li, and Y.~Li.
\newblock {\em Hadoop high availability through metadata replication}.
\newblock In Proceedings of the first international workshop on Cloud data management. ACM, New York, 37-44, 2009.

\bibitem{Wang:2009:KHG:1645164.1645176}
Jianwu Wang, Daniel Crawl, and Ilkay Altintas.
\newblock Kepler + hadoop: A general architecture facilitating data-intensive applications in scientific workflow systems.
\newblock In {\em Proceedings of the 4th Workshop on Workflows in Support of Large-Scale Science}, WORKS '09, pages 12:1--12:8, New York, NY, USA, 2009. ACM.

\bibitem{White2013NineData}
Ethan~P. White, Elita Baldridge, Zachary~T. Brym, Kenneth~J. Locey, Daniel~J. McGlinn, and Sarah~R. Supp.
\newblock Nine simple ways to make it easier to (re)use your data.
\newblock {\em PeerJ PrePrints}, 1:e7v2, 2013.

\bibitem{white2012hadoop}
Tom White.
\newblock {\em Hadoop: The definitive guide}.
\newblock " O'Reilly Media, Inc.", 2012.

\bibitem{Woodman2015WorkflowProvenance}
Simon Woodman, Hugo Hiden, and Paul Watson.
\newblock Workflow provenance: An analysis of long term storage costs.
\newblock In {\em Proc. of WORKS}, pages 1--9, 2015.

\bibitem{Woodman2011AchievingVersioning}
Simon Woodman, Hugo Hiden, Paul Watson, and Paolo Missier.
\newblock Achieving reproducibility by combining provenance with service and workflow versioning.
\newblock In {\em Proceedings of the 6th Workshop on Workflows in Support of Large-scale Science}, WORKS '11, pages 127--136, New York, NY, USA, 2011. ACM.

\bibitem{wu2012}
Qishi Wu, Mengxia Zhu, Yi~Gu, Patrick Brown, Xukang Lu, Wuyin Lin, and Yangang Liu.
\newblock A distributed workflow management system with case study of real-life scientific applications on grids.
\newblock {\em Journal of Grid Computing}, 10(3):367--393, Sep 2012.

\bibitem{A_Leaf_Recognition}
S.~G. Wu, F.~S. Bao, E.~Y. Xu, Y.~X. Wang, Y.~F. Chang, and C.~L. Shiang.
\newblock A leaf recognition algorithm for plant classification using probabilistic neural network.
\newblock In {\em IEEE 7th International Symposium on Signal Processing and Information Technology}, Cario, Egypt, 2007.

\bibitem{Visualization_and_Adaptive}
X.~Yang, S.~Liu, K.~Feng, S.~Zhou, and X.~H. Sun.
\newblock Visualization and adaptive subsetting of earth science data in hdfs: A novel data analysis strategy with hadoop and spark.
\newblock {\em IEEE International Conferences on Big Data and Cloud Computing (BDCloud).}, 1:89--96, 2016.

\bibitem{Yuan:2010:DPS:1838759.1838814}
Dong Yuan, Yun Yang, Xiao Liu, and Jinjun Chen.
\newblock A data placement strategy in scientific cloud workflows.
\newblock {\em Future Gener. Comput. Syst.}, 26:1200--1214, 2010.

\bibitem{Yuan2011On-demandSystems}
Dong Yuan, Yun Yang, Xiao Liu, and Jinjun Chen.
\newblock On-demand minimum cost benchmarking for intermediate dataset storage in scientific cloud workflow systems.
\newblock {\em J. Parallel Distrib. Comput.}, 71:316--332, 2011.

\bibitem{zaharia2010spark}
Matei Zaharia, Mosharaf Chowdhury, Michael~J Franklin, Scott Shenker, and Ion Stoica.
\newblock Spark: Cluster computing with working sets.
\newblock {\em HotCloud}, 10(10-10):95, 2010.

\bibitem{zhang2012confucius}
Jia Zhang, Daniel Kuc, and Shiyong Lu.
\newblock Confucius: A tool supporting collaborative scientific workflow composition.
\newblock {\em IEEE Transactions on Services Computing}, 7(1):2--17, 2012.

\bibitem{Zhang2011Recommend-as-you-go:Reuse}
Jia Zhang, Wei Tan, Alexander John, Ian Foster, and Ravi Madduri.
\newblock {Recommend-as-you-go: A novel approach supporting services-oriented scientific workflow reuse}.
\newblock In {\em Proc. of SCC}, pages 48 -- 55, 2011.

\bibitem{zhang2013pear}
Jiajie Zhang, Kassian Kobert, Tom{\'a}{\v{s}} Flouri, and Alexandros Stamatakis.
\newblock Pear: a fast and accurate illumina paired-end read merger.
\newblock {\em Bioinformatics}, 30(5):614--620, 2013.

\bibitem{zhao2004semantically}
Jun Zhao, Carole Goble, Robert Stevens, and Sean Bechhofer.
\newblock Semantically linking and browsing provenance logs for e-science.
\newblock In {\em Semantics of a Networked World. Semantics for Grid Databases}, pages 158--176. 2004.

\bibitem{zhao2011opportunities}
Yong Zhao, Xubo Fei, Ioan Raicu, and Shiyong Lu.
\newblock Opportunities and challenges in running scientific workflows on the cloud.
\newblock In {\em 2011 International Conference on Cyber-Enabled Distributed Computing and Knowledge Discovery}, pages 455--462. IEEE, 2011.

\bibitem{Zheng:2015:ICW:2755979.2755984}
Charles Zheng and Douglas Thain.
\newblock Integrating containers into workflows: A case study using makeflow, work queue, and docker.
\newblock In {\em Proceedings of the 8th International Workshop on Virtualization Technologies in Distributed Computing}, VTDC '15, pages 31--38, New York, NY, USA, 2015. ACM.

\bibitem{Zimmermann:2004:MVH:998675.999460}
Thomas Zimmermann, Peter Weisgerber, Stephan Diehl, and Andreas Zeller.
\newblock Mining version histories to guide software changes.
\newblock In {\em Proc. of ICSE}, pages 563--572, 2004.

\bibitem{772961}
M.~{zur Muhlen}.
\newblock Evaluation of workflow management systems using meta models.
\newblock In {\em Proceedings of the 32nd Annual Hawaii International Conference on Systems Sciences. 1999. HICSS-32. Abstracts and CD-ROM of Full Papers}, volume Track5, pages 11 pp.--, Jan 1999.

\bibitem{Exploration_of_modularity}
A.~M. Şuti, M.~V.~D. Brand, and T.~Verhoeff.
\newblock Exploration of modularity and reusability of domain-specific languages: an expression dsl in metamod.
\newblock {\em Computer Languages, Systems and Structures}, 1:1477--8424.

\end{thebibliography}
\end{document}